\documentclass[12pt]{article}
\usepackage{undertilde,subfigure}
\usepackage{mathrsfs}
\usepackage{natbib,graphicx,setspace,lscape,longtable,epstopdf,xr}
\usepackage{mathrsfs,amsmath,amsthm,amssymb,color,subfigure}
\usepackage{natbib,epsfig,graphicx,pdfpages}%, mathabx, verbatim}

\usepackage{float}
\usepackage[ruled]{algorithm2e}
\usepackage{caption}
\usepackage{multirow}
\usepackage{mathtools}
\DeclarePairedDelimiter\floor{\lfloor}{\rfloor}
\DeclareMathOperator*{\argmax}{argmax} % thin space, limits underneath in displaysc
\DeclareMathOperator*{\argmin}{argmin} % thin space, limits underneath in displays
\DeclareMathOperator*{\median}{median} % thin space, limits underneath in displays

\bibpunct{(}{)}{;}{a}{,}{,}

\AtBeginDocument{%
\addtolength\abovedisplayskip{-0.2\baselineskip}%
\addtolength\belowdisplayskip{-0.2\baselineskip}%
}

\newtheorem{theorem}{Theorem}
\newtheorem{lemma}{Lemma}
\newtheorem{proposition}{Proposition}

\newtheorem{definition}{Definition}

\newcommand{\csection}[1]
{\begin{center}
	\stepcounter{section}
	{\bf\large\arabic{section}. #1}
\end{center}
}

\def\1{{\bf 1}}

\def\nk{n_{\rm kt}}
\def\br{\rm br}

\def\beq{\begin{equation}}
\def\eeq{\end{equation}}
\def\beqr{\begin{eqnarray}}
\def\eeqr{\end{eqnarray}}
\def\beqrs{\begin{eqnarray*}}
\def\eeqrs{\end{eqnarray*}}
\def\bet{\begin{theorem}}
\def\eet{\end{theorem}}
\def\bel{\begin{lemma}}
\def\eel{\end{lemma}}
\def\bep{\begin{proposition}}
\def\eep{\end{proposition}}
\def\bg{\begin{figure}[tbph]\begin{center}}
	\def\eg{\end{center}\end{figure}}

\def\bc{\begin{center}}
\def\ec{\end{center}}

\def\wt{\widetilde}
\def\wh{\widehat}

\def\ol{\overline }

\def\E{\mathbb{E}}
\def\bB{\mathbf{B}}

\def\mN{\mathcal{N}}

\def\mG{\mathbb G}
\def\mR{\mathbb{R}}
\def\mI{\mathcal I}
\def\mS{\mathbb S}
\def\mH{\mathcal H}
\def\mM{\mathcal M}

\def\mS{\mathcal S}

\newcommand{\balpha}{\boldsymbol{\alpha}}
\newcommand{\bbeta}{\boldsymbol{\beta}}

\def\0s{\mathbf{0}}
\def\wh{\widehat}

\def\ol{\overline }

\def\E{\mathbb{E}}
\def\bB{\mathbf{B}}

\def\mN{\mathcal{N}}

\def\mG{\mathscr{G}}
\def\mR{\mathbb{R}}
\def\mI{\mathcal I}
\def\mS{\mathbb S}
\def\mH{\mathcal H}
\def\mM{\mathcal M}

\def\mS{\mathcal S}

\def\mC{\mathcal{C}}

\def\bse{\begin{eqnarray*}}
\def\ese{\end{eqnarray*}}
\def\be{\begin{eqnarray}}
\def\ee{\end{eqnarray}}
\def\bsq{\begin{equation*}}
\def\esq{\end{equation*}}
\def\bq{\begin{equation}}
\def\eq{\end{equation}}

\def\B{{\bf B}}

\def\Ct{{C_1}}

\def\k{{\bf k}}

\def\e{{\bf e}}

\def\H{{\bf H}}
\def\I{{\bf I}}

\def\x{\bm{x}}

\def\w{\bm{w}}

\def\0{{\bf 0}}

\def\x{{\bf x}}
\def\Z{{\bf Z}}

\def\wh{\widehat}

\def\bphi{\boldsymbol\phi}
\def\bvarphi{\boldsymbol\varphi}
\def\bpi{\boldsymbol\pi}

\def\bpsi{\boldsymbol\psi}

\def\bOmega{\boldsymbol\Omega}
\def\bg{\boldsymbol\gamma}

\def\bv{\mathbf v}
\def\zero{\mathbf 0}

\def\boxit#1{\vbox{\hrule\hbox{\vrule\kern6pt\vbox{\kern6pt#1\kern6pt}\kern6pt\vrule}\hrule}}

\newcommand{\EE}{\mbox{$\mathbb E$}}

\newcommand{\llog}{\mathrm{log}}

\def\bSigma{\boldsymbol{\Sigma}}
\newcommand{\ut}[1]{\underset{\widetilde{}}{#1}}
\newcommand{\btheta}{\boldsymbol{\theta}}
\def\mbR{\mathbb{R}}
\newcommand{\bfTheta}{\boldsymbol{\Theta}}
\def\bTheta{\boldsymbol{\Theta}}

\def\mbG{\mathbb{G}}
\def\mbW{\mathbb{W}}
\newcommand{\bPhi}{\boldsymbol{\Phi}}
\newtheorem{assumption}[definition]{Assumption}

\def\LIC{\mbox{LIC}}

\def\bw{\mathbf{w}}
\def\w{\mathbf{w}}

\usepackage[normalem]{ulem}
\begin{document}

\begin{center}
	{\bf\Large
		Group Network Hawkes Process
	}\\
	
	\bigskip

	Guanhua Fang$^1$,
	Ganggang Xu$^2$,
	Haochen Xu$^1$,
	Xuening Zhu$^1$, and
	Yongtao Guan$^{3,4}$
	
	{\it\small
		$^1$Fudan University, Shanghai, China;		
		$^2$University of Miami, USA;
$^3$The Chinese University of Hong Kong, Shenzhen, China;
$^4$Shenzhen Research Institute of Big Data, Shenzhen, China

	}
	
\end{center}

\begin{footnotetext}[1]
	{Guanhua Fang and Ganggang Xu are joint first authors.
		Xuening Zhu ({\it xueningzhu@fudan.edu.cn}) is the corresponding author.
	}
\end{footnotetext}

\begin{singlespace}
	\begin{abstract}
		
		In this work, we study the event occurrences of individuals interacting in a network.
		To characterize the dynamic interactions among the individuals,
		we propose a group network Hawkes process (GNHP) model whose network structure is observed and fixed.
		In particular, we introduce a latent group structure among individuals to account for the heterogeneous user-specific characteristics. A maximum likelihood approach is proposed to simultaneously cluster individuals in the network and estimate model parameters. A fast EM algorithm is subsequently developed by utilizing the branching representation of the proposed GNHP model. Theoretical properties of the resulting estimators of group memberships and model parameters are investigated under both settings when the number of latent groups $G$ is over-specified or correctly specified. A data-driven criterion that can consistently identify the true $G$ under mild conditions is derived. Extensive simulation studies and an application to a data set collected from Sina Weibo are used to illustrate the effectiveness of the proposed methodology.\\
		
		
	\end{abstract}
\end{singlespace}

%	\vspace{10em}
\baselineskip=26pt

\section{INTRODUCTION}

Point process models have gained increasing popularity for modeling activities observed on various networks, examples include posting activities in online social networks, corporation transactions in a financial network, and neuron spikes in a brain network. The main goal of this work is to develop a new modeling framework for random event times observed on a network consisting of heterogeneous nodes. While the proposed framework is applicable for the analysis of general event time data, we describe our model in the context of a motivating dataset collected from Sina Weibo (the largest Twitter type social media platform in mainland China).

\begin{figure}[H]
\centering
\begin{center}
	\subfigure{\includegraphics[scale=0.20]{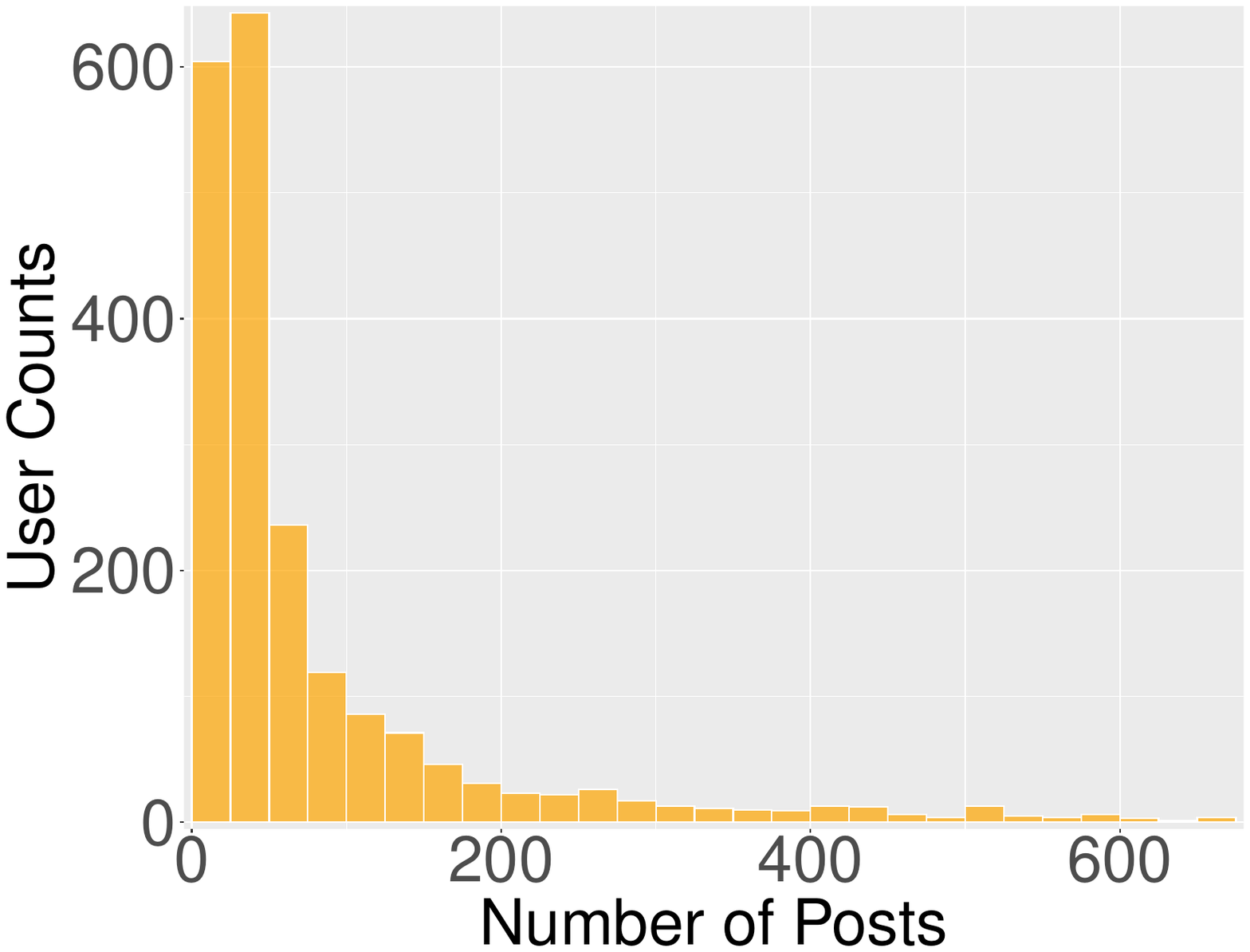}}
	\subfigure{\includegraphics[scale=0.20]{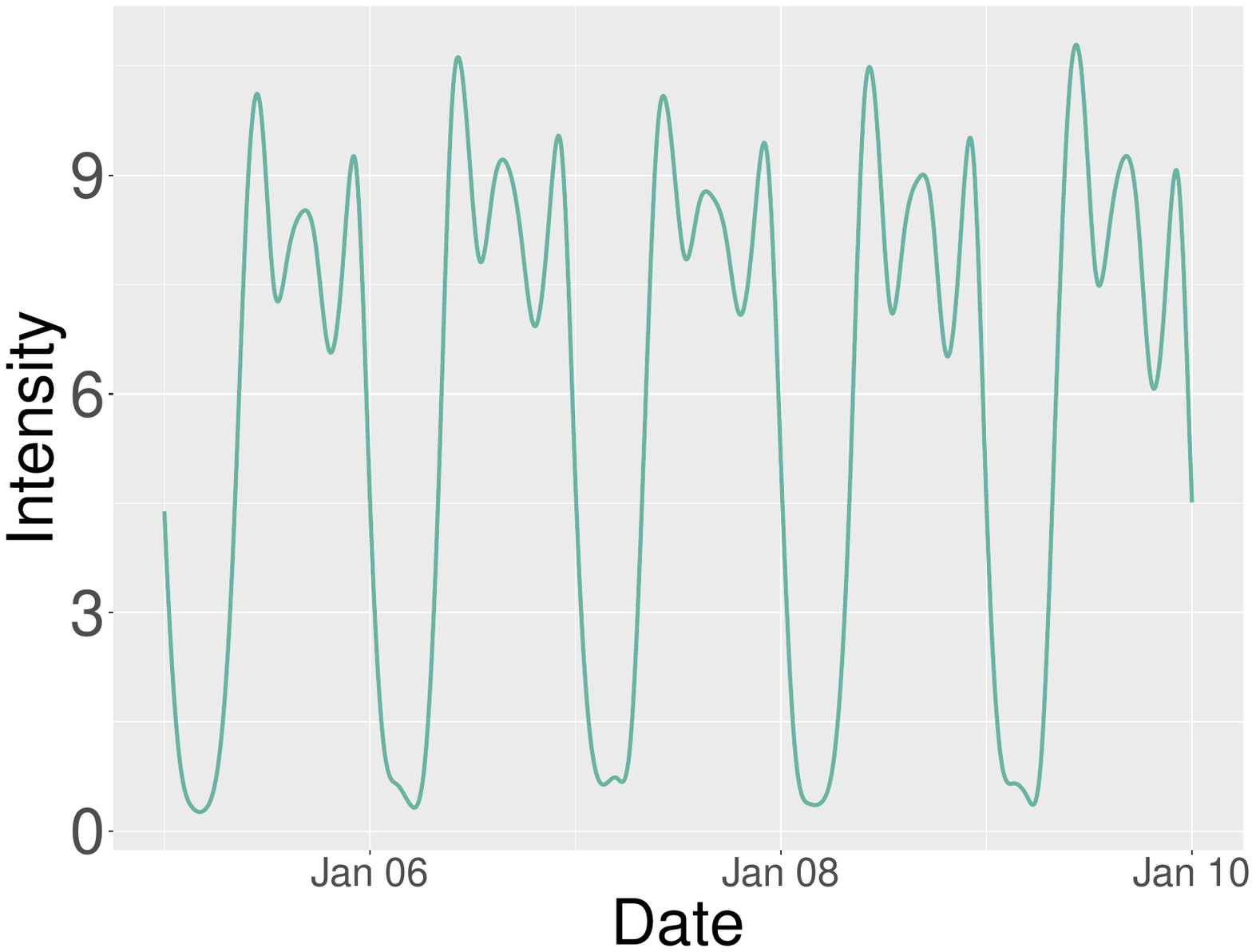}}
	\subfigure{\includegraphics[scale=0.20]{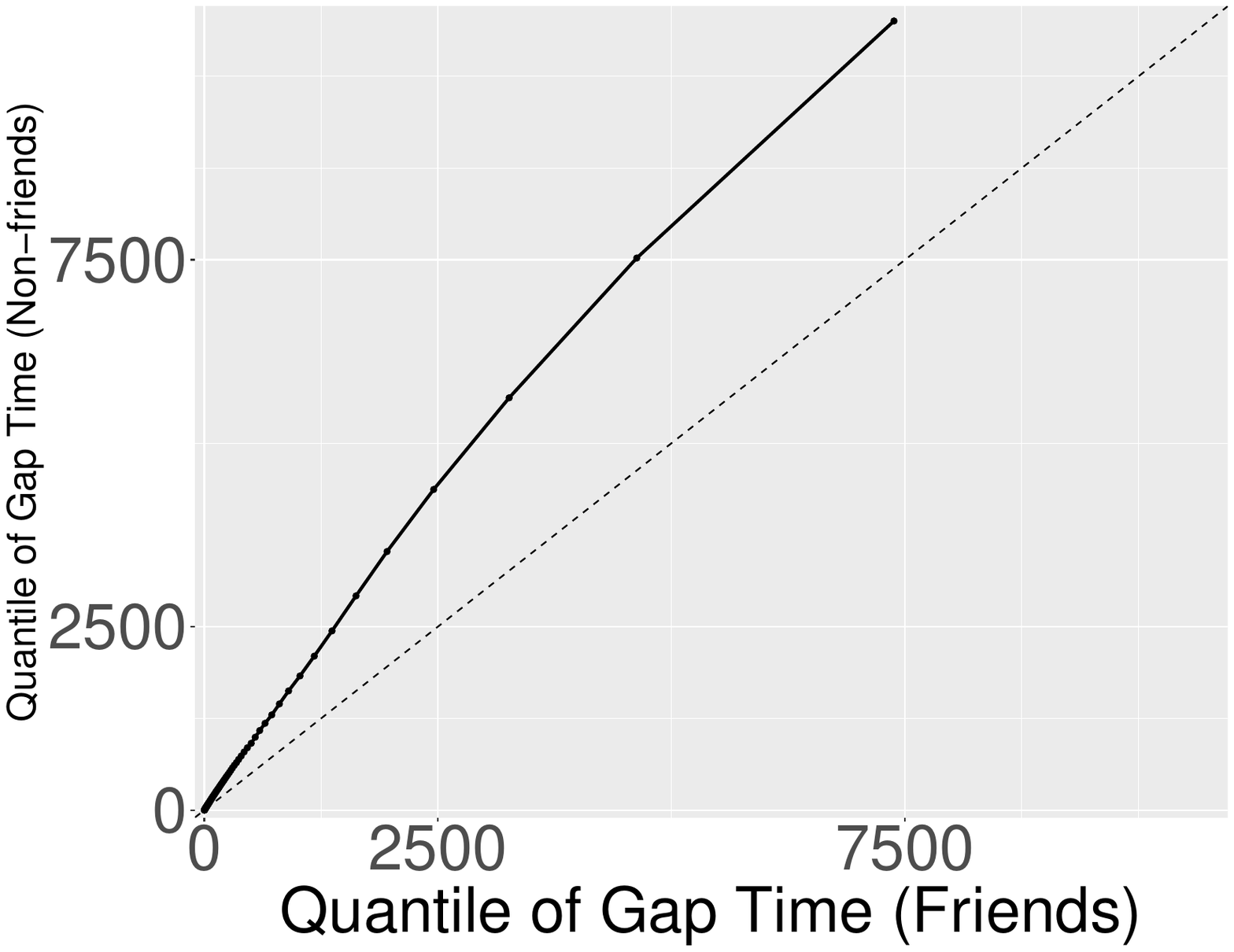}}
\end{center}
\vspace{-1.5em}
\caption{\small
	Left: histogram of users' post counts.
	Middle: estimated overall intensity function of all users' posting times.
	Right: QQ-plot of gap times among friends and among random picked non-friends.}
\label{time_interval_hist}
\end{figure}

The dataset contains posting times of 2,038 Sina Weibo users from January 1st to January 15th, 2014.
{Figure~\ref{time_interval_hist} shows that user posting patterns be highly heterogeneous and complex. Firstly, the distribution of user post counts reveals great variability in users' activities levels: while most users had less than 200 posts during the study period, a small portion of users were much more active.
	Secondly, the averaged overall intensity of posting times aggregated over all users demonstrates the existence of apparent daily periodic patterns.
	Lastly, quantiles of gap times between a user's posting time and the closest posting time from his/her connected friends and randomly picked non-friends are drastically different.} In particular, the quantiles of gap times among friends are consistently smaller than those among the non-friends, suggesting that a user's activities were heavily influenced by his/her friends.  %were heterogeneous both across users and within a day, clustered, and influenced by activities of other users that the user was connected to. The posting patterns were complex and dynamic. We will develop a flexible class of models that incorporate all these features.}

{A popular model for event time data such as our motivating example is the multivariate Hawkes process \citep{hawkes1971spectra}, which has been widely used to model event times of multiple types in {a variety of fields}, such as criminology \citep{Linderman2014}, finance \citep{bacry2013modelling}, information diffusion \citep{farajtabar2017coevolve}, and social studies \citep{zhou2013learning,Fox2016Modeling}. In the network setting, the first line of existing research aims at recovering the unknown network structure using the observed event time data, see, e.g., \cite{zhou2013learning,xu2016learning,achab2018uncovering,bacry2020sparse}. In contrast, another line of research takes the network structure as given knowledge and incorporates it into the modeling of the event time data, examples include \cite{Fox2016Modeling, farajtabar2017coevolve} and \cite{zarezade2018steering}. In the second framework, model parameters were assumed to be node-specific, and therefore the number of parameters grows at least linearly with the number of nodes. This may be problematic if many nodes in the network produce only scarce event times, which was observed in Figure \ref{time_interval_hist}. Section~\ref{sec:ework} gives a more detailed comparison between the proposed model and existing works.

%\begin{figure}[H]
%\centering
%\includegraphics[width = 0.7\textwidth]{userpost_total.pdf}
%\caption{\small The histogram of total number of posts for $m = 2038$ users in the Sina Weibo dataset.} \label{userpost_total1}
%\end{figure}

{In this work, we propose a group network Hawkes process (GNHP) to model the network heterogeneity by introducing a latent group structure among the network nodes. We assume that nodes in the same group share similar node-wise characteristics and that interaction patterns between any two connected nodes are determined by their group memberships. For each latent group, the background intensity varies over time and is nonparametrically approximated by spline basis functions. The proposed model is more parsimonious than existing models where all network nodes are considered as different \citep[e.g.,][]{Fox2016Modeling,farajtabar2017coevolve,zarezade2018steering}, but the latent group structure coupled with the observed network structure still allows us to build sufficiently flexible multivariate Hawkes process network models. Furthermore, {the proposed GNHP model admits an equivalent branching process structure that enables us to develop easily interpretable numerical measures to quantify interactions within the network,} see Section~\ref{sec::brach} for more details. The branching structure also allows us to develop a computationally efficient EM algorithm for model estimation, see Section~\ref{sec:em}. Lastly, the estimated group memberships cluster network nodes into several subgroups in a data-driven manner, offering further insights into the network activity dynamics. Therefore, the proposed GNHP model is an important addition to the existing toolbox for analyzing event time data observed on a network.
	
	{Our work also makes important theoretical contributions to the literature. {The estimation of the  (multivariate) Hawkes process is most commonly conducted through the maximum likelihood estimation \citep{ogata1988statistical,mohler2011self, chen2013inference,zhou2013learning}.} While \cite{ogata1978asymptotic} and \cite{chen2013inference} studied the asymptotic property of the maximum likelihood estimator (MLE) for a univariate Hawkes process, to the best of our knowledge, however, little is known about the limiting behaviors of the MLE for a high-dimensional multivariate Hawkes process {such as the proposed GNHP. In this work, we establish the consistency results for the MLE of the model parameters and the latent group memberships for the GNHP, with the number of groups being possibly over-specified.}  When the number of groups is correctly specified, we establish the asymptotic normality of the MLE. Finally, we show that a likelihood-based {information} criterion (LIC) can consistently select the number of groups. A similar theoretical framework has been considered in the panel data literature \citep{su2016identifying,liu2020identification,zhu2022simultaneous}, although the treatments of point process data and panel data are very different. In this sense, our work also bridges a gap between the point process and the panel data literature.

		The rest of the article is organized as follows. In Section 2, we introduce the GNHP model together with its branching structure representation and compare the proposed model to existing works. In Section 3, we detail our estimation procedure, including the computationally efficient EM algorithm. {Theoretical properties of resulting estimators are investigated} in Section 4. Simulation results under different network settings are presented in Section 5. In Section 6, we apply the proposed model to the Sina Weibo dataset. The article is concluded
		with a brief discussion in Section 7. Additional simulation resulats and all technical details are left to online supplementary material.

		\section{GROUP NETWORK HAWKES PROCESS}\label{sec::model}
		
		\subsection{Background on Hawkes Process}

		Denote by $0\le t_1\le t_2\le \ldots\le t_n\le T$ a realization of a temporal point process in $[0,T]$ and let $N(t) = \sum_{k = 1}^n {I}(t_k < t)$ be the associated counting process, where ${I}(\cdot)$ is an indicator function. Let $\mH_t = \{t_k: t_k < t\}$ be the process history up to time $t$, and the
		conditional intensity function of a point process is defined as
		$
		\lambda(t|\mH_t) = \underset{\Delta \rightarrow 0}{\mathrm{lim}} \Delta^{-1}{\EE \left[ N(t + \Delta)-N(t)  | \mH_t \right] }.\nonumber
		$
		The classical Hawkes process  model \citep{hawkes1971spectra} assumes  that $\lambda(t|\mH_t)$ takes the form
		\beq
		\label{hawkes-classical}
		\lambda(t|\mH_t) = \mu + \sum_{t_k \in \mH_t} f(t - t_k),
		\eeq
		where $\mu>0$ is a background rate of events
		and $f(\cdot)$ is a non-negative triggering function. The Hawkes process is considered as ``self-exciting" since the past events in $\mH_t$ contribute to the instantaneous intensity $\lambda(t|\mH_t)$ at time $t$ through $f(\cdot)$. The triggering function controls the dependence range and strength between the intensity at time $t$ and the past events, and  popular choices include the exponential kernel \citep{hawkes1971spectra} and the power-law kernel \citep{ogata1988statistical}.
		
		{ {\sc \bf Notations.}
			For a vector $\bv = (v_1,\cdots, v_p)^\top\in \mR^p$, let $\|\bv\|_2 = (\sum_{i = 1}^pv_i^2)^{1/2}$
			and $\|\bv\|_\infty = \max_j |v_j|$ denote the $L_2$-norm and the infinity norm of  $\bv$, respectively.
			For a function $h(t)\in \mR^1$ with $t\in [0,T]$, define
			$\|h(\cdot)\|_T = \{T^{-1}\int_0^Th(t)^2dt\}^{1/2}$
			and $\|h(\cdot)\|_\infty = \sup_{t\in [0,T]}|h(t)|$.
			For a matrix $\H = (h_{ij})\in \mR^{m\times n}$, define the row norm $\|\H\|_\infty =
			\max_i (\sum_j |h_{ij}|)$ and $L_1$ norm $\|\H\|_1 =
			\sum_i (\sum_j |h_{ij}|)$.
			For any set $\mS$, define $\mS^n = \{\bv = (v_1,\cdots, v_n)^\top: v_i\in \mS\}$ as collection of vectors of length $n$, whose elements are in $\mS$. Finally, we denote $[G]=\{1,\cdots,G\}$ and $\1=(1,\cdots,1)^\top$.

		}
		\subsection{Network Hawkes Process with Latent Group Structures}
		\label{sec::NHP}
		
		{In this section, we extend the classical Hawkes process to the network setting with a latent group structure. Consider a network with $m$ nodes, where
			{the relationships among the nodes are represented by an adjacency matrix $A = (a_{ij})\in\mR^{m\times m}$, with $a_{ij} = 1$ if the $i$th node {follows} the $j$th node and  $a_{ij} = 0$ if otherwise.
				By convention, we do not allow self-connected nodes, i.e., $a_{ii} = 0$.} To account for potential heterogeneity of network nodes, we assume that the nodes in the network belong to $G$ latent groups where nodes within the same group share the same node-specific characteristics and interactions with nodes from other groups.
			
			Denote $\mG = (g_1,\cdots,g_m)^\top\in\mR^m$ as the latent group membership vector of all nodes, where  $g_i\in [G]$ for $i=1,\cdots,m$. For node $i$, let $0\le t_{i1}\le t_{i2}\le \cdots \le t_{in_i}\le T$ be the observed $n_i$ event times and  $N_i(t) = \sum_{k = 1}^{n_i}  I (t_{ik}\le t)$ be the associated counting process. {Given the group membership vector $\mG$ and event history $\mH_t =\cup_{i=1}^m \mH_{i,t}$  with $ \mH_{i,t}=\{t_{ik}: t_{ik} < t; 1 \leq k \leq n_i\}$, the proposed GNHP model assumes that the conditional intensity for the $i$th node is of the form
				\begin{align}
					\label{intensity0}
					\lambda_{i}(t|\mG,\mH_t) =
					\mu_{g_{i}}(t)
					+ \beta_{g_i} \sum_{t_{ik}\in \mH_{i,t}} f_b(t - t_{ik};\eta_{g_i})
					+ \sum_{j = 1}^{m} \phi_{g_{i}g_{j}} \frac{a_{ij}}{d_i} \sum_{t_{jl} \in \mH_{j,t}}  f_b(t - t_{jl};\gamma_{g_i}),
				\end{align}
				for $1\le i\le m$, where $\mu_{g}(\cdot)$ is the background intensity,
				$f_b(\cdot;\eta)$
				is a triggering function  governed by parameter $\eta$,
				$d_i = \sum_{j=1}^ma_{ij}$ is the out-degree  of node $i$   \citep{Zhu2017Network}, and
				$\{\beta_g, \gamma_g, \eta_g, \phi_{gg'}\}$ are unknown group-level
				parameters for $g,g'\in [G]$.
				{Following \cite{chen2017multivariate}, we assume the support of $f_b(t;\gamma)$ is $[0,b]$.
					For example, $f_b(\cdot;\gamma)$ can be the truncated exponential kernel
					\be
					\label{truncEXP}
					f_b(t;\gamma)=
					\gamma\left[1-\exp(-b\gamma)\right]^{-1}\exp(-\gamma t) I(t\leq b), \quad \text{ for any } t\ge 0.
					\ee
					Our theoretical investigation allows the truncation range $b\to\infty$ as $m,T\to\infty$.} For identifiability of parameters $\beta_g$'s and $\phi_{gg'}$'s, we assume that $\int_{0}^\infty f_b(t;\gamma)dt = 1$ for any given $\gamma$. In addition, we assume that $\|\partial^k f_b(\cdot;\gamma)/\partial \gamma^k\|_\infty <\infty$ for $1\le k\le 3$.

				% 			\begin{remark}
					% The use of a truncated triggering function such as~\eqref{truncEXP} is convenient for our theoretical investigations. In particular, our theory allows $b\to\infty$ as $m, T\to\infty$, which enables us to directly study the impacts of the triggering function on the asymptotic properties of the estimated model. From a practical standpoint, the choice of $b$ has a rather limited impact on the estimation of $\gamma$, provided that $b$ is sufficiently large. Therefore, the use of $f_b(t;\gamma)$ is much less restrictive than the bounded-support assumption (i.e., $b$ is finite) commonly used in the existing literature on multivariate Hawkes process theory, see, e.g., \cite{hansen2015lasso,chen2017multivariate}.
					% 			\end{remark}

				The proposed conditional intensity in (\ref{intensity0}) can be decomposed into the following three parts.
				
				\vspace*{-3mm}
				\begin{itemize}
					\itemsep-0.5em
					\item {\sc Background intensity} $\mu_{g_i}(t)$. This describes the overall activity pattern of the node $i$.
					\item  {\sc Momentum intensity} $\beta_{g_i}\sum_{t_{ik} \in \mH_{i,t}}f_b(t-t_{ik};\eta_{g_i})$. It models ``self-exciting" influence of its own past events on the occurrence of a new event at $t$ at the node $i$.
					\item {\sc Network intensity} $d_i^{-1}\sum_{j = 1}^{m} \phi_{g_{i}g_{j}} a_{ij} \sum_{t_{jl} \in \mH_{j,t}}  f_b(t - t_{jl};\gamma_{g_i})
					$. This models influences from past events of other nodes on the occurrence of a new event at $t$ at node $i$.
				\end{itemize}
				The out-degree $d_i$ is used to prevent the inflation of the {\it Network Intensity} when $d_i\to\infty$, which is commonly done in literature \citep{Zhu2017Network,zhu2019network}.
				The $\phi_{g_ig_j}$'s in the {\it Network Intensity} represent the average network influences from the connected nodes on the $i$th node.
				Finally, we remark that the triggering functions
				in the {\it Momentum Intensity} and the {\it Network Intensity} do not necessarily share the same form.
				%{In addition, if the $i$th node follows the $j$th node (i.e., $a_{ij} = 1$),
					%then the weight is decided by both their group connection strength (i.e., $\alpha_{g_ig_j}$)
					%and the out-degree of the $i$th node (i.e., $d_{i}$). }

				The network dependence of the proposed GNHP can be characterized by the transition matrix $\B = (b_{ij})\in\mR^{m\times m}$, where $b_{ij} = \phi_{g_ig_j}d_i^{-1}a_{ij}
				+ \beta_{g_i}I(i=j)$, for $i,j=1,\cdots,m.$ Detailed properties of $\B$ will be further explored in the next subsection based on the following assumption.
				\begin{assumption}\label{assum:station}
					Assume $\|\bB\|_{\infty} \leq c_B < 1$,
					where
					$c_B$ is  a positive constant.
				\end{assumption}	
				Assumption~\ref{assum:station} is a sufficient condition for stability of a multivariate Hawkes process and has been widely used in the literature, see, e.g., \cite{hansen2015lasso,chen2017multivariate}. In the next subsection, we show that for the GNHP model, it is a sufficient condition to ensure that the expected number of offspring events triggered by a parent event at any network node is finite.

				\subsection{Branching Structure of the GNHP Model}
				\label{sec::brach}
				\citet{Hawkes1974A} provides an equivalent branching structure representation for the classical Hawkes process~\eqref{hawkes-classical},  which classifies the observed events into two disjoint processes: a {\it parent Poisson process} with a rate $\mu$,
				and {\it offspring processes} triggered by past events. The branching structure of the classical Hawkes process is illustrated in the left panel of Figure \ref{branching-fig}.
				
				\begin{figure}[ht!]
					\centering
					\includegraphics[width = 0.7\textwidth]{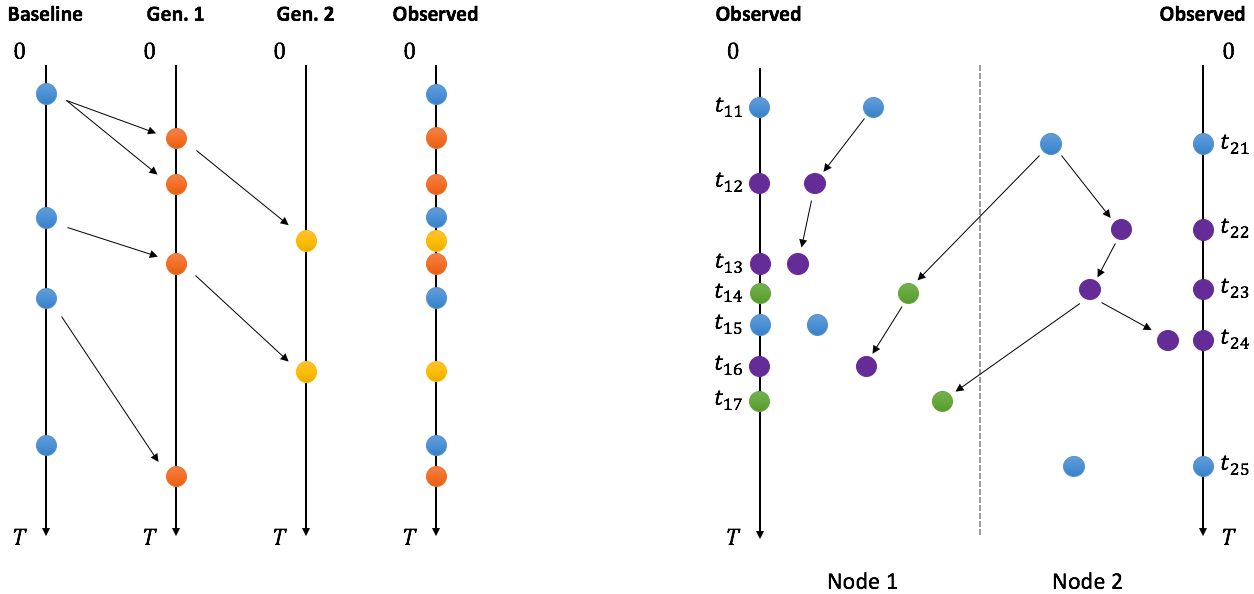}
					\caption{{\small Branching structures of the classical Hawkes process (left) and the GNHP (right).
							Left: parent events (blue circle) and two generations of offspring events; Right:
							a two-node network where Node 1 follows Node 2.
							Parent events from both nodes (blue circles), event times triggered by their own past events (purple circles), event times in Node 1 triggered by past events from Node 2 (green circles).} } \label{branching-fig}
				\end{figure}
				
				Following~\cite{rasmussen2013bayesian} and \cite{Halpin2013Modelling}, a branching structure representation can be derived for the GNHP by treating the aggregated point process from all nodes, denoted by $N^{\rm pool}$, as a marked point process  on $[0,T]$, with the mark for any $t\in N^{\rm pool}$ being the node index where the event occurs. Specifically, event times in $N^{\rm pool}$ can be categorized into two types: the parent and offspring events.  Let
				$\mM_i^{parent}$
				be the set of parent events from node $i$ and $\mM_{ik}^{fam}=\mM_{ik}^{off}\cup\{t_{ik}\}$, where $\mM_{ik}^{off}$ is the {set} of offspring events generated from a parent event $t_{ik}\in \mM_i^{parent}$. Note that $\mM_{ik}^{fam}$ may contain event times from other nodes due to the network interactions. 	The branching structure is defined as {follows} and  illustrated in Figure \ref{branching-fig}.
				\vspace*{-3mm}
				\begin{enumerate}
					\itemsep-0.5em
					\item For any $1\leq i\le m,$ the parent events $\mM_i^{parent}$ follows a Poisson process with an intensity $\mu_{g_i}(t)$, and all Poisson processes are independent.
					\item Each parent event $t_{ik}\in \mM_i^{parent}$ generates a {set} of offspring events $\mM_{ik}^{off}$, and all resulting $\mM_{ik}^{fam}$'s are independent. Event times in each $\mM_{ik}^{fam}$ are generated recursively as follows.
					\vspace*{-3mm}
					\begin{enumerate}
						\itemsep-0.5em
						\item Generation $0$, denoted as $\mM_{ik}^{gen_0}$, consists of only $t_{ik}$, i.e., $\mM_{ik}^{gen_0}=\{t_{ik}\}$.
						\item Having generated offspring event times up to $n$ generations, $\mM_{ik}^{gen_0},\cdots,\mM_{ik}^{gen_n}$, each event time in $\mM_{ik}^{gen_n}$ generates a series of event times of the $n+1$ generation. Specifically, suppose $t_{jl}\in \mM_{ik}^{gen_n}$ is an event time located in the $j$th node, then on the time interval $[t_{jl},T]$, it generates (i) a Poisson process with intensity  $\beta_{g_j} f_b(t - t_{jl};\eta_{g_j})$ on the node $j$; and (ii) a Poisson process with intensity $\phi_{g_{i'}g_{j}} d_{i'}^{-1} f_b(t - t_{jl};\gamma_{g_{i'}})$ on the node $i'$ for any $i'\ne j$ where $a_{i'j}=1$. All Poisson processes are independent.
						\item Obtain $\mM_{ik}^{fam}=\cup_{n=0}^{\infty}\mM_{ik}^{gen_n}$.
					\end{enumerate}
				\end{enumerate}
				%Figure \ref{branching-fig} illustrates the branching structure of the GNHP model with a network of two nodes.

				%						To establish the estimation consistency, we first define a network dependence measure.
				%			Let
				%			$
				%			\tau_m = \|I - (1+\delta)B^\top\one\|_\infty
				%			$
				%			with $\delta = (c_B^{-1}+1)/2$.
				%			{\red Here $\tau_m$ quantifies the largest column summation of $B$,
					%				which could be treated as a measure of  the largest influential power of the nodes in the network.
					%				As we discuss in details in Appendix C (Lemma \ref{lem_E_Ni}--\ref{fam_num}), $\tau_m$ is closely related to the network dependence among the nodes.
					%				Particularly, larger $\tau_m$ implies stronger dependence level in the network. Hence, it is a critical value we need to control in dealing the network dependence.}
				\begin{theorem}
					\label{thmfam} Denote by $\#_{i}(\mS)$ the total number of events occurring at a subset of network nodes $\mS\subset \{1,\cdots,m\}$ that are offspring events of a parent event originated from the $i$th node on $[0,\infty)$. Then under Assumption~\ref{assum:station}, one has that
					\be
					\label{eq:thmfam}
					E\left[\#_{i}(\mS)\right]=\e_{\mS}^\top(\I-\B)^{-1} \e_i,\text{ for any } 1\leq i\leq m,
					\ee
					where $\e_{\mS}$ is a vector with $1$ for entries whose indexes are in $\mS$ and $0$ elsewhere, and $\e_i=\e_{\{i\}}$.
				\end{theorem}
				The proof is given in the online supplement.
				Theorem~\ref{thmfam} provides a useful tool to quantify the network interactions. Examples include
				
				\vspace*{-3mm}
				\begin{itemize}
					\itemsep-0.5em
					\item {\sc Node-to-node influence}: the $(j,i)$th entry of $(\I-\B)^{-1}$ gives the average number of events at node $j$ that are triggered by a parent event at the $i$th node (set $\mS=\{j\}$).

					\item {\sc Node-to-network influence:} by setting $\mS=\{1,\cdots,m\}$, we obtain the sum of the $i$th column of $(\I-\B)^{-1}$  as the average number of events in the entire network that are directly/indirectly triggered by a parent event at the $i$th node, i.e., $E\{\#_i(\mS)\} = E\{|\mM_{ik}^{fam}|\}$.
					
					\item {\sc Dynamic Group-to-Group influence:}
					Let $\mS_g$ be the collection of node indexes in group $g$.
					Denote
					$\#(\mS_{g}, \mS_{g'}, [t_1,t_2])$ as the
					total number of events from group $g'$ triggered by parent events from group $g$ occurring within $[t_1,t_2]$, {which} reflects the influential power of group $g$ on group $g'$. Based on Theorem~\ref{thmfam}, it can be verified that $E\Big\{\#(\mS_g,\mS_{g'},[t_1,t_2])\Big\} = \e_{\mS_{g'}}^\top(\I-\B)^{-1} \e_{\mS_g}\int_{t_1}^{t_2}\mu_g(t)dt$. We can subsequently define the limiting case as follows
					\begin{align}
						{\rm GIF}_{gg'}(t)=\lim_{\Delta\to0}{\Delta^{-1}}E\Big\{\#(\mS_g,\mS_{g'},[t,t+\Delta])\Big\} = \e_{\mS_{g'}}^\top(\I-\B)^{-1} \e_{\mS_g}\mu_g(t),\quad t\in[0,T],\label{E_Sgg_count}
					\end{align}
					which is more convenient for graphical illustrations.
				\end{itemize}

				\subsection{Comparisons with Existing Literature}
				\label{sec:ework}
				There has been much work on multivariate Hawkes process \citep[e.g.,][]{zhou2013learning,bacry2013modelling,chen2017multivariate}, and models for the conditional intensity can be generally expressed as
				\beq
				\lambda_{i}(t|\mG,\mH_t) = \mu_i + \sum_{j = 1}^m \sum_{t_{jl} \in \mH_{j,t}} \zeta_{ij}(t-t_{jl}), \quad i=1,\cdots,m,\label{multi_hawkes}
				\eeq
				where $\mu_i$ is the background rate at node $i$ and $\zeta_{ij}(\cdot)$ is some transfer function between node $j$ and  node $i$. Key differences among existing multivariate Hawkes process models center around constructions of $\zeta_{ij}(\cdot)$'s. A popular modeling strategy is to assume that	
				$\zeta_{ij}(\cdot) = \theta_{ij}f(\cdot;\gamma)$, $i,j=1,\cdots,m$,
				with some parametric $f(\cdot;\gamma)$. Such a model involves a total of $O(m^2)$ parameters, limiting its suitability to applications with a relatively small $m$ \citep[e.g.,][]{bacry2013modelling}. When modeling a large network, one needs to impose some sparse structure on $\theta_{ij}$'s, in which case a nonzero $\theta_{ij}$ implies that node $i$ is directly influenced by node $j$, and estimated $\theta_{ij}$'s may help recover the latent network structure \citep[e.g.,][]{xu2016learning,bacry2020sparse}. While most work in this line lack rigorous theory, there has been some recent
				theoretical studies of such models when $m$ is diverging, see, e.g., \cite{hansen2015lasso},
				\cite{chen2017multivariate} and \cite{cai2020latent}.
				
				{There is also a large body of literature on the identification of network node clusters \citep[e.g.,][]{zhao2012consistency, amini2013pseudo}, which is commonly referred to as ``community detection".  We remark that the concept of ``community" in the community detection literature is fundamentally different from the ``groups" in the proposed GNHP model. In community detection, the network adjacent matrix $A$ is assumed to be a random matrix consisting of Bernoulli random variables, and the identification of node communities $\mathbf c$ critically depends on the conditional probability $P(A|\mathbf c)$. However, the adjacent matrix $A$ in the GNHP is considered deterministic and there is no probability associated with it. In contrast, the groups in the GNHP are formed by maximizing the likelihood of event times collected from all network nodes, and nodes in the same group are forced to share similar node-specific characteristics such as $\mu_{g_i}(\cdot)$'s and $\beta_{g_i}$'s. Although some recent works in community detection literature utilize nodal features to help identify communities \citep{yan2021covariate, zhang2021directed, weng2022community}, the central piece of these models remains to be $P(A|\mathbf c)$. The difference is most manifested in the extreme case when all nodes are isolated from each other (i.e., $A$ consists of all $0$'s), while the GNHP is still valid by manually setting $\phi_{g_ig_j}$'s as $0$, the community detection can no longer be performed since there is no network anymore. We note another recent work \cite{matias2018semiparametric} also considers latent group structures when extending the stochastic block model for recurrent interaction events in continuous time. The key difference between their work and ours lies in that they require that each observed event time is associated with a label indicating this event is an interaction between which two nodes. However, such information is not available in event times modeled by the GNHP, which instead focuses on modeling events that occur on individual nodes but may have some correlations. In fact, as suggested by the branching structure in Section~\ref{sec::brach},  identifying the triggering source of an event in the GNHP is the most challenging task. Similarly, in the extreme case when there are only self-activities on each node but no interaction between any pair of nodes, \cite{matias2018semiparametric} is not applicable but the GNHP model remains valid with $\phi_{g_ig_j}$'s set to $0$. }

				{The second} line of research focuses on {analyzing the network activities through} parameterizations utilizing a known network structure, \citep[e.g.,][]{Fox2016Modeling,farajtabar2017coevolve,zarezade2018steering}. For instance, \cite{Fox2016Modeling} models email communications in a network by assuming
				$\zeta_{ij}(t) = \theta_i a_{ij}\gamma_i \exp(-\gamma_it)$,
				where $\theta_i$'s and $\gamma_i$'s are unknown parameters.	As a result, the total number of unknown parameters is reduced to $O(m)$. Similar but more complex models were studied in \cite{farajtabar2017coevolve} and \cite{zarezade2018steering}.
				
				Our proposed GNHP model falls into the second line of research with four distinct features: (a) the background intensities $\mu_i(\cdot)$'s are allowed to be time-varying and nonparametrically approximated using spline basis functions, which provides greater modeling flexibility. (b) the {\it Momentum Intensity} in model~\eqref{intensity0} has a clear interpretation given the network structure. (c) the latent group structure imposed on network nodes not only accounts for commonly observed network heterogeneity in a natural way but also effectively reduces the number of parameters. (d) the estimated group memberships for network nodes may provide further insights on the network activities using various influence measures discussed in Section~\ref{sec::brach}}.

			{
				
				\section{MODEL ESTIMATION}
				
				\subsection{Background Intensity Approximation}		
				As illustrated in Figure~\ref{time_interval_hist}, human activities such as social media posting often exhibit periodic patterns. We, therefore, assume that the background intensity of a given node takes a periodic form, for which there exists a finite $\omega>0$ such that
				$\mu_{g}(t)=\mu_{g}(t+l\omega)$, $t\in[0,T-l\omega]$, $l=0,1,2,\cdots,$ for any $g\in[G]$.
				For our motivating example, it is natural to choose $\omega= 1$ day (or $24$ hours) to account for daily posting behaviors. Without assuming a restrictive parametric form for $\mu_g(\cdot)$'s, we approximate $\mu_{g}(\cdot)$ using periodic splines.  Let $\k_{\nk}(\cdot) = (k_1(\cdot),\cdots, k_{\nk }(\cdot))^\top$ be a collection of basis functions defined on $[0,\omega]$ and $\x_{\nk}(t)=\sqrt{\nk }\k_{\nk}(t-\floor{t/\omega}{\omega})$ for any $t\in[0,T]$, where $\floor{a}$ is the largest integer less than or equal to $a$, then $\mu_g(\cdot)$ is approximated by
				\beq
				\mu_{g}(t)=\w_g^\top\x_{\nk}(t), \quad t\in[0,T],
				\label{mu}
				\eeq
				where $\w_g$, $g\in[G]$, are the coefficient vectors that need to be estimated. For our theoretical investigation, we require the following assumption for the spline basis functions.
				
				\begin{assumption}\label{assum:basis}
					Assume that there exists a constant $R>0$ such that $k_1(\cdot),\cdots, k_{\nk }(\cdot)$  satisfy: (a)  $\|k_j(\cdot)\|_{\infty} \leq R$ and $ \int_0^\omega  |k_j(t)|dt = O(\nk ^{-1})$; (b) $\int_0^\omega |k_j(t)k_{j+l}(t)|dt = O(\nk ^{-1})$, and
					{$k_j(t)k_{j+l}(t) = 0$ for $l > J$ and any $t\in[0,\omega]$,} where $J \geq 0$ is a finite integer. Denote by $\mu^0_g(\cdot)$ the true background function of group $g$, and assume that for some constant $\nu>0$, it holds that
					\beq
					\max_{1\le g\le G}\inf_{\bw_g\in \mR^{\nk }, \|\bw_g\|_\infty\leq R/\sqrt{\nk }}\big\| \mu_g^{0}(\cdot) - \bw_g^\top\x_{\nk}(\cdot)\big\|_{\infty} = O(\nk ^{-\nu}).\label{approx_error}
					\eeq
				\end{assumption}
				Assumption~\ref{assum:basis} requires that the true background intensity functions can be approximated sufficiently well by a linear combination of spline basis functions, which is mild for many spline families. In our numerical examples, we choose $k_1(\cdot),\cdots, k_{\nk }(\cdot)$ as the $r$th-order B-spline basis functions with equally spaced knots on $[0,\omega]$, which meet Assumption~\ref{assum:basis} \citep{zhou1998local}.
				\subsection{Maximum Likelihood Estimation}
				
				For any $g\neq g'$, denote $\mbW\subset\mbR^{\nk}$, $\bTheta\subset\mbR^{+3}$ and  $\bPhi \subset\mbR^+$ as the parameter spaces for parameters in $\w_g$, $\btheta_g=(\beta_g,\eta_g,\gamma_g)^\top$, and $\phi_{gg'}$'s, respectively.
				Here $\mbR^+$ denotes $[0,\infty)$.
				Denote
				by $\mbG \equiv[G]^m$ as the parameter space for the
				membership vector $\mG=(g_1,\cdots,g_m)^\top$.
				For our theoretical investigation, we impose the following assumption on the parameter spaces.
				\begin{assumption}\label{assum:para_space}
					There exist $R>0,C>0$ such that (a) $\sup_{\w\in\mbW}\|\w\|_{\infty}\leq R/\sqrt{\nk}$ and $\inf_{\w\in\mbW}\inf_{t\in [0,T]}\w^\top\x_{\nk}(t)\ge C$; (b) $\sup_{\btheta\in\bfTheta}\|\btheta\|_{\infty}\leq R$; and (c) $\sup_{\phi\in\bPhi}|\phi|\leq R$.
				\end{assumption}
				
				Let $\mN_i=\{j: a_{ij}= 1\}$ be the set of $d_i$ neighboring nodes  of node $i$, $\mG_i = (g_{j_1} ,\cdots,g_{j_{d_i}})^\top$ be the corresponding group membership vector of all nodes in $\mN_i$, and the re-scaled interaction parameter $\bvarphi_{g_i,\mG_i}=d_i^{-1/2}(\phi_{g_ig_{j_1}},\cdots,\phi_{g_ig_{j_{d_i}}})^\top $, $i=1,\cdots,m$. The term $d_i^{-1/2}$ in $\bvarphi_{g_i,\mG_i}$ is convenient for our theoretical study but is of no practical importance. We can see that the conditional intensity~\eqref{intensity0} of node $i$ only depends on $\w_{g_i}^\top\x_{\nk}(\cdot)$, $\btheta_{g_i}$ and $\bvarphi_{g_i,\mG_i}$, and hence can be rewritten as
				\begin{align}
					\label{intensity}
					\hspace{-1em}\lambda_i(t|\w_{g_i},\btheta_{g_i},\bvarphi_{g_i,\mG_i},{\mH_t}) =
					\w_{g_i}^\top\x_{\nk}(t)
					+ \beta_{g_i} \hspace{-0.75em}\sum_{t_{ik}\in\mH_{i,t}}\hspace{-0.75em} f_b(t - t_{ik};\eta_{g_i})
					+ \sum_{j = 1}^{m}\hskip-0.25em \frac{a_{ij} \phi_{g_{i}g_{j}}}{d_i}\hspace{-0.75em} \sum_{t_{jl}\in\mH_{j,t}}\hspace{-0.75em}  f_b(t - t_{jl};\gamma_{g_i}),
				\end{align}
				for $ i=1,\cdots,m.$ Denote the parameter vectors $\ut\w=(\w_1^\top,\cdots,\w_G^\top)^\top\in\mbW^G$, $\ut\btheta=(\btheta_1^\top,\cdots,\btheta_G^\top)^\top\in\bTheta^G$, and $\ut\bphi=(\phi_{11},\cdots,\phi_{1G},\phi_{21},\cdots,\phi_{G,G})^\top\in \bPhi^{G^2}$. Consequently, the scaled log-likelihood function (divided by $mT$) of the proposed GNHP can be shown to have the form
				\be
				\label{comp_lik}
				\ell(\ut\w,\ut{\btheta},\ut\bphi,\mG|\mH_T)=\frac{1}{m}\sum_{i=1}^{m}\ell_i(\w_{g_i},\btheta_{g_i},
				\bvarphi_{g_i,\mG_i}|\mH_T),\text{ where }
				\ee
				\[
				\ell_i(\w_{g_i},\btheta_{g_i},\bvarphi_{g_i,\mG_i}|\mH_T)=\frac{1}{T}
				\left\{\sum_{k=1}^{n_i}\log\left[\lambda_i(t_{ik}|\w_{g_i},\btheta_{g_i},\bvarphi_{g_i,\mG_i},{\mH_{t_{ik}}})\right] -
				\int_0^T\lambda_i(t|\w_{g_i},\btheta_{g_i},\bvarphi_{g_i,\mG_i},{\mH_t})dt\right\}.
				\]
				
				For ease of presentation, from now on, we denote $\ut\bpsi=(\ut\w^\top,\ut\btheta^\top,\ut\bphi^\top,\mG^\top)^\top$ and the MLE of the parameters are then obtained by
				\be\label{mle}
				\wh{\ut\bpsi}\equiv\Big(\wh{\ut\w}^\top,\wh{\ut{\btheta}}^\top,\wh{\ut\bphi}^\top,\wh\mG^\top\Big)^\top=\argmax_{\ut\w,\ut{\btheta},\ut\bphi,\mG}\ell\Big(\ut\w,\ut{\btheta},\ut\bphi,\mG|\mH_T\Big), \text{ where } \wh\mG=(\wh g_1,\dots,\wh g_m)^\top.
				\ee

				\subsection{An EM Algorithm}
				\label{sec:em}
				Direct maximization of~\eqref{comp_lik} is a non-concave problem with a large number of parameters, which can be computationally challenging. In this subsection, we propose a more efficient algorithm by taking advantage of the branching structure given in Section~\ref{sec::brach}.
				
				For the $k$th event of node $i$ that occurs at time $t_{ik}$,
				define $Z_{ik} = (j,l)$,
				where $j = l = 0$ indicates a parent event,
				and otherwise means the $k$th event is triggered by the $l$th event from node $j$ at time $t_{jl}$. Given $\Z_i=(Z_{i1},\cdots,Z_{in_i})^\top$, event times of node $i$ can be categorized as:
				\begin{enumerate}
					\item Parent events with $Z_{ik}=(0,0)$, which constitute a realization of a Poisson process on $[0,T]$ with an intensity $\mu_{g_i}(\cdot)$. The log-likelihood  can then be written as
					\be
					\label{lik-P}
					\ell_{\br,i}^{(1)}\left(\bw_{g_i}|\mH_T,\Z_i\right)=\sum_{k = 1}^{n_i}I\left[Z_{ik} = (0,0)\right]\log \left[\w_{g_i}^\top\x_{\nk}(t_{ik})\right]-\int_0^{T}\left[\w_{g_i}^\top\x_{\nk}(t)\right] dt.
					\ee
					\item All $t_{ik}$'s triggered by a past event of node $i$, i.e., $Z_{ik}=(i,l)$ for some $1\le l<k$, which form a realization of a Poisson process on $[t_{il},T]$ with an intensity $\beta_{g_i}f_b(t-t_{il};\eta_{g_i})$ for any $t\in[t_{il},T]$. The joint log-likelihood of all such events is of the form
					\be
					\label{lik-m}
					\hspace{-1em}\ell_{\br,i}^{(2)}\left(\beta_{g_i},\eta_{g_i}|\mH_T,\Z_i\right)= \sum_{l = 1}^{n_i-1}\bigg\{ {\sum_{ k = l+1}^{n_i}}I\left[Z_{ik} = (i,l)\right] \llog \left[\beta_{g_i}f_b(t_{ik}-t_{il};\eta_{g_i})\right]- \beta_{g_i} \int_{t_{il}}^{T} f_b(t-t_{il};\eta_{g_i}) dt\bigg\}.
					\ee
					\item All $t_{ik}$'s triggered by a past event of a node $j\in\mN_i$, i.e., $Z_{ik}=(j,l)$ for some $1\le l<n_j$ and $j\in\mN_i$, which form a realization of a Poisson process on $[t_{jl},T]$ with an intensity $d_i^{-1}\phi_{g_ig_j}f_b(t-t_{jl};\gamma_{g_i})$ for any $ t\in[t_{jl},T]$. The joint log-likelihood of all such events becomes
					\be
					\label{lik-n}
					\begin{split}
						\ell_{\br,i}^{(3)}\left(\gamma_{g_i},\bvarphi_{g_i,\mG_i}|\mH_T,\Z_i\right)= \sum_{j \in\mN_i} \sum_{l = 1}^{n_j}\bigg\{\sum_{k = 1}^{n_i} I\left[Z_{ik} = (j,l)\right] \llog &\left[\frac{\phi_{g_ig_j}}{d_i}f_b(t_{ik}-t_{jl};\gamma_{g_i})\right]\\
						&-\frac{\phi_{g_ig_j}}{d_i}\int_{t_{jl}}^{T} f_b(t-t_{jl};\gamma_{g_i})dt\bigg\}.
					\end{split}
					\ee
				\end{enumerate}
				Consequently, when $\Z=(\Z_1^\top,\cdots, \Z_m^\top)^\top$ is observed, by the branching structure given in Section~\ref{sec::brach}, the complete log-likelihood for the proposed GNHP is then of the form
				\begin{align}
					\ell_{\br}\left(\ut\bpsi|\mH_T,\Z\right) =\sum_{i = 1}^m
					\left[\ell_{\br,i}^{(1)}\left(\bw_{g_i}|\mH_T,\Z_i\right)+
					\ell_{\br,i}^{(2)}\left(\beta_{g_i},\eta_{g_i}|\mH_T,\Z_i\right)+
					\ell_{\br,i}^{(3)}\left(\gamma_{g_i},\bvarphi_{g_i,\mG_i}|\mH_T,\Z_i\right)\right].\label{log-lik}
				\end{align}
				
				The MLE~\eqref{mle} can then be obtained by an EM algorithm using the complete likelihood~\eqref{log-lik}, and similar approaches have been used in \cite{Veen2008Estimation,Halpin2013Modelling,Fox2016Modeling}. Since the likelihood~\eqref{comp_lik} is non-concave, the initial values of the group memberships and model parameters are of crucial importance. We propose to use a combination of the K-means algorithm and a stochastic EM algorithm to generate sensible initial values.
				The details are given in the Section A of the supplementary material.

			\subsection{Selection of Number of Groups}	
			Denote the true number of latent groups as $G_0$. Theorems~\ref{thm1}-\ref{thm_member} in Section~\ref{sec:theory} suggest that under suitable conditions the MLE of model parameters are consistent for their theoretical counterparts and all node memberships can be correctly estimated as long as $G\geq G_0$. However, according to Theorem~\ref{thm_normal}, the asymptotic normality of $\wh{\ut{\btheta}}$ and $\wh{\ut\bphi}$ depends on the assumption that $G=G_0$. Therefore, it is of practical interest to develop a data-driven criterion to determine $G_0$.
			
			With a slight abuse of notation, we denote
			$\wh{\ut \w}^{(G)}, \wh{\ut \btheta}^{(G)}, \wh{\ut \bphi}^{(G)}, \wh\mG^{(G)}$ as the MLE obtained in~\eqref{mle} for a given $G$. We consider the following likelihood based criterion function,
			\beq
			\LIC(G) = \ell\Big(\wh{\ut \w}^{(G)}, \wh{\ut \btheta}^{(G)}, \wh{\ut \bphi}^{(G)}, \wh\mG^{(G)}|\mH_T\Big) - \lambda_{mT}G,\label{LIC}
			\eeq
			where $\ell(\cdot|\mH_T)$ is as defined in~\eqref{comp_lik} and $\lambda_{mT}>0$ is a tuning parameter depending on $m$ and $T$.
			The optimal $G$ is selected by $\wh G =\arg\max_G \LIC(G)$. If $\lambda_{mT}$ satisfies the rate condition in Theorem~\ref{thm_LIC}, then under suitable conditions, one can select $\wh G=G_0$ with probability tending to $1$.
			%In practice, we choose $\lambda_{mT}$  as $ $

			\section{THEORETICAL PROPERTIES}
			\label{sec:theory}		
			We now investigate the asymptotic properties of the MLE~\eqref{mle}. Denote $\mu_g^0(\cdot)$ as the true background intensity of group $g$ and $\mu_g^*(\cdot)=\w_g^{*\top}\x_{\nk}(\cdot)$ as the best spline approximation to $\mu_g^0(\cdot)$ with $\w_g^*=\argmin_{\w\in\mbW}\|\w^\top\x_{\nk}(\cdot)-\mu_g^0(\cdot)\|_{\infty}$, $g\in[G]$.  By Assumption~\ref{assum:basis}, one has that
			\be
			\label{muapprox}
			\max_{1\le g\le G}\|\mu_g^*(\cdot)-\mu_g^0(\cdot)\|_{\infty}=O(\nk^{-\nu}).
			\ee
			Denote $\beta_g^0,\gamma_g^0,\phi_{gg'}^0$ as the true values of $\beta_g,\gamma_g,\phi_{gg'}$ respectively, for  $g'\neq g$, $g,g'\in[G]$.  Let $\mG^0=(g_1^0,\cdots,g_m^0)^\top$ be the true membership vector, where $g_i^0\in[G_0]$. Note that $G$ and $G_0$ may be different in our theoretical framework. Correspondingly, $\ut{\btheta}^0%=(\btheta_1^{0\top},\cdots,\btheta_G^{0\top})^\top
			$,
			$\ut\bphi^0 %= (\phi_{11}^0,\phi_{12},\cdots, \phi_{1G}^0,\phi_{21}^0,\cdots,\phi_{GG}^0)^\top
			$,
			$\mG_i^0 $, and $\bvarphi_{g_i^0,\mG_i^0}^0$, $i=1,\cdots,m$, are defined by replacing parameters with the true values in their definitions.

			\subsection{Technical Assumptions}
			\label{sec:assump}
			Denote $\ut\bpsi=(\ut\w^\top,\ut\btheta^\top,\ut\bphi^\top,\mG^\top)^\top$ and define the function $\overline{\ell}(\ut\bpsi)=\E[\ell(\ut\bpsi|\mH_T)]$, which is of the form
			\be
			\label{ellbar}
			\overline{\ell}(\ut\bpsi)=\frac{1}{m}\sum_{i=1}^{m}\overline
			\ell_i(\w_{g_i},\btheta_{g_i},\bvarphi_{g_i,\mG_i}), \text{ and } \overline\ell_i(\w_{g_i},\btheta_{g_i},\bvarphi_{g_i,\mG_i})=\E\left[\ell_i(\w_{g_i},\btheta_{g_i},\bvarphi_{g_i,\mG_i}|\mH_T)\right].
			\ee		
			Denote by $\bpsi_i=(\w^\top, \btheta^\top, \bvarphi_i^\top)^\top$, $\bpsi_i^*=\left(\w_{g_i^0}^{*\top},\btheta_{g_i^0}^{0\top},
			\bvarphi_{g_i^0,\mG_i^0}^{0\top}\right)^\top$, and a parameter space $\bOmega_{i} = \{\bpsi_i\in\mbW\times\bTheta\times \bPhi_i\}$ with $\bPhi_i=\{\phi/\sqrt{d_i}:\phi\in\bPhi\}^{d_i}$.
			{
				To establish the parameter estimation consistency,
				we require the following assumptions.
				\begin{assumption}\label{assum:iden}
				There exist a sequence $\tau_m>0$ and a constant $\Ct>0$ such that for sufficiently large $T$, it holds that
					$\inf_{\bpsi_i\in \bOmega_{i}}\left[\overline\ell_i(\bpsi_i^*)-\overline\ell_i(\bpsi_i)\right]\geq{\min\{\frac{\|\bpsi_i^*-\bpsi_i\|^2}{\tau_m},\Ct\}}$ for any $i=1,\cdots,m$.
				\end{assumption}	
				\begin{assumption}\label{assum:li}
					The functions $\overline\ell_i(\w,\btheta,\bvarphi_i)$'s are Lipschitz continuous in the sense that there exist a constant $M>0$ such that
					$\sup_{1\leq i\leq m}\sup_{\w,\w'\in\mbW,\btheta,\btheta'\in\bTheta,\bvarphi_i,\bvarphi_i'\in\bPhi_i}
					\frac{\left|\overline\ell_i(\w,\btheta,\bvarphi_i)-
						\overline\ell_i(\w',\btheta',\bvarphi_i')\right|}{\|\w-\w'\|^2+
						\|\btheta-\btheta'\|^2+\|\bvarphi_i-\bvarphi_i'\|^2}\leq M$.
				\end{assumption}	
				
				\begin{assumption}\label{assum:true}
					There exists a $c_0>0$ such that  $\inf_{g\neq g'\in[G_0]}\left[\left\|\w_{g}^*-\w_{g'}^*\right\|+
					\left\|\btheta_g^0-\btheta_{g'}^0\right\| \right]>c_0$.
				\end{assumption}
				
				\begin{assumption}\label{assum:pi}
					Define
					$\pi_{g,m}=\frac{1}{m}\sum_{i=1}^{m}I(g_i^0=g)$ and
					$\pi_{gg',m}=\frac{1}{m}\sum_{i=1}^m$ $\sqrt{\frac{1}{d_i}\sum_{j\in\mN_i} {I(g_i^0=g,g_j^0=g')}}$, $g,g'\in [G_0]$.
					Assume that as $m\to\infty$, $\pi_{g,m}\to\pi_g$ and
					$\pi_{gg',m}\to\pi_{gg'}$ and that there exists a constant $c_{\pi}>0$ such that $\min_{g,g'\in[G_0]}\min\{\pi_g,\pi_{gg'}\}\ge c_{\pi}$.
				\end{assumption}
				
							{			
				Assumption \ref{assum:iden} is a condition to ensure the identifiability of model parameters, which essentially requires two things: (1) $\bpsi_i^*$ is a global maximizer $\overline\ell_i(\bpsi_i)$, and $\overline\ell_i(\bpsi_i^*)$ is at least $C_1$ greater than the function values at other local maximizers;
 (2) $\overline\ell_i(\bpsi_i)$ is locally concave in a neighborhood of $\bpsi_i^*$, where it holds that $\lambda_{\min}(-\ddot{\overline\ell}_i(\bpsi_i))\ge 2\tau_m^{-1}$ with $\ddot{\overline\ell}_i(\cdot)$ being the negative hessian matrix of $\overline\ell_i(\cdot)$. To see the second part, note that a second order Taylor expansion of $\overline\ell_i(\bpsi_i)$ around $\bpsi_i=\bpsi_i^*$ yields that $\overline\ell_i(\bpsi_i^*)-\overline\ell_i(\bpsi_i)\approx-\frac{1}{2}(\bpsi_i-\bpsi_i^*)^\top\left[\ddot{\overline\ell}_i(\bpsi_i^{'})\right](\bpsi_i-\bpsi_i^*) $ for some $\bpsi_i^{'}$ in a neighborhood of $\bpsi_i^*$. To further justify part (2), some straightforward calculation yields that
$
-\ddot{\overline\ell}_i(\bpsi_i^*)\approx H_{iT}(\bpsi_i^*) = \E\left[\frac{1}{T}\int_0^T \frac{\dot \lambda_i(t|\bpsi_i^*,\mH_t)\dot \lambda_i^\top(t|\bpsi_i^*,\mH_t)}{\lambda_i(t|\bpsi_i^*,\mH_t)}dt\right]$,
where $\dot \lambda_i(\cdot|\bpsi_i,\mH_t)=\partial \lambda_i(\cdot|\bpsi_i,\mH_t)/\partial \bpsi_i$. By definition, $H_{iT}(\bpsi_i)$ is a positive semi-definite matrix of dimension $\nk+d_i+3$,  whose upper left $\nk\times \nk$ submatrix is a banded matrix due to Assumption~\ref{assum:basis}. Therefore, $\lambda_{\min}(H_{iT}(\bpsi_i))$ is mainly determined by $d_i$, and may approach $0$ when $d_i$ diverges. To account for the case that the network becomes denser ($d_i$'s increase),
%Assumption~\ref{assum:iden} allows
we allow for $\tau_m\to\infty$ at a slow rate as $m, T$ increases in later theoretical discussions. In Section~B.1 of the supplementary material, we provide some empirical evidence for Assumption~\ref{assum:iden} by simulation.
					}

				Assumption \ref{assum:li} imposes a smoothness condition on the expected node-wise log-likelihood, which is reasonable for the problem under consideration. Assumption~\ref{assum:true} asserts that at least one of the background intensity or the node-specific parameter vector are well-separable between any two latent groups, which is a reasonable assumption for a wide range of applications such as the social network data studied in Section~\ref{sec:data}. Finally, Assumption \ref{assum:pi} requires that there is a sufficient number of nodes in each latent group and that there is a sufficient number of connected nodes between any two different latent groups.
				%\blue{Similar conditions are assumed in group panel data literature as \cite{su2016identifying,su2018identifying,liu2020identification}.}
				
			}
			\subsection{Model Estimation Consistency {When $G\ge G_0$}}
			
	{The membership estimation accuracy when $G\ge G_0$ is evaluated by the following measure
			\beq
			\rho_{mT}=\frac{1}{m}\sum_{g=1}^{G}\sum_{i=1}^{m}I\left(i\in\wh\mC_g, g_i^0\neq \chi(g)\right), \label{eq:mem_error_rate}
			\eeq
			where $\chi(g) = \arg\max_{g'\in [G_0]}\sum_{i = 1}^mI\left(i\in \wh \mC_g, g_i^0=g'\right)$ gives the true group label
			of the majority nodes in $\wh \mC_g = \{i: \wh g_i = g\}$, for $g\in[G]$. Note that $1-\rho_{mT}$ is referred to as the cluster ``purity" in the machine learning literature \citep{schutze2008introduction}, and is well defined even when $G>G_0$.
}			
						
The following theorem states the estimation consistency of the MLE~\eqref{mle} when $G\ge G_0$.
				
			\begin{theorem}
				\label{thm1}
				%Define $\tau_m = \big\|\1^\top[\I-(1+\delta)B]^{-1}\big\|_{\infty}$,
				%where $\delta = (c_B^{-1}-1)/2$ with $c_B$ defined in Assumption~\ref{assum:station}.
				Assume Assumptions~1--7, $G\geq G_0$, and that $\tau_m x_{mT}=o(1)$ with $x_{mT}=\sqrt{\frac{1}{T}b(\nk+d_{\max})\log(mT)[\log(\max\{b,\nk,\log(mT)\})]^3}$,
				{where $d_{\max} = \max\limits_{1\le i\le m} d_i$ and 	$\bar d = \frac{1}{m}\sum_{i = 1}^m d_i$}
				\begin{singlespace}
			 \hspace{-1.5em}
		are maximum and average out-degrees. Then it holds that as $m,T\to\infty$,
				\begin{enumerate}
					\item[(a).]			$\rho_{mT}= O_p\left(e_{mT}^{1/2}+\tau_m e_{mT}
					\right)$, with $e_{mT}=\sqrt{\frac{1}{T}b(\nk+\overline d)\log(mT)[\log(\max\{b,\nk,\log(mT)\})]^3}$;
					\item[(b).]
					$\frac{1}{m}\sum_{i=1}^{m}\|\wh\mu_{\wh g_i}(\cdot)-\mu_{g_i^0}^0(\cdot)\|_T=O_p\left(e_{mT}^{1/6}+\tau_m^{1/2}e_{mT}^{1/4}+\nk^{-\nu}\right)$, where $\wh g_i$ is the estimated membership of node $i$, $i=1,\cdots,m$, and $\wh\mu_g(\cdot)=\wh\w_g^\top\x_{\nk}(\cdot)$ for $g\in[G]$;
					\item[(c).]  $\frac{1}{m}\sum_{i=1}^{m}\big\|\wh\btheta_{\wh g_i}-\btheta_{g_i^0}^0\big\|+\frac{1}{m}\|\wh\B-\B^0\|_1=O_p\left(e_{mT}^{1/6}+\tau_m^{1/2}e_{mT}^{1/4}\right)$, where $\wh\B$ and $\B^0$ are the estimated/true $\B$ in Assumption~\ref{assum:station}, respectively.
				\end{enumerate}
							\end{singlespace}
			\end{theorem}
			The proof is given in the Supplementary Material.

			Theorem~\ref{thm1} asserts that even if $G$ is over-specified, both the clustering error $\rho_{mT}$ and the parameter estimation errors converge to $0$, and the convergence rate is primarily determined by $T$ rather than the number of nodes $m$. The convergence rates are also negatively impacted by large values of $b$ and $\ol d$, where the former is the triggering function range that controls the strength of serial dependence among offspring of the same parent, and the latter represents the level of connectivity among network nodes. Furthermore, the quantity $\tau_m$ in Assumption~\ref{assum:iden} is also affected by $d_i$'s and may also grow as the network becomes denser, leading to a slower convergence rate.  Particularly, it is of important practical interest to consistently estimate the transition matrix $\B^0$, since the estimator $\wh\B$ can be used to study the network interactions following Section~\ref{sec::brach}.
			We remark that Theorem~2 is unlikely to hold when $G<G_0$, in which case nodes from different groups are forced into the same group, resulting in biased parameter estimators.

			%Furthermore, by Theorem~\ref{thmfam}, $\tau_{m}$ is roughly (when $\delta=0$) the average number of offspring events on the network that are triggered by a parent event occurring on the most influential node. In this sense, $\tau_m$ can be viewed as a measure of the network connectivity level.
			%			A larger $\tau_m$ would potentially result in more clustered event times and it would require a larger $T$ to offset such an effect for model estimation.  Condition~\eqref{assum:T_consist} requires that $\tau_m$ cannot exceed the order of $O(T^{1/8})$ as $m$ increases, which is an upper bound  new to the literature.

			\subsection{Selection Consistency of Number of Groups}
	In this subsection, we study the selection consistency of the LIC criterion proposed in (\ref{LIC}).
	To this end, we show  in
	the following Theorem that $\wh G$ selected by maximizing LIC estimates $G_0$ consistently
	when the penalty parameter $\lambda_{mT}$ is appropriately chosen.
	\bet\label{thm_LIC}
	Assume Assumptions 1-7 and $x_{mT}\tau_m = o(1)$ with $x_{mT}$ as in Theorem~\ref{thm1}. % (\ref{assum:T_consist}),
	%and (\ref{assum:nb}).
If $\lambda_{mT}\tau_m=o(1)\text{ and }
{\lambda_{mT}^{-1}\sqrt{\frac{1}{T}b(\nk+\overline d)\log(mT)[\log(\max\{b,\nk,\log(mT)\})]^3}=o(1)}$,
	then $P(\wh G = G_0) \to 1$.
	\eet
	The proof is given in the Supplementary Material.	
	
	Theorem~\ref{thm_LIC} requires that
	$\lambda_{mT}$ converges to zero but not too fast.
	In our numerical examples, we set
	$\lambda_{mT} = ({15 T})^{-1} \left(\median_{1\le i\le m}n_i\right)^{0.6}\bar d^{0.25}$
	and verify its finite sample performances in details in~Section~\ref{sec:sim}. Such a choice ensures that $\wh G$ does not depend on the unit of $T$.

	% \xggrev{
		% For some $1/2<\delta_1<1$, $\delta_2>0,\delta_3>0$ and $C>0$, we can consider
		% \[
		% \begin{split}
			% \lambda_{mT}&=\frac{1}{T}\left([\log (mT)]^{3}\tau_{m}^{2}b\sqrt{T\bar d\left[{\nkb}\log(\nk\bar d){+} \log (m)\right]}\right)\approx \frac{C}{T} T^{\delta_1}\bar d^{\delta_2}\left[\log\left(\sum_{i=1}^{m}n_i\right)\right]^{\delta_3}\\
			% &\approx \frac{C}{T} \left(\frac{1}{m}\sum_{i=1}^{m}n_i\right)^{\delta_1}\bar d^{\delta_2}\left[\log\left(\sum_{i=1}^{m}n_i\right)\right]^{\delta_3}.
			% \end{split}
		% \]
		% one can try multiple values of $\delta_1,\delta_2,C$ through simulation
		% }			

}		
			
			\subsection{Convergence Rates and Asymptotic Normality {When $G=G_0$}}
			\label{sec:norm}
	We now study the convergence rates and asymptotic normality of the model parameter estimators when $G = G_0$,  { which requires some modifications of the MLE~\eqref{mle} as follows.	
	
				{\bf Membership refinements.} The MLE~\eqref{mle} maximizes the overall log-likelihood function~\eqref{comp_lik}, but not necessarily each node-specific likelihood $\ell_i\big(\w_g,\btheta_{g},\bvarphi_{g,\mG_i}|\mH_T\big)$, $i=1,\cdots,m.$ As a result, if $\ell_i\big(\wh\w_{\wh g_i},\wh\btheta_{\wh g_i},\wh\bvarphi_{\wh g_i,\wh\mG_i}|\mH_T\big)$ is too low for some $i$, its membership estimate $\wh g_i$ may be incorrect. To address this issue, we propose a membership refinement strategy. {One way to check whether $\wh g_i$ results in a too small $\ell_i\big(\wh\w_{\wh g_i},\wh\btheta_{\wh g_i},\wh\bvarphi_{\wh g_i,\wh\mG_i}|\mH_T\big)$ is to compare it to the profile likelihood $\ell_i^{\rm p}(g|\mH_T)=\sup_{\varphi_i\in\bPhi_i}\ell_i\big(\wh\w_{g},\wh\btheta_{g},\bvarphi_i|\mH_T\big)$. If for some $\wt g$ such that $\ell_i^{\rm p}(\wt g|\mH_T)$ is much greater than $\ell_i\big(\wh\w_{\wh g_i},\wh\btheta_{\wh g_i},\wh\bvarphi_{\wh g_i,\wh\mG_i}|\mH_T\big)$, it may be a sign to relabel the node $i$ to group $\wt g$. Furthermore, instead of maximizing over all $\bvarphi_i\in\bPhi_i$, our theoretical investigation suggests that, for any given $g$, it suffices to} define the profile likelihood function as $\ell_i^{\rm p}(g|\mH_T)=\ell_i\big(\wh\w_g,\wh\btheta_{g},\wh\bvarphi_i^{\rm p}(g)|\mH_T\big)$, where
				$$
				\wh\bvarphi_i^{\rm p}(g)=\argmax_{\bvarphi_i\in\left\{\wh\bvarphi_{g',\mG_i}:  g'\in[G],\mG_i\in [G]^{d_i}\right\}}\ell_i\big(\wh\w_g,\wh\btheta_{g},\bvarphi_i|\mH_T\big),\quad \text{ for } g\in [G].
				$$
				Denote  $\wh g_i^{\dag}=\argmax_{1\leq g\leq G} \ell_i^{\rm p}(g|\mH_T)$. The refined membership of node $i$ is then defined as
				\vskip -2em
				\begin{singlespace}
				\be
				\label{grefine}
				\begin{split}
					&\wh g_i^r=\begin{cases} \wh g_i,&\text{ if } \ell_i^{\rm p}(\wh g_i^{\dag}|\mH_T)-\ell_i\left(\wh\w_{\wh g_i},\wh\btheta_{\wh g_i},\wh\bvarphi_{\wh g_i,\wh\mG_i}|\mH_T\right)\leq 0,\\
						\wh g_i^{\dag} ,&\text{ if } \ell_i^{\rm p}(\wh g_i^{\dag}|\mH_T)-\ell_i\left(\wh\w_{\wh g_i},\wh\btheta_{\wh g_i},\wh\bvarphi_{\wh g_i,\wh\mG_i}|\mH_T\right)> 0.
					\end{cases}\\
				\end{split}
				\ee	
			\end{singlespace}
		With this modification, we define the estimated groups $\wh \mC_g^r = \{i: 1\le i\le m, \text{ and }\wh g_i^r = g\}$, $g\in [G]$, and the true groups $\mC_{g'}^0 = \{i: 1\le i\le m, \text{ and } g_i^0 = g'\}$, $g'\in [G_0]$. When $G= G_0$, a stronger version of Theorem~\ref{thm1} can be established as following.

\bet\label{thm_member}
Assume $G= G_0$, Assumptions~1--7, and that $\tau_m^2x_{mT}=o(1)$ with $x_{mT}$ as defined in
Theorem~\ref{thm1}. Then, as $m,T\to\infty$, it holds that
\begin{enumerate}
\item[(a).] $\sup_{1\leq i\leq m}\left[\big\|\wh\mu_{\wh g_i^r}(\cdot)-\mu_{g_i^0}^0(\cdot)\big\|_T+\big\|\wh\btheta_{\wh g_i^r}-\btheta_{g_i^0}^0\big\|\right]=O_p(x_{mT}^{1/3}+\tau_m x_{mT}^{1/2}+\nk^{-\nu})$ where $\wh g_i^r$ is the estimated membership of node $i$ after refinement, $i=1,\cdots,m$;
\item[(b).] if we further assume that $\tau_m^4x_{mT}=o(1)$, then for each $g\in [G]$,
there exists $g'\in [G_0]$ such that
$P(\wh \mC_g^r=\mC_{g'}^0) \to 1$ as $m,T\to\infty$.
\end{enumerate}
\eet
The proof is given in the Supplementary Material.

Theorem \ref{thm_member} (a) states that when $G=G_0$, the background intensity $\mu_{g_i}^0(\cdot)$ and the node-specific parameters $\btheta_{g_i^0}^0$ can be consistently estimated for all network nodes. Theorem \ref{thm_member} (b) shows that, after some label permutation, all nodal memberships can be correctly identified with a probability tending to 1, which is crucial to establishing asymptotic normality of the parameter estimators.

Denote the refitted estimator $\big(\wh{\ut\w}^{r\top},\wh{\ut\btheta}^{r\top}, \wh{\ut\bphi}^{r\top}\big)^\top= \argmax_{\ut \w,\ut\btheta, \ut\bphi}\ell(
\ut \w,\ut\btheta, \ut\bphi, \wh\mG^r|\mH_T)$  with $\wh \mG^r = (\wh g_1^r, \wh g_2^r,\cdots, \wh g_m^r)^\top $ being the refined membership estimator.	Due to 	Theorem \ref{thm_member} (b), for ease of presentation, we drop the notation $\mG_0$ whenever there is no ambiguity. In particular, the model parameters that need to be estimated become $\ut\bpsi=(\ut\w^\top,\ut{\btheta}^\top,\ut\bphi^\top)^\top$, whose true values are $\ut\bpsi^*=(\ut\w^{*\top},\ut{\btheta}^{0\top},\ut\bphi^{0\top})^\top$. For any given $\ut\bpsi$,  we view the conditional intensity function~\eqref{intensity} as a function of $\ut\bpsi$ and denote it as $\lambda_i(t|\ut\bpsi,\mH_t)$ for any $t\in[0,T]$ and $i=1,\cdots,m$. Define the matrix
			\be
			\label{hess}
			\H_{mT}(\ut\bpsi) = \frac{1}{mT}{\sum_{i=1}^{m}}\int_0^T \E\left[\frac{\dot \lambda_i(t|\utilde\bpsi,\mH_t)\dot \lambda_i^\top(t|\utilde\bpsi,\mH_t)}{\lambda_i(t|\utilde\bpsi,\mH_t)}\right]dt,
			\ee
			where %$\lambda_i^0(\cdot|\mH_t)$ is the true conditional intensity and
$\dot \lambda_i(\cdot|\ut\bpsi,\mH_t)=\partial \lambda_i(\cdot|\ut\bpsi,\mH_t)/\partial\ut\bpsi$, $i=1,\cdots,m$.
To establish the asymptotic normality result, we assume the following condition
for $\H_{mT}(\ut\bpsi)$.

			\begin{assumption}\label{assum:convex}
				There exist positive constants $\epsilon,\tau_{\min},\tau_{\max}$ such that for any $\|\ut\bpsi-\ut\bpsi^*\|\leq \epsilon$, one has that $\lambda_{\min}\Big[\H_{mT}(\ut\bpsi)\Big]\ge {\tau_{\min}}$ {and $\lambda_{\max}\Big[\H_{mT}(\ut\bpsi)\Big]\le {\tau_{\max}}$} for sufficiently large $m,T$.
			\end{assumption}
			Lemma~D.7 in the Supplementary Material shows that, under suitable conditions, $\H_{mT}(\ut\bpsi)$	is asymptotically equivalent to the negative Hessian matrix of $\overline{\ell}(\ut\bpsi)$ in~\eqref{ellbar}. Therefore, Assumption~\ref{assum:convex} essentially assumes that $\overline{\ell}(\ut\bpsi)$  is locally concave in a neighborhood of $\bpsi^*$, which is a mild assumption similar to Assumption~\ref{assum:iden}. In Section~B.2 of the supplementary material, we provide more discussions on  Assumption~\ref{assum:convex} through some simulation studies.		
			
			The following theorem establishes the convergence rates of the background intensity estimators
			and the asymptotic normality of the model parameters.
			
			\bet\label{thm_normal}
			Assume Assumptions 1-8 and and that $\tau_m^2x_{mT}=o(1)$ with $x_{mT}$ as defined in
				Theorem~\ref{thm1}. Then if $G=G_0$, the following holds.
			\begin{enumerate}
				\item[(a).] If  ${\nk}/{\sqrt{mT}}=o(1)$ and $\nu>1/2$,
				then as $m,T\rightarrow \infty$, one has that
				\begin{align}
					\max_{1\le g\le G}\big\|\wh\mu_g^r(\cdot) - \mu_g^0(\cdot)\big\|_T
					=   O_p\left({\sqrt{{\nk}/mT}}+\nk^{-\nu}\right).\label{conv_rate}
				\end{align}
				\item[(b).] Denote $\wh{\ut\balpha}^r=(\wh{\ut\btheta}^{r\top},\wh{\ut\bphi}^{r\top})^\top$, ${\ut\balpha}^0=({\ut\btheta}^{0\top},{\bphi}^{0\top})^\top$,
				and let $\bSigma_{\alpha}=\lim_{(m,T)\rightarrow\infty}\mI_{\alpha}\H_{mT}^{-1}(\ut\bpsi^*)\mI_{\alpha}^\top$ with $\mI_\alpha = (\zero_{(3G+G^2)\times \nk}, \I_{(3G+G^2)\times (3G+G^2)})$. If it further holds that $ {\sqrt{mT}}/{\nk^{\nu}} = o(1)$,
				%\beq
				%\frac{(mT)^{1/2}}{\nk^r{\sqrt{\tau_m b}\log(mT)}} = o(1), { \text{ and } \frac{
						%		\nkb^4\tau_m^3 b^3[\log(mT)]^4}{mT}=o(1)},\label{normal_cond}
				%\eeq
				{then
					$\sqrt{mT}\big(\wh{\ut\balpha}^r - \ut\balpha^0\big)\xrightarrow{d} N(\0, \bSigma_{\alpha})$ as $m,T\to\infty$.}
			\end{enumerate}

			\eet
			The proof is given in the Supplementary Material.
			
			Theorem~\ref{thm_normal} part (a) gives an upper bound of the convergence rate of the nonparametric background intensity estimators that consists of two parts.
			{The first part
				$\nk/mT$ is due to the estimation variance, and the second part is introduced by the basis approximation error.}
			%although \eqref{conv_rate} may not be the sharp upper bound. The convergence rate in~\eqref{conv_rate} slows down when $b$ and $\tau_m$ increase, which is reasonable because larger $b$ and $\tau_m$ indicate a greater level of dependence among observed event times on the network.
			Part (b) gives sufficient conditions under which  the $\sqrt{mT}$-convergence rate for $\wh{\ut\balpha}^r$
			and the asymptotic normality can be established.
			The result is proved by using a martingale central limit theorem \citep{fleming2011counting}.
			{Particularly it requires that $\sqrt{mT}/\nk^{\nu} = o(1)$ so that the approximation bias is dominated by the estimation variance, a popular strategy used in the nonparametric inference literature.
				In a simple setting with $\max\{b, \tau_m, d_{\max}\}<\infty$, Theorem~\ref{thm_normal} (b) requires $\nk$ to satisfy
				$(mT)^{1/(2\nu)}\ll \nk\ll \left({T}/[\log(mT)\{\log(\log(mT))\}^3] \right)^{1/(1+\delta)}$ for some $\delta>0$, where $a_{mT}\ll b_{mT}$ means $a_{mT}/b_{mT}\to 0$ as $m,T\to\infty$. As one can see, a larger $\nu$, which suggests a smoother background intensity, will make the above condition more likely to hold.}
			A plug-in estimator of the covariance matrix $\bSigma_{\alpha}$ can be constructed straightforwardly using the definition of $\H_{mT}(\ut\bpsi^*)$ in~\eqref{hess} for  statistical inferences.
}		
		
		\section{Simulation Studies}
		\label{sec:sim}
		{
			In this section, we conduct simulation studies to evaluate the numerical performance of the proposed GNHP model.
			The following two types of network structures are considered:
			
{\sc Stochastic Block Model (SBM).}
This model is widely used in the community detection literature. The network consists of $m$ nodes belonging to $3$ blocks and each node is randomly assigned a block label with a probability of $1/3$. An edge between two nodes is generated with a probability $0.3 m^{-0.3}$ if they are in the same block, or with a probability of $0.3 m^{-0.8}$ otherwise.

{\sc Power-law Network.} {Such a network resembles the commonly observed social network structure where most nodes have few followers while a small fraction of nodes have a large number of followers. For each node $i$, the number of randomly picked followers is $ 4 f_i$, where $f_i$ follows the power-law distribution $P(f_i = f) = c f^{-2}, 0 \leq f \leq m$, where $c$ is the normalizing constant. }
			
			{For each type of network, we random assign each node to $G_0=3$ latent groups with } group proportions as
			$\bpi_{m} = (\pi_{1,m}, \pi_{2,m}, \pi_{3,m})^\top =(0.3, 0.4, 0.3)^\top$.
			To mimic typical daily activities of social network users, the background intensity of each latent group takes a periodic form
\[
\mu_g(t) = C_g \sum_{h = 1}^{H_g} b_{h,g} \exp\{-{ (t - a_{h,g})^2}/{w_{h,g}} \} \quad g\in [G_0], ~ t \in [0, \omega].
\]
			Triggering functions of all groups are of the form   $f(t,\gamma)\propto \gamma\exp(-\gamma t)\mathbf 1\{t \leq b\}$ with $b$ = 5 hours.
			Model parameters (except $C_g$'s and $b_{h,g}$'s) are listed in Table \ref{tab:base_inten}, and the true background intensities are illustrated in Figure~\ref{base_intense_sim}, {which have the same shapes of the estimated background intensities in our Sina Weibo data analysis (see Figure~\ref{base_intense}), where the} common period is set as $\omega=24$ hours and $C_g$ is chosen such that $\int_0^{\omega} \mu_g(t) dt$ is set at the targeted values in Table \ref{tab:base_inten}.
			
			\vspace{1em}
			
   \begin{minipage}{\textwidth}
	\hspace{-2em}
	\begin{minipage}[b]{0.49\textwidth}
		\scalebox{0.6}{
			\begin{tabular}{c|c|cc|cc}
				\hline
				\hline
				& \multicolumn{1}{c | }{Background intensity }&  \multicolumn{2}{c | }{Momentum}&  \multicolumn{2}{c  }{Network}\\
				$g$ & $\int_0^{\omega} \mu_g(t) dt$ & $\beta_g$& $\eta_g$ & $\gamma_g$ & $(\phi_{g1},\phi_{g2},\phi_{g3})$\\
				\hline
				1 &  2.5 & 0.5& 3 & 2& (0.4, 0.1, 0.1) \\
				2 &  1 & 0.4& 2 & 4& (0.6, 0.4, 0.5) \\
				3 &  0.5 & 0.7& 4 & 1& (0.15, 0.2, 0.1)\\
				\hline
				\hline
			\end{tabular}
		}
	%	\vspace{2em}
		\captionof{table}{Parameter setting for $G_0 = 3$ groups.}\label{tab:base_inten}
	\end{minipage}
	\hspace{-0.5em}
	\begin{minipage}[b]{0.49\textwidth}
		\includegraphics[trim=0cm 0cm 0cm 2cm,clip,scale=0.25]{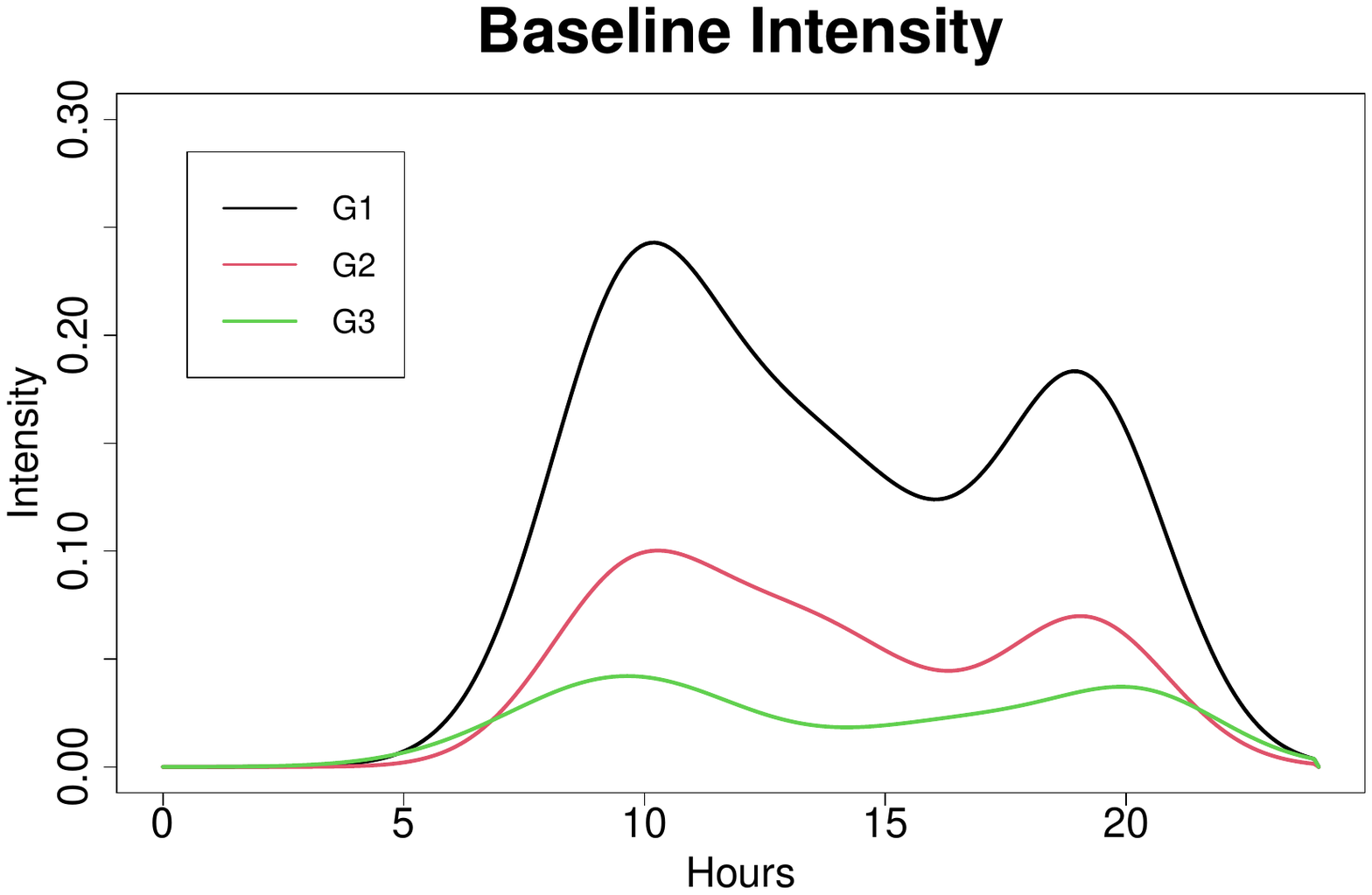}
		\vspace{-2em}
		\captionof{figure}{True background intensities.}\label{base_intense_sim}
	\end{minipage}
\end{minipage}

			\vspace{1.5em}

			Data are simulated from model~\eqref{intensity0} with $ m \in \{100, 200\}$ and $ T \in \{5 \omega, 10 \omega, 20 \omega, 40 \omega\}$. B-splines are used to approximate the background intensity $\mu_g(\cdot)$'s, with $\nk=11$ equally spaced internal knots between $[0,\omega]$. For each $(m,T)$, summary statistics are computed based on $K=1000$ simulation runs.  For the $k$th simulation run, we denote the estimators as $\{\wh \mu_g^{(k)}(\cdot), \wh \beta_g^{(k)},
			\wh \eta_g^{(k)}, \wh \gamma_g^{(k)}, \wh\phi_{gg'}^{(k)}\}$ and the group membership estimator as
			$ \wh\mC^{(k)}=\{\wh g_i^{(k)}: 1\le i\le m\}$. The group membership estimation accuracy rate is then computed  by
			GAR$ = K^{-1}\sum_{k=1}^K$ GAR$^{(k)}$ with GAR$^{(k)}=1-\rho_{mT}^{(k)}$, where $\rho_{mT}^{(k)}$ is obtained by applying \eqref{eq:mem_error_rate} to $\wh\mC^{(k)}$.

			\subsection{Estimation Accuracy When $G=G_0$}
			\label{sec:simu:G0}
			When $G=G_0$, parameter estimation accuracy can be evaluated by the root mean squared error (RMSE) of estimates after some label switching.
			The estimation accuracy $\mu_g(\cdot)$ is evaluated by
			$K^{-1}\sum_{k=1}^K\int |\wh \mu_g^{(k)}(t) - \mu_g(t)|dt$.
			For comparison, we also present the estimation accuracy of the ``Oracle" estimator for which the true group memberships of all nodes are known and fixed when finding the MLE. All simulation results are summarized in Figure~\ref{fig:baseline:plot} and Tables~\ref{tab:sbm:correctG}-\ref{tab:pl:correctG}.

			%		\begin{figure}[H]
				%			\centering
				%			\includegraphics[width = 0.75\textwidth]{images/baseline_est.pdf}
				%			\caption{\small Estimation of baseline functions for three classes via cubic Bsplines with standard deviations.} \label{base_intense_sim}
				%		\end{figure}
			%	
			
\begin{figure}[ht!]
	\begin{center}
		\subfigure{\includegraphics[trim=0cm 0cm 0cm 0cm,clip,scale=0.31]{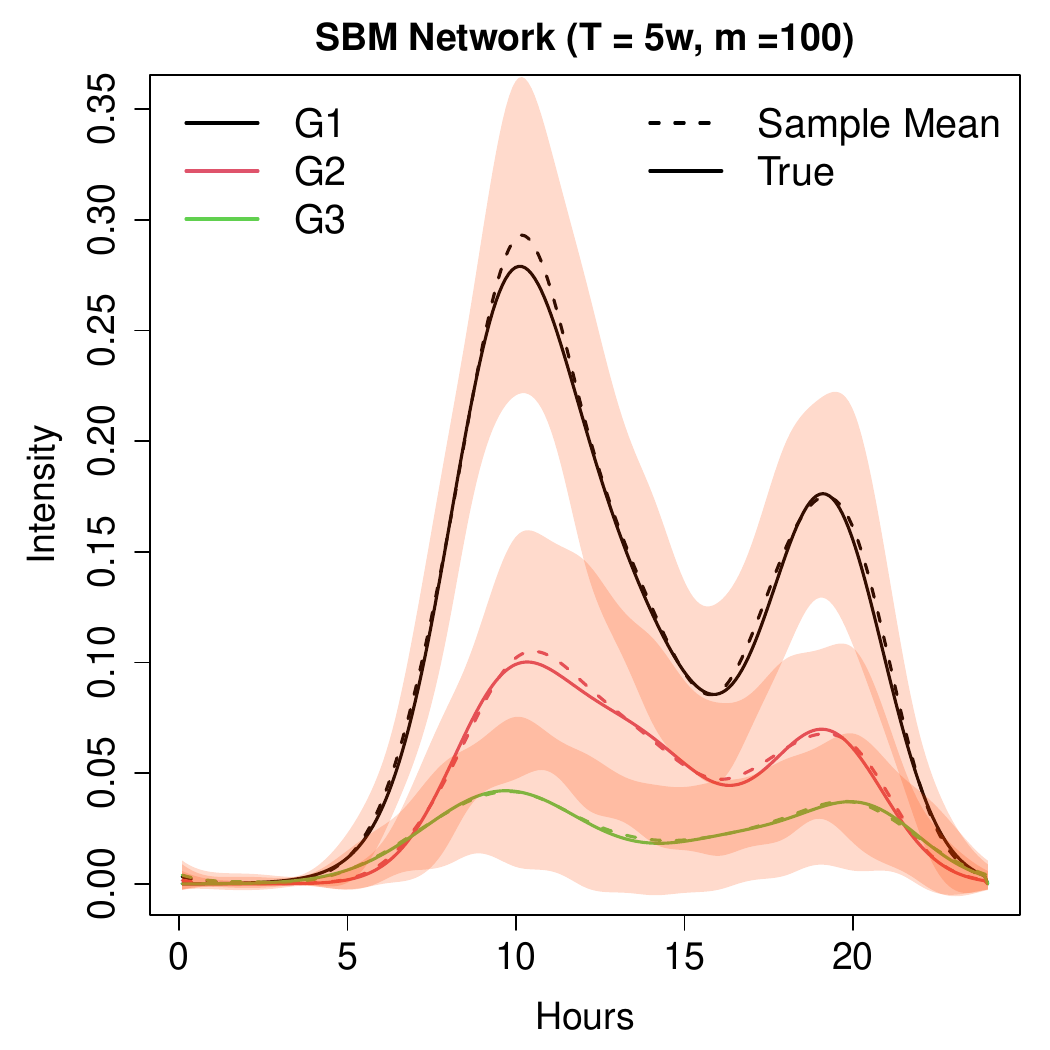}}
		\subfigure{\includegraphics[trim=0cm 0cm 0cm 0cm,clip,scale=0.31]{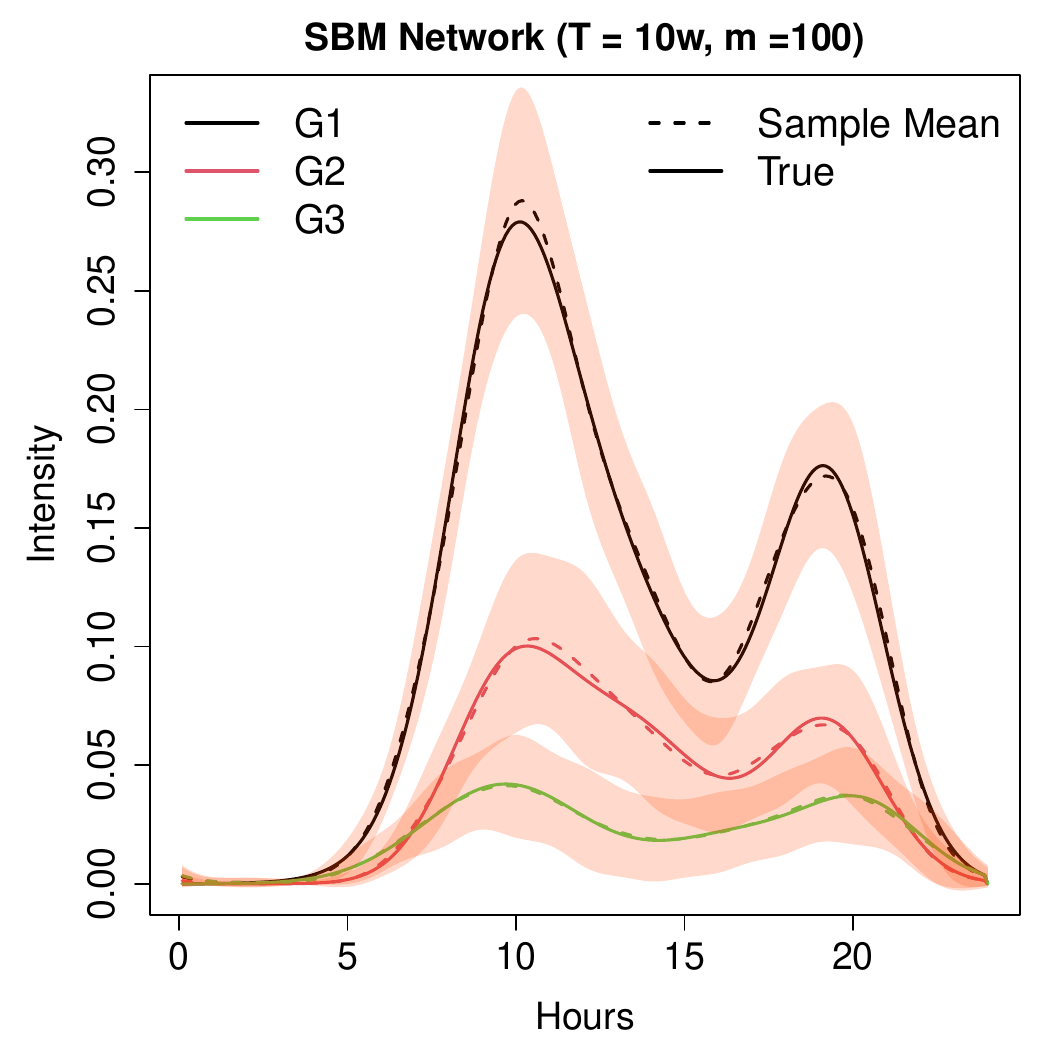}}
		\subfigure{\includegraphics[trim=0cm 0cm 0cm 0cm,clip,scale=0.31]{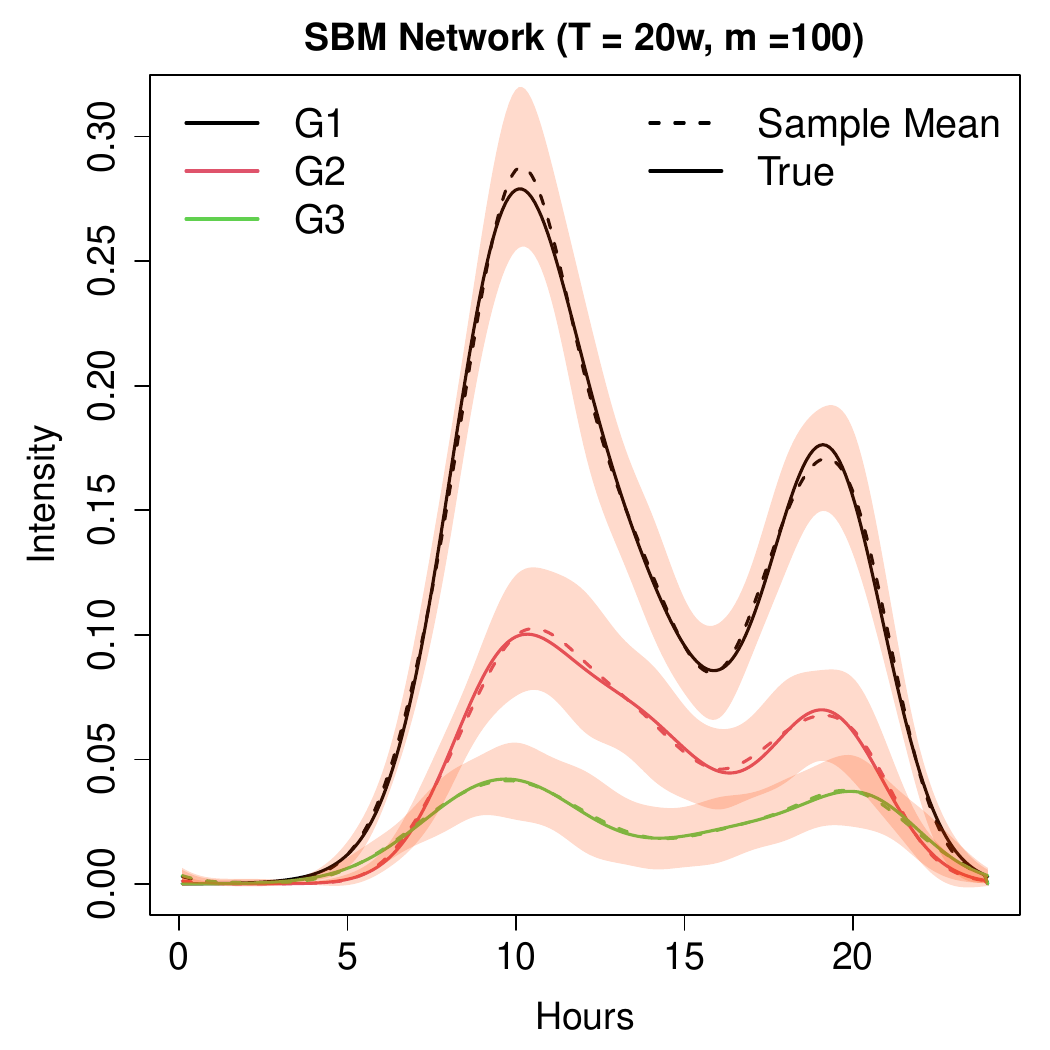}}
				\subfigure{\includegraphics[trim=0cm 0cm 0cm 0cm,clip,scale=0.31]{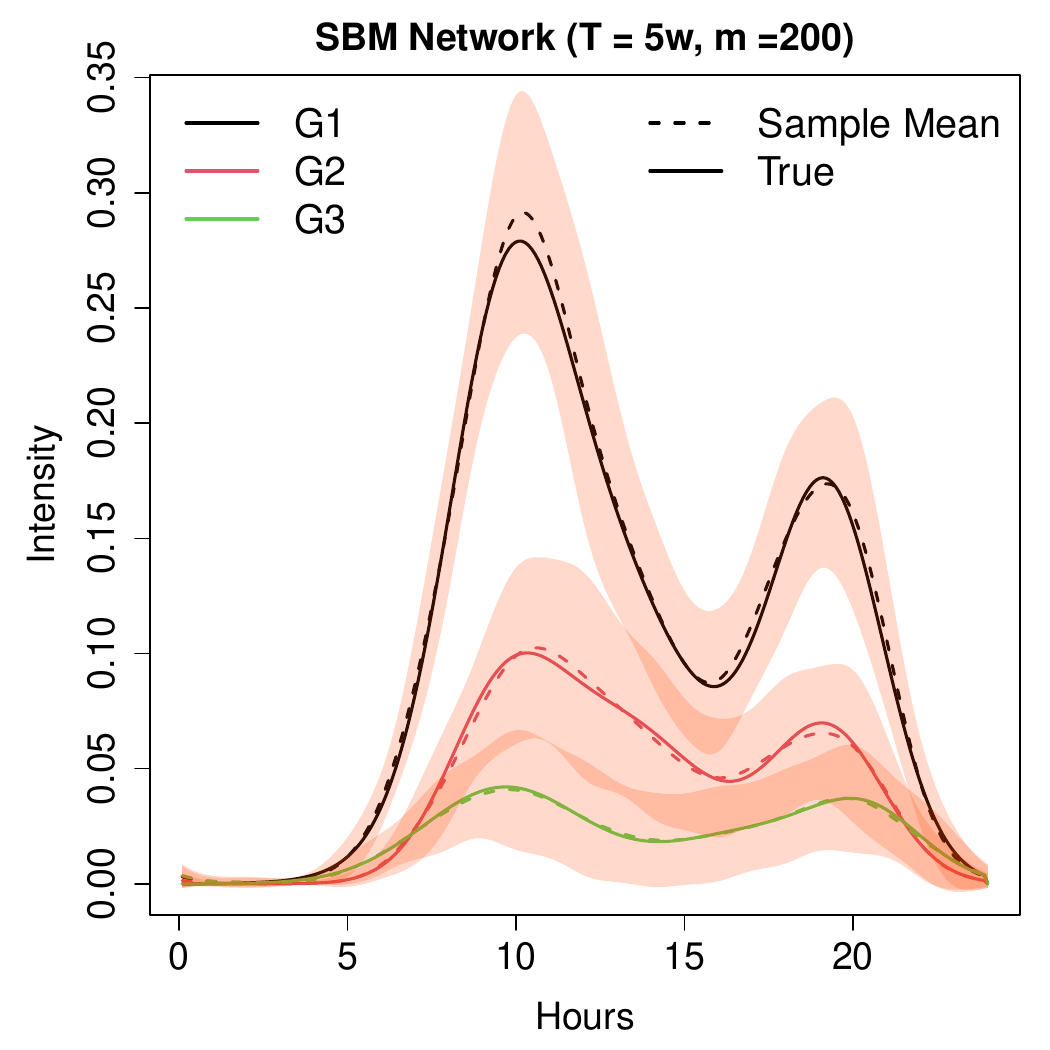}}
		\subfigure{\includegraphics[trim=0cm 0cm 0cm 0cm,clip,scale=0.31]{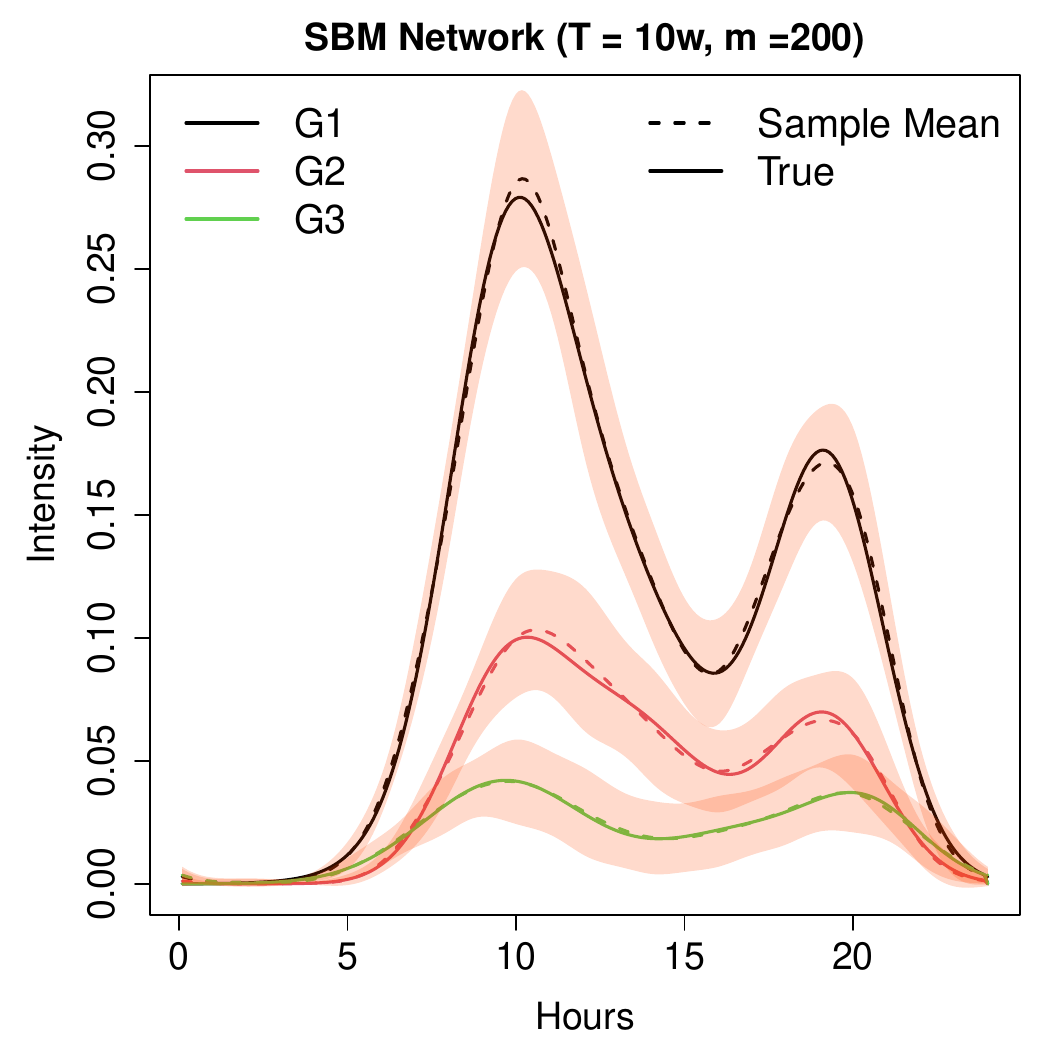}}
		\subfigure{\includegraphics[trim=0cm 0cm 0cm 0cm,clip,scale=0.31]{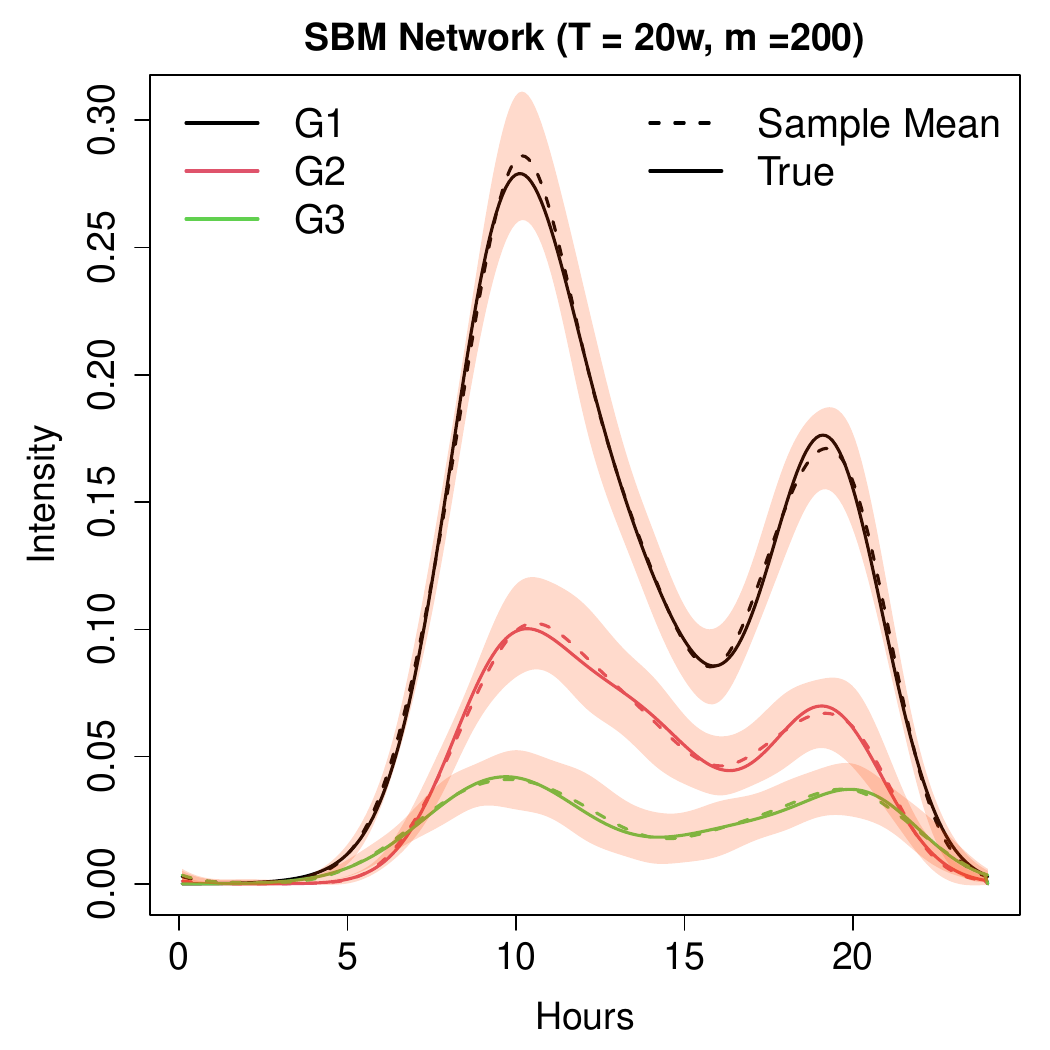}}
	\end{center}
\vskip -2em
	\caption{The mean estimated background intensities with $\pm 1.96$ times sample standard errors.}\label{fig:baseline:plot}
\end{figure}

{From Figure~\ref{fig:baseline:plot}, we observe that the estimated background intensities become closer to the true background intensities as either $m$ or $T$ increases, which supports our theoretical findings in Theorems~\ref{thm1} (b) and~\ref{thm_normal} (a).  Results for the Power-law network are similar and given in Section~\ref{sec:power-sim} of the supplementary material. From Tables~\ref{tab:sbm:correctG}-\ref{tab:pl:correctG}, we can see that when both $m$ and $T$ are small, the group memberships of a proportion of network nodes may be incorrectly estimated in both network settings. However, in all case scenarios, the GARs of the GNHP model are much better than those of the intensity-based K-means algorithm, which is used to provide initial membership estimates for the GNHP estimation, see Section~A.2.1 of the supplementary material for details. As $m$ and/or $T$ increases, the GARs of the GNHP gradually approach $1$. This is consistent with our theoretical findings in Theorem~\ref{thm_member}. }  Consequently, as $m$ or $T$ increases, the estimation accuracy of the model parameters improves steadily and approaches that of the ``Oracle" estimator. Overall, the simulation results support the estimation consistency of the proposed GNHP model when the group number $G$ is correctly specified.

\begin{table}[ht]
	\centering
	\caption{Estimation Accuracy of GNHP for SBM network when $G=G_0$.}\label{tab:sbm:correctG}
	\scalebox{0.58}{
		\begin{tabular}{cc|ccccccc|cc}
			\hline
			\hline
			\multicolumn{10}{c}{$m = 100$} \\
			\hline
			& &$\mu$ ($\times10^{-3}$) & $\beta$($\times10^{-3}$)& $\eta$ ($\times10^{-3}$)& $\gamma$ ($\times10^{-3}$)&  & $\phi$ ($\times10^{-3}$)& & GAR (s.e.) & GAR (s.e.)\\
			$T$  &  &GNHP (Oracle)& GNHP (Oracle)& GNHP (Oracle)& GNHP (Oracle)&  & GNHP (Oracle) & &(GNHP with & (Intensity-based \\
			& $g/g'$ &-&-&-&-& 1 & 2&3 &refinement) &K-means) \\
			\hline
			\multirow{3}{*}{5$\omega$}&1&340 (309)&16.3 (15.9)&199 (189)&212 (189)&29.1 (26)&45.3 (32.1)&29.6 (26.3)&\multirow{3}{*}{95 (2.3)}&\multirow{3}{*}{53 (5.1)}\\
			&2&271 (249)&24.7 (22.4)&228 (211)&357 (336)&40.3 (39.3)&55.4 (51.1)&52.8 (48.9)&&\\
			&3&192 (187)&26.5 (25.3)&431 (407)&183 (177)&21.7 (21.2)&34 (32.3)&29.1 (28.7)&&\\
			\hline
			\multirow{3}{*}{10$\omega$}&1&229 (228)&10.7 (10.7)&129 (127)&133 (127)&18.3 (17.9)&24.6 (23.2)&19.5 (18.3)&\multirow{3}{*}{99 (0.93)}&\multirow{3}{*}{59 (4.2)}\\
			&2&184 (181)&15.4 (15.2)&148 (149)&237 (219)&28.2 (27.6)&35.7 (35.8)&34.4 (33.9)&&\\
			&3&132 (132)&18 (18.4)&283 (284)&110 (109)&14.4 (14.4)&21.4 (21.9)&18.7 (18.5)&&\\
			\hline
			\multirow{3}{*}{20$\omega$}&1&165 (164)&7.31 (7.24)&92.7 (92.1)&93.5 (91)&13 (12.3)&15.4 (15.1)&12.3 (11.6)&\multirow{3}{*}{100 (0.2)}&\multirow{3}{*}{64 (3.0)}\\
			&2&131 (129)&11.2 (11.7)&95.3 (95.6)&161 (164)&19.1 (19.8)&24.5 (24.3)&24 (24.2)&&\\
			&3&95.8 (96.2)&12.5 (13.4)&207 (212)&78 (78.6)&10.3 (10.2)&15.8 (16)&13.6 (13.1)&&\\
			\hline
			\multirow{3}{*}{40$\omega$}&1&126 (126)&5.19 (5.35)&65.6 (63.7)&69.1 (67.2)&9.06 (9.32)&11.1 (11.2)&9.47 (9.66)&\multirow{3}{*}{100 (0)}&\multirow{3}{*}{66 (2.5)}\\
			&2&93.8 (93.3)&7.77 (8.21)&66.5 (66.4)&110 (109)&13.3 (13.2)&16.7 (17.4)&15.8 (16.1)&&\\
			&3&69 (69)&8.71 (8.64)&140 (143)&55.1 (54.5)&7.05 (6.96)&11.2 (10.2)&9.49 (9.32)&&\\
			\hline
			\multirow{3}{*}{5$\omega$}&1&272 (250)&12.2 (11.2)&148 (133)&218 (199)&36 (26)&29.9 (26.1)&25.3 (22.4)&\multirow{3}{*}{94 (1.7)}&\multirow{3}{*}{49 (3.2)}\\
			&2&205 (182)&16.3 (14.3)&141 (129)&257 (236)&27.8 (27.3)&35.4 (32.7)&37.1 (34.7)&&\\
			&3&152 (146)&18.2 (18.1)&307 (293)&144 (141)&18.3 (18)&28.4 (26.6)&20.5 (20)&&\\
			\hline
			\multirow{3}{*}{10$\omega$}&1&181 (179)&8.58 (8.4)&99.7 (98.5)&138 (136)&19.2 (18.1)&17.8 (17.4)&17.7 (17.5)&\multirow{3}{*}{99 (0.72)}&\multirow{3}{*}{55 (3.4)}\\
			&2&136 (133)&10.5 (10.3)&95.3 (91.8)&167 (166)&19.6 (19.1)&23.9 (23.3)&24.7 (24.4)&&\\
			&3&106 (106)&13.1 (13.3)&196 (190)&98.1 (101)&12.2 (12)&19.3 (19.5)&14.1 (14.2)&&\\
			\hline
			\multirow{3}{*}{20$\omega$}&1&133 (133)&5.65 (5.74)&69.5 (67.8)&98.5 (96.5)&12.7 (12.9)&12 (12.3)&11.3 (11.6)&\multirow{3}{*}{100 (0.17)}&\multirow{3}{*}{64 (4.4)}\\
			&2&98.7 (97.7)&6.91 (7)&66.2 (68)&115 (115)&12.8 (13.1)&17.2 (17.1)&17.2 (17.4)&&\\
			&3&77 (77.9)&8.81 (8.72)&144 (146)&64.6 (63.4)&8.98 (8.94)&13.7 (13.7)&10 (9.86)&&\\
			\hline
			\multirow{3}{*}{40$\omega$}&1&103 (103)&3.99 (4.01)&50.5 (50.1)&69.5 (69.3)&9.17 (9.21)&8.49 (8.44)&8 (7.91)&\multirow{3}{*}{100 (0)}&\multirow{3}{*}{71 (4.4)}\\
			&2&72.3 (72.7)&5.01 (5.05)&47.4 (47.5)&82.2 (82.3)&9.37 (9.41)&12.2 (11.9)&12.2 (12.2)&&\\
			&3&56.8 (56.9)&6.33 (6.32)&102 (101)&45.2 (44.7)&6.26 (6.24)&9.47 (9.33)&6.85 (6.77)&&\\
			\hline
			\hline
		\end{tabular}
	}
\end{table}

\begin{table}[ht]
	\centering
	\caption{Estimation Accuracy of GNHP for Power Law  network when $G=G_0$.}\label{tab:pl:correctG}
	\scalebox{0.58}{
		\begin{tabular}{cc|ccccccc|cc}
			\hline
			\hline
			\multicolumn{10}{c}{RMSE TABLE, PL setting.  } \\
			\hline
			\multicolumn{10}{c}{$m=100$} \\
			\hline
			& &$\mu$ ($\times10^{-3}$) & $\beta$($\times10^{-3}$)& $\eta$ ($\times10^{-3}$)& $\gamma$ ($\times10^{-3}$)&  & $\phi$ ($\times10^{-3}$)& & GAR (s.e.) & GAR (s.e.)\\
$T$  &  &GNHP (Oracle)& GNHP (Oracle)& GNHP (Oracle)& GNHP (Oracle)&  & GNHP (Oracle) & &(GNHP with & (Intensity-based \\
& $g/g'$ &-&-&-&-& 1 & 2&3 &refinement) &K-means) \\			
			\hline
			\multirow{3}{*}{5$\omega$}&1&395 (349)&17.8 (15.5)&213 (193)&410 (362)&61 (50.4)&43.2 (38.8)&51.5 (46)&\multirow{3}{*}{93 (2.8)}&\multirow{3}{*}{49 (4.3)}\\
			&2&340 (290)&23.5 (20.2)&213 (191)&449 (405)&50.4 (46.6)&55.5 (48.4)&74.4 (67.4)&&\\
			&3&240 (231)&26.1 (25.4)&420 (388)&292 (283)&38 (37.8)&46.3 (42.4)&53.8 (51.8)&&\\
			\hline
			\multirow{3}{*}{10$\omega$}&1&249 (246)&12.4 (11.8)&140 (135)&237 (225)&32.3 (31.2)&28.8 (25.8)&31.6 (31.5)&\multirow{3}{*}{99 (1.3)}&\multirow{3}{*}{55 (4.5)}\\
			&2&208 (202)&15 (14.3)&144 (137)&289 (272)&33.6 (32.7)&35.5 (33.8)&45.4 (45.3)&&\\
			&3&161 (163)&17.2 (17)&296 (279)&153 (159)&26.5 (26)&26.1 (27.3)&34.3 (35.6)&&\\
			\hline
			\multirow{3}{*}{20$\omega$}&1&181 (179)&7.5 (8.13)&103 (96.9)&141 (151)&22.4 (22.4)&19.4 (19)&23.2 (22.6)&\multirow{3}{*}{100 (0.31)}&\multirow{3}{*}{63 (5.6)}\\
			&2&149 (147)&9.89 (9.7)&92.2 (91.5)&208 (194)&22.9 (22.4)&22.5 (22.3)&34.7 (33.3)&&\\
			&3&115 (117)&12.5 (12.1)&201 (197)&108 (106)&17.5 (17.8)&18.8 (19.4)&23.7 (24.8)&&\\
			\hline
			\multirow{3}{*}{40$\omega$}&1&133 (134)&5.71 (5.63)&70.5 (68.9)&119 (117)&16.2 (16.4)&13.5 (13.9)&16 (15)&\multirow{3}{*}{100 (0.059)}&\multirow{3}{*}{74 (6.9)}\\
			&2&107 (106)&6.84 (6.89)&69 (67.7)&137 (137)&16.4 (16.7)&16.5 (16.4)&23.4 (23.8)&&\\
			&3&83.1 (84.7)&9.29 (8.85)&143 (146)&72.9 (73)&12.6 (12.3)&14.2 (13.9)&17.3 (17.7)&&\\
			\hline
			\multirow{3}{*}{5$\omega$}&1&296 (254)&12.3 (9.92)&136 (118)&216 (191)&35.2 (25.5)&43.2 (32.1)&32.4 (30.6)&\multirow{3}{*}{94 (1.9)}&\multirow{3}{*}{51 (3.6)}\\
			&2&281 (252)&17 (15)&164 (139)&331 (298)&32.1 (30.3)&45.7 (42.5)&56.2 (51)&&\\
			&3&195 (188)&17.7 (17.2)&299 (287)&186 (181)&23 (22.5)&37.5 (37.6)&35.7 (34.1)&&\\
			\hline
			\multirow{3}{*}{10$\omega$}&1&191 (187)&7.25 (6.99)&85 (83.3)&137 (133)&20.5 (18.8)&26.2 (23.2)&21.3 (20.6)&\multirow{3}{*}{99 (0.9)}&\multirow{3}{*}{59 (3.4)}\\
			&2&186 (181)&10.5 (10.4)&103 (97.4)&212 (202)&21.1 (20.9)&29.5 (29.3)&38.8 (37.5)&&\\
			&3&139 (138)&12.9 (12.8)&214 (209)&117 (118)&16.1 (16.1)&24.8 (24.8)&23.9 (23.7)&&\\
			\hline
			\multirow{3}{*}{20$\omega$}&1&138 (137)&4.87 (4.96)&60.1 (58.9)&98.8 (100)&12.2 (12.1)&15.5 (15.9)&14.2 (14.4)&\multirow{3}{*}{100 (0.2)}&\multirow{3}{*}{67 (3.2)}\\
			&2&128 (128)&7.43 (7.42)&68.5 (68.9)&146 (146)&14.4 (14.6)&21.7 (21.3)&26 (25.5)&&\\
			&3&99.4 (99.1)&8.7 (8.61)&139 (141)&75.6 (76.7)&11.4 (11.1)&17.7 (18)&16.4 (16.4)&&\\
			\hline
			\multirow{3}{*}{40$\omega$}&1&110 (110)&3.23 (3.49)&39.4 (41.6)&82.8 (82.3)&9.34 (9.18)&11.2 (11.3)&10.4 (10.4)&\multirow{3}{*}{100 (0.027)}&\multirow{3}{*}{75 (4.2)}\\
			&2&97 (95.3)&5.1 (5.11)&51.3 (49.6)&117 (110)&10.4 (10.5)&15.1 (15)&17.2 (18)&&\\
			&3&72.6 (71.9)&6.09 (6.16)&91.5 (93.9)&52.5 (54)&7.56 (7.8)&11.7 (12.5)&12.2 (11.8)&&\\
			\hline
		\end{tabular}
	}
\end{table}

			\subsection{Coverage Probability When $G=G_0$}
			\label{sec:simu:G0:cover}
			We next investigate the quality of statistical inference for the GNHP model by evaluating the coverage probabilities of the $95\%$  confidence intervals for model parameters, derived through the limiting distribution  given by Theorem~\ref{thm_normal} (b). For instance, the $95\%$ confidence interval for $\beta_g$ is given by CI$_{\beta_g}^{(k)} =
			(\wh \beta_g^{(k)} - 1.96\wh \sigma_{\beta_g}^{(k)}, \wh \beta_g^{(k)} + 1.96\wh \sigma_{\beta_g}^{(k)})$, where $\wh \sigma_{\beta_g}^{(k)}$ is the square root of the corresponding diagonal entry
			of the estimated covariance matrix $\bSigma_{\alpha}$. Empirical coverage probabilities based on $1000$ simulation runs are summarized in Table~\ref{tab:sbm:coverage}.

\begin{table}[ht]
	\centering
	\caption{{\color{black} Empirical coverage probabilities ($\%$) for the GNHP model parameters.}}\label{tab:sbm:coverage}
	\scalebox{0.65}{
		\begin{tabular}{cc|cccccc|cccccc|cccccc|cccccc}
			\hline
			&&\multicolumn{12}{c|}{ SBM network  }&\multicolumn{12}{c}{ Power Law network  } \\
			& & \multicolumn{6}{c}{ $m = 100$  }
			& \multicolumn{6}{c|}{ $m = 200$  }&\multicolumn{6}{c}{ $m = 100$  }
			& \multicolumn{6}{c}{ $m = 200$  } \\
			\hline
			$T$  &   & $\beta_g$ & $\eta_g$ & $\gamma_g$ &  & $\phi_{gg'}$ &  & $\beta_g$ & $\eta_g$ & $\gamma_g$ &  & $\phi_{gg'}$ &  & $\beta_g$ & $\eta_g$ & $\gamma_g$ &  & $\phi_{gg'}$ &  & $\beta_g$ & $\eta_g$ & $\gamma_g$ &  & $\phi_{gg'}$ &\\
			& $g/g'$  &-& - & - & 1 & 2 & 3 & -& - & - & 1 & 2 & 3  & -& - & - & 1 & 2 & 3 & -& - & - & 1 & 2 & 3\\
			\hline
			
			\multirow{3}{*}{5$\omega$}&1&93.8&93.4&91.5&92.1&85.5&91.3&93.3&94.9&91.8&82.2&91.8&93.5&94.4&95.2&93.1&84.8&92.6&84.4&89.2&93.9&90.8&86.8&85.3&92.6\\
			&2&91.5&90.6&90.8&93.4&90.6&91.3&90.8&90&92.9&93.3&94.9&92.7&92.6&90.5&92.2&92.2&90.9&93.1&88.9&90.3&91.1&91.6&90.5&91.1\\
			&3&95&92.1&93.3&94&92.5&94.4&94.9&93.3&93.1&94.9&94.1&93.7&92.2&95.2&95.7&92.6&91.3&92.2&94.7&94.2&92.1&94.7&91.8&91.6\\
			\hline
			\multirow{3}{*}{10$\omega$}&1&94.6&95.2&94.8&94.8&92.5&94.4&95.9&94.4&94.4&93&93.2&93.8&94.5&96.2&94.5&91.8&91.2&94&94.5&95.4&94.8&89&93.3&92\\
			&2&96&93.7&94.2&91.5&93.5&92.9&94&93.4&94.6&96.5&96.1&94.8&97.3&95.6&94&94&91.2&96.7&95.1&93.6&93.6&92.7&96.3&93.3\\
			&3&95.6&95.8&95.4&94.4&96.6&94.4&93.4&94.8&92.6&96.7&93.8&95.7&94&96.7&96.2&92.9&94&92.9&93.3&93.9&93.6&92.7&95.4&95.7\\
			\hline
			\multirow{3}{*}{20$\omega$}&1&96.6&93.8&94.2&93.4&95.8&96&95.5&95.1&95.5&95.5&95.7&94.6&95&95.6&95&92.5&93.1&95.6&96.6&95.2&92.2&95.2&94.2&95.6\\
			&2&93.4&94.8&94.8&95.8&93.8&92.8&95.5&93.6&95.3&96.1&93.8&95.1&92.5&96.2&93.1&94.4&96.2&92.5&96.9&93.5&94.5&95.6&94.5&93.9\\
			&3&94.4&93.8&95&93.6&94&93.8&95.1&93.1&95.7&94.2&91.9&94.9&97.5&92.5&91.9&95&95.6&95&91.8&96.6&95.9&95.6&95.6&93.5\\
			\hline
			\multirow{3}{*}{40$\omega$}&1&95&93.2&94.8&95.8&95.6&95&95.8&93.8&93.4&94&94.8&95.6&93.5&91.6&90.9&93.5&94.8&96.8&95.4&95.4&87.7&92.3&92.8&92.3\\
			&2&93.6&95.2&94.4&94&95.6&96&94.2&94.2&94&94.6&95.6&94&92.2&92.9&91.6&92.2&97.4&93.5&95.4&91.3&90.8&92.8&94.9&94.9\\
			&3&94.4&94.2&94.4&95.4&93&95.2&93.4&95.4&94&95.6&94.8&94.6&92.2&92.9&96.8&95.5&92.2&94.2&94.4&97.4&91.8&95.4&94.4&94.9\\
			\hline
			\hline
			%%%%%%%%%%%%%%%%%%%%%%%%%%%%%%%%%%%%%%%%%%%%%%%%%%%%%
		\end{tabular}
	}
\end{table}
			Table~\ref{tab:sbm:coverage} shows that the coverage probabilities have some departure from the nominal $95\%$ when $m=100$ and $T=5\omega$ in both network settings, which is not surprising since on average only around 5\% and 7\% of group memberships are correctly estimated in these two settings. As $m$ and $T$ increase, the empirical coverage probabilities for almost all parameters approach the nominal level, which supports our theoretical findings in Theorem~\ref{thm_normal}.

			\subsection{Estimation Accuracy with a Mis-specified $G$}
			\label{sec:simu-overG}
			When $G$ is mis-specified, we define
			PD$_{\bbeta} =\median_{1\le k\le K}\big\{\frac{1}{m}\sum_{i=1}^m\big|\beta_{g_i^0}^0 - \wh \beta_{\wh g_i^{(k)}}^{(k)}\big|\big\}$,
			and PD$_\eta$ and PD$_\gamma$ are similarly defined. For the estimated background intensities, we define PD$_\mu= \median_{1\le k\le K}\big\{\frac{1}{m}\sum_{i=1}^m \|\wh \mu_{\wh g_i^{(k)}}(t) - \mu_{g_i^0}^0(t)\|_T\big\}$, and
			for the network effects, we evaluate the estimation accuracy of the transition matrix $\bB$ using PD$_{\bB}$ = $ \median_{1\le k\le K}\Big\{\frac{1}{m}\sum_{i=1}^m\sum_{j=1}^m a_{ij}|\wh b_{ij}^{(k)} - b_{ij}^{0}|\Big\}$.
			To identify the true number of groups, we use $\lambda_{mT} = ({15 T})^{-1} \left(\median_{1\le i\le m}n_i\right)^{0.6}\bar d^{0.25}$ for the LIC in (\ref{LIC}), and report the selection error rate as
			SR$(G) = K^{-1}\sum_{k = 1}^K I(\wh G^{(k)} = G)$, where $\wh G^{(k)}$ maximizes the LIC in the $k$th simulation. {To study the effect of the membership refinement algorithm in Section~\ref{sec:norm}, we also compare the GARs with and without membership refinements. In the refinement step, we randomly sample $1,000$ $\bpsi_i$'s to find the refined membership~\eqref{grefine} if the candidate set size exceeds $1,000$.
		}

\begin{table}[ht!]
	\centering
	\caption{Estimation accuracy and selection rate of $G$ when $G$ is miss-specified.}\label{tab:sbm:overG}
	{\color{black}
		\scalebox{0.42}{
			\begin{tabular}{cc|rrrrrccc|rrrrrccc|rrrrrccc|rrrrrccc}
				\hline
				\hline
				&&\multicolumn{16}{c|}{SBM network }&\multicolumn{16}{c}{Power Law network } \\
				\hline
				& & \multicolumn{8}{c}{ $m = 100$ } &
				\multicolumn{8}{c|}{ $m = 200$ }&\multicolumn{8}{c}{ $m = 100$ } &
				\multicolumn{8}{c}{ $m = 200$ } \\
				\hline
				$T$ & $G$ &$\mu$ & $\beta$ & $\eta$ & $\gamma$ & $\bB$ & SR & GAR & GAR &$\mu$ & $\beta$ & $\eta$ & $\gamma$ & $\bB$ & SR & GAR & GAR &$\mu$ & $\beta$ & $\eta$ & $\gamma$  & $\bB$ & SR & GAR & GAR &$\mu$ & $\beta$ & $\eta$ & $\gamma$ & $\bB$ & SR & GAR & GAR \\
				\cline{3-7}			 \cline{11-15} 			 \cline{19-23} 			 \cline{27-31}
				& & \multicolumn{5}{c}{$(\times 10^{-3})$}&(\%) & (\%) & (w.o.r)&
				\multicolumn{5}{c}{$(\times 10^{-3})$ }&(\%) & (\%) & (w.o.r)  &\multicolumn{5}{c}{$(\times 10^{-3})$ }&(\%) & (\%) &  (w.o.r) &
				\multicolumn{5}{c}{$(\times 10^{-3})$ }&(\%) & (\%) & (w.o.r)  \\
				\hline
\multirow{6}{*}{5$\omega$}&2&33.5&74.2&842&737&117&0&70.6(2.4)&70.5(2.2)&33.6&77&864&763&118&0&65.6(1.8)&64.7(2)&37&81.1&916&815&121&63&64.7(2.6)&64(2.6)&33.6&74.4&797&788&113&95&70.5(2.2)&69.2(2.9)\\
&3&19.8&30.2&389&301&51.5&100&94.6(2.3)&93.8(2.6)&16.7&24.9&319&285&49.1&100&94.6(1.6)&93.7(1.6)&24.4&34.4&424&442&72.4&37&92.8(2.9)&90.7(3.5)&19.6&27.7&310&336&56.6&5&93.8(1.9)&91.6(2.2)\\
&4&28.4&42.2&549&428&73.1&0&92.7(3.2)&91.1(3.4)&26.2&39.3&472&436&71.4&0&91.6(2.3)&91.1(2.4)&32.5&46.6&566&620&98.8&0&89.8(3.7)&86.6(4)&30.7&40.4&459&485&80.5&0&91.6(3)&89.3(2.8)\\
&5&33.8&46.1&614&490&83.8&0&91.4(3.2)&89.1(4)&30.5&46.4&540&535&84.2&0&90.5(3.1)&89.7(2.6)&38.4&52.7&630&751&115&0&88.7(3.9)&84.9(4.6)&36.9&45.9&524&569&94.8&0&90.6(3.1)&87.8(3)\\
&6&36.4&52.4&670&566&94.6&0&90.4(3.4)&87.2(4.9)&33.5&51.6&607&592&89.5&0&90.1(3)&88.6(3)&42.2&58.1&702&907&132&0&87.8(3.8)&83.7(5)&41&49.9&580&638&106&0&89.8(3.3)&87(2.9)\\
&oracle&14.8&17.8&235&218&31.9& - & - - & - - &11.4&13&171&182&25.6& - & - - & - - &16.7&19.8&233&299&43.9& - & - - & - - &13.5&13.7&161&212&30.1& - & - - & - - \\
\hline
\multirow{6}{*}{10$\omega$}&2&30.6&67.8&709&701&113&0&72.6(1.5)&72(1.9)&31.4&73.1&748&740&116&0&67.1(1.6)&66.3(1.9)&32.8&75&787&773&117&4&67.2(1.7)&66.5(2.2)&30.8&70&706&768&109&68&72.2(1.1)&71.5(1.7)\\
&3&11.4&14.5&188&152&25.7&100&99.2(0.91)&97.5(1.5)&9.36&11.7&135&147&21.9&100&99(0.7)&97(1)&13.3&15.9&205&235&36.1&96&98.5(1.2)&96(1.9)&11&12.3&149&156&27.5&32&98.6(0.84)&95.9(1.3)\\
&4&17&21.6&263&228&38.6&0&98.5(1.5)&97.1(1.6)&14.9&20.5&228&266&41&0&97.4(1.6)&96.4(1.4)&19.1&23.8&283&345&57.3&0&96.7(2.3)&94.7(2.3)&18&23&234&262&44.8&0&97.5(1.7)&95.3(1.9)\\
&5&20.5&24.8&308&275&46.8&0&98.2(1.4)&96.3(1.9)&18.7&24.5&268&279&46.6&0&97.4(1.7)&95.9(1.5)&22.4&27.4&332&399&66.8&0&96.5(2.2)&94.3(2.3)&22.9&25&274&312&56.8&0&96.9(1.9)&94.9(1.7)\\
&6&22.3&28.8&350&299&53&0&97.8(1.8)&95.5(2.3)&20.7&27.2&318&331&50.8&0&97.1(1.8)&95.8(1.5)&23.5&30.2&365&450&70.7&0&96.5(2.5)&93.5(2.7)&25.2&25.5&305&326&62.4&0&96.9(1.8)&94.3(2.2)\\
&oracle&10.7&13.6&178&135&22.5& - & - - & - - &8.35&9.55&113&129&17.8& - & - - & - - &11.7&13.4&173&213&30.3& - & - - & - - &9.9&9.06&116&130&21.3& - & - - & - - \\
\hline
\multirow{6}{*}{20$\omega$}&2&29.6&66.1&681&690&112&0&72.7(1.5)&71.6(2.7)&30.4&70.9&724&729&116&0&67.7(0.93)&66.9(1.9)&31.2&73.1&743&757&116&0&68(0.98)&67(2.2)&29.9&68.6&690&753&108&1&72.4(0.44)&71.7(2.5)\\
&3&7.9&9.48&116&109&15&99&99.9(0.26)&97.7(1.3)&6.18&6.75&91.6&95.7&12.5&100&99.9(0.2)&97.6(1.1)&8.67&9.72&118&155&21&100&99.9(0.31)&97.4(1.6)&7.28&6.71&82&101&15&99&99.9(0.25)&97.2(1.5)\\
&4&11&13.9&154&148&22.6&1&99.8(0.64)&97.8(1.4)&8.79&11.8&127&136&22.1&0&99.3(0.79)&97.9(1)&11.5&14.6&169&209&33.8&0&99.3(1.1)&97(1.7)&10.7&12.1&130&153&25.8&0&99.3(0.93)&97.1(1.3)\\
&5&12.6&16.2&178&168&27.9&0&99.8(0.48)&97.6(1.4)&11.5&13.6&161&184&26.8&0&99.3(0.85)&98(0.89)&13.4&17&195&225&39.1&0&99.2(1)&96.8(1.9)&14&13.7&153&185&32.1&0&99.2(0.93)&97.1(1.4)\\
&6&13.7&17.3&193&181&30.7&0&99.7(0.53)&97.2(1.4)&13.2&15.8&164&186&30.7&0&99.4(0.71)&97.5(0.99)&14.2&18.6&202&236&42.4&0&99.1(1.1)&96.3(1.7)&15.7&15.2&174&208&35.2&0&99.3(0.77)&96.9(1.2)\\
&oracle&7.87&9.35&115&108&14.8& - & - - & - - &6.09&6.44&90.6&92.5&12.4& - & - - & - - &8.6&9.69&116&153&20.6& - & - - & - - &7.22&6.59&77.4&98.9&14.7& - & - - & - - \\
\hline
\multirow{6}{*}{40$\omega$}&2&29.1&65&667&680&112&0&72.4(2.3)&71.6(3.5)&30&70.2&715&723&115&0&67.9(0.85)&67.1(2)&30.6&71.5&724&754&116&0&68.1(0.6)&67.9(1.4)&29.5&68.2&679&735&108&0&72.5(0.3)&72.4(0.87)\\
&3&5.82&6.93&81.7&73.8&10.9&99&100(0)&97.8(1.6)&4.49&4.43&61.3&60.6&8.73&100&100(0)&98.5(1.2)&6.33&6.52&80.3&108&15.6&100&100(0)&98.3(1.6)&5.73&4.47&53.4&79.4&10.3&100&100(0)&98.6(1.1)\\
&4&7.64&8.56&101&89.7&15.4&1&100(0.12)&98.2(1.3)&5.75&5.52&70.7&80.5&12.2&0&100(0.17)&98.7(0.94)&7.75&8.18&95&125&21.4&0&99.9(0.26)&98.5(1.3)&7.19&6.46&70.3&109&14.9&0&99.9(0.22)&98.2(1.3)\\
&5&8.93&10&114&97.8&18.7&0&100(0.11)&98(1.3)&7.29&7.61&103&105&16.4&0&99.9(0.18)&98.8(0.82)&8.42&9.82&117&144&24.7&0&100(0.25)&98.4(1.4)&8.94&7.17&81.8&123&19.3&0&100(0.17)&98.6(0.96)\\
&6&9.32&11.4&121&108&21.4&0&100(0.14)&98.1(1.2)&8.3&9.05&110&111&19.9&0&100(0.13)&98.6(0.93)&9.29&9.47&117&157&27.5&0&99.9(0.26)&98.4(1.1)&10&8.53&103&133&22.2&0&99.9(0.2)&98.1(1)\\
&oracle&5.83&6.91&80.8&73.7&10.8& - & - - & - - &4.51&4.46&62.3&60.9&8.71& - & - - & - - &6.27&6.52&81.1&109&15.6& - & - - & - - &5.73&4.32&53.6&81.7&10.5& - & - - & - - \\
\hline				\hline
			\end{tabular}
		}
	}
\end{table}

			{Since the refinement step is rather time consuming, especially when $G$ is large, we only provide} summary statistics based on $K=200$ simulation runs in Table~\ref{tab:sbm:overG}, where we can see that
			even when $G$ is over-specified, the consistency result still holds.
			For instance, PD$_\beta$ drops from approximately 42.2$\times 10^{-3}$  to 20.5$\times 10^{-3}$ as $(m,T)$
			increases from $(100, 5 \omega)$ to $(200, 10 \omega)$ with $G = 4$, which corroborates with the result in Theorem \ref{thm1}. When $G>3$, estimation errors are generally larger than those of the case with $G=G_0$, which is reasonable due to the additional estimation variability introduced by an overly large $G$.
			Furthermore, the proposed LIC can correctly select the true number of groups with a probability tending to $1$ as $(m, T)$ increases, which supports our conclusion of Theorem~\ref{thm_LIC}. {Lastly, the GAR after the refinement tends to be slightly better than the GAR without the refinement, suggesting it is beneficial to perform membership refinement.}
		}
	
{
\subsection{Computation Times}\label{sec::comp_cost}
We now study the computation times of the proposed EM algorithm outlined in Section~\ref{sec:em}. For each initial membership estimate, we perform 100 Stochastic EM iterations, 50 EM iterations with fixed memberships, and then 50 EM iterations.  Fixing $m=100$, computation times (in seconds) of this algorithm on a cluster of Intel Xeon Gold 6126 CPUs with 2.6 Ghz are summarized in Figure~\ref{fig-time}. We can see that the CPU time grows approximately linearly as $T$ and the total number of events increase. The CPU time grows slightly faster than linearly as $G$ increases since the number of parameters is of the order $G\nk+3G+G^2$. Results are similar for the Power-law Network and are thus given in Section~B.4 of the supplementary material. To increase the chance of finding the global optimal, for each simulation run, we use $50$ initial membership estimates following the intensity-based K-means algorithm (using different starting values, supplement with random membership estimates if less than 50 distinct membership estimates are generate). Since the computations with different initial values can be easily paralleled, the overall computation times of the proposed EM algorithm appears to be reasonable for practical use.
		\begin{figure}[ht]
	\centering
	\includegraphics[width = 0.325\textwidth]{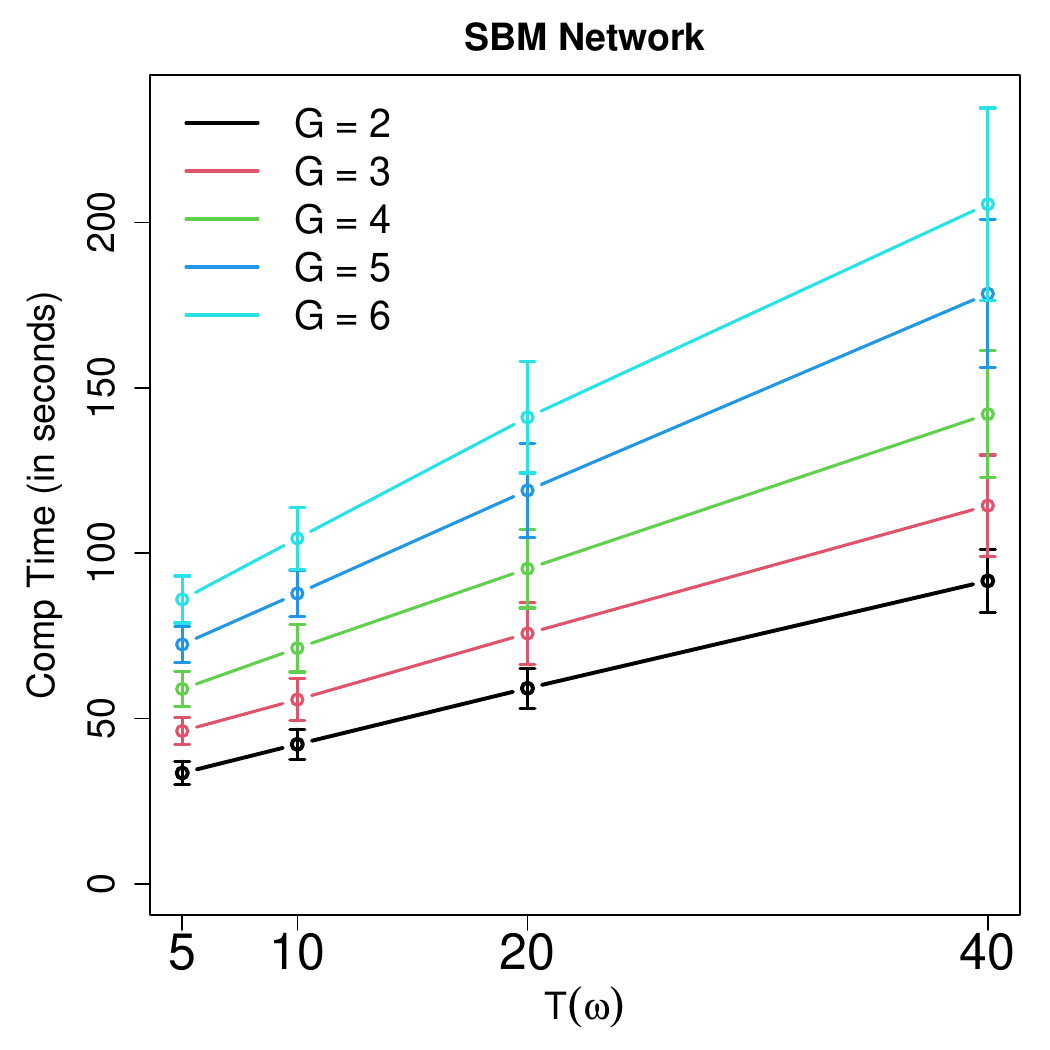}
	\includegraphics[width = 0.325\textwidth]{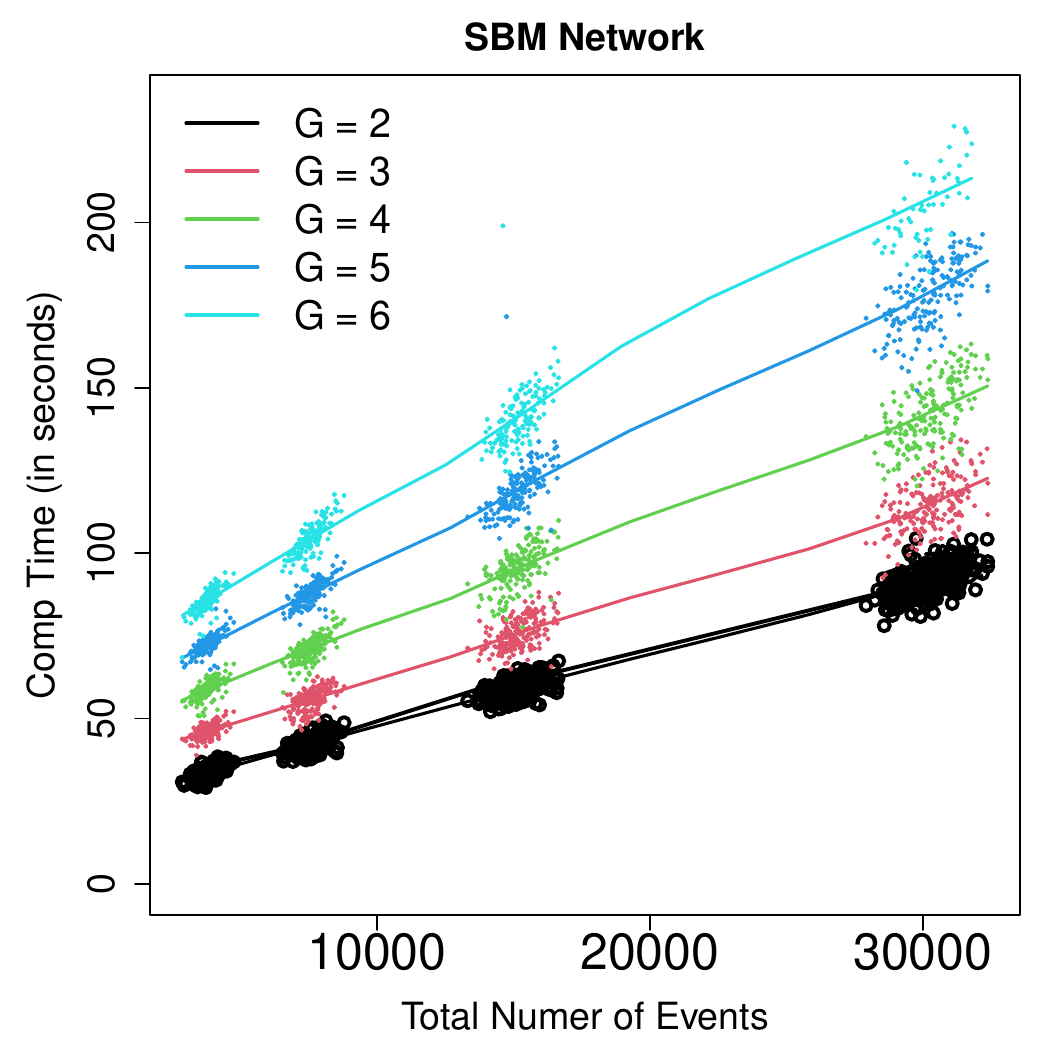}
	\includegraphics[width = 0.325\textwidth]{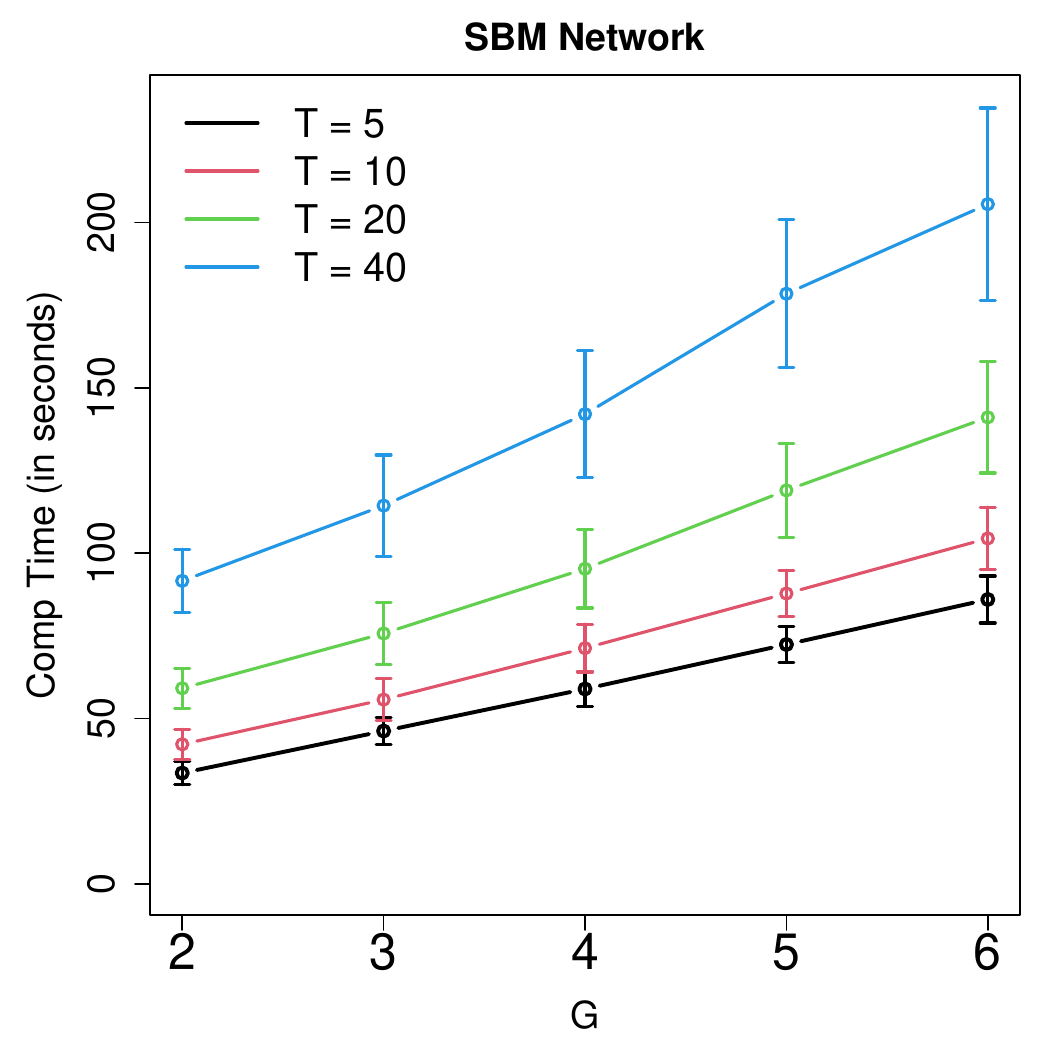}
	\caption{Computation times of the proposed EM algorithm in various settings.}\label{fig-time}
\end{figure}

}

		\section{EMPIRICAL STUDY: A SINA WEIBO DATASET}
		
		\label{sec:data}
		
		We apply the  GNHP model to a dataset collected from Sina Weibo,
		the largest Twitter type social media in China, where we collect posting time stamps of $m = 2,038$ users from January 1st to 15th, 2014, resulting in a $T = 600$ hours. { Details of the data collection process is given in Section~B.5 of the Supplementary Material.}
Figure \ref{degrees} gives two sample Weibo posts by James Cameron with posting times. The network adjacency matrix $A$ is constructed using the following-followee relationships among users, which gives the network density $\sum_{i,j} a_{ij}/m(m-1) = 2.7 \%$, suggesting a highly sparse network.
		Distributions of in-degrees and out-degrees of the network are given in Figure \ref{degrees}, where we can see that
		the in-degrees tend to be more skewed than the out-degrees. This phenomenon is typical for a social network platform, where a few influential users may have a large number of followers but most users do not follow too many other users.

		\begin{figure}[htp]
			\begin{center}
				\subfigure{\includegraphics[scale=0.25]{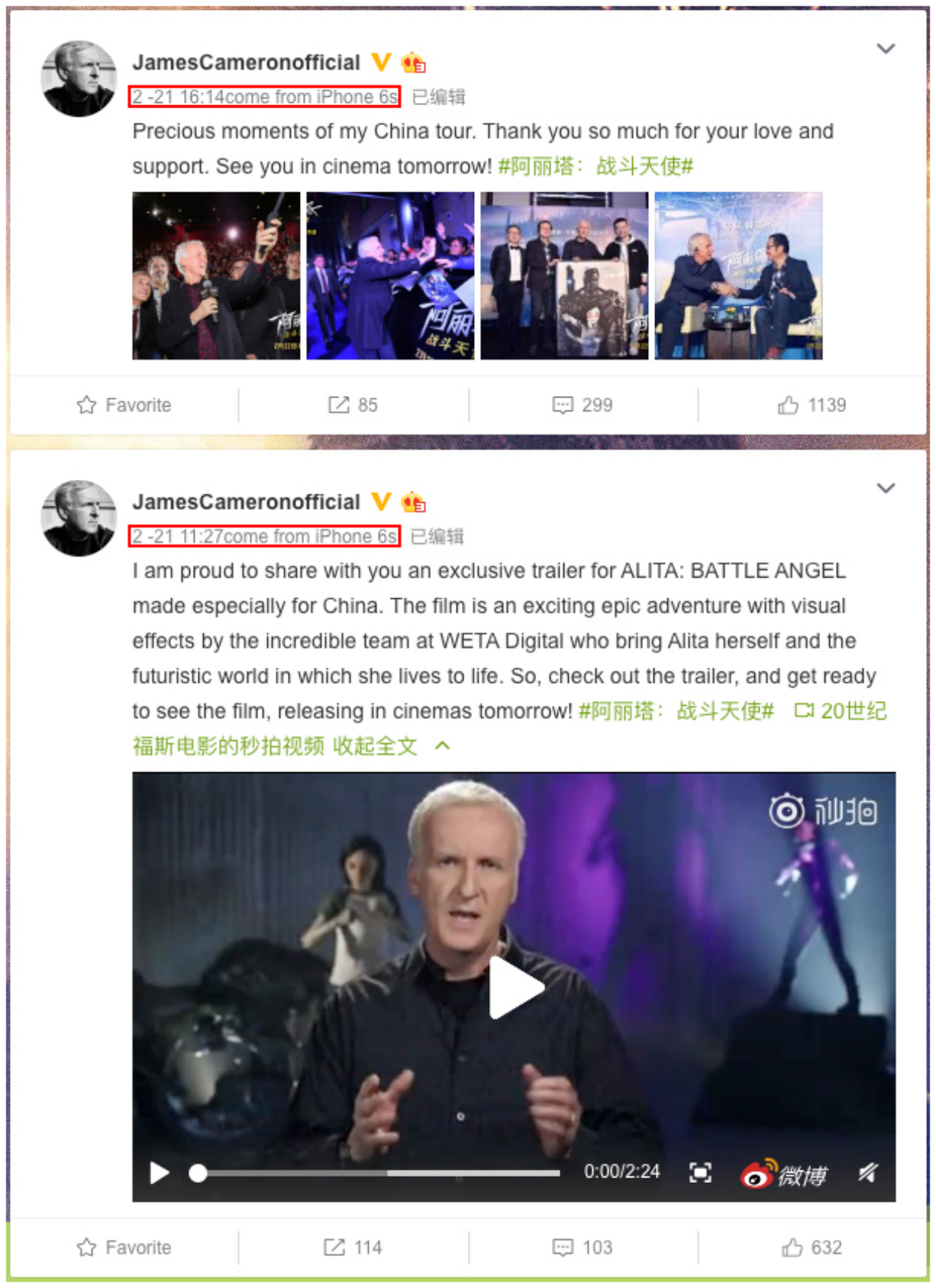}}
				\subfigure{\includegraphics[scale=0.25]{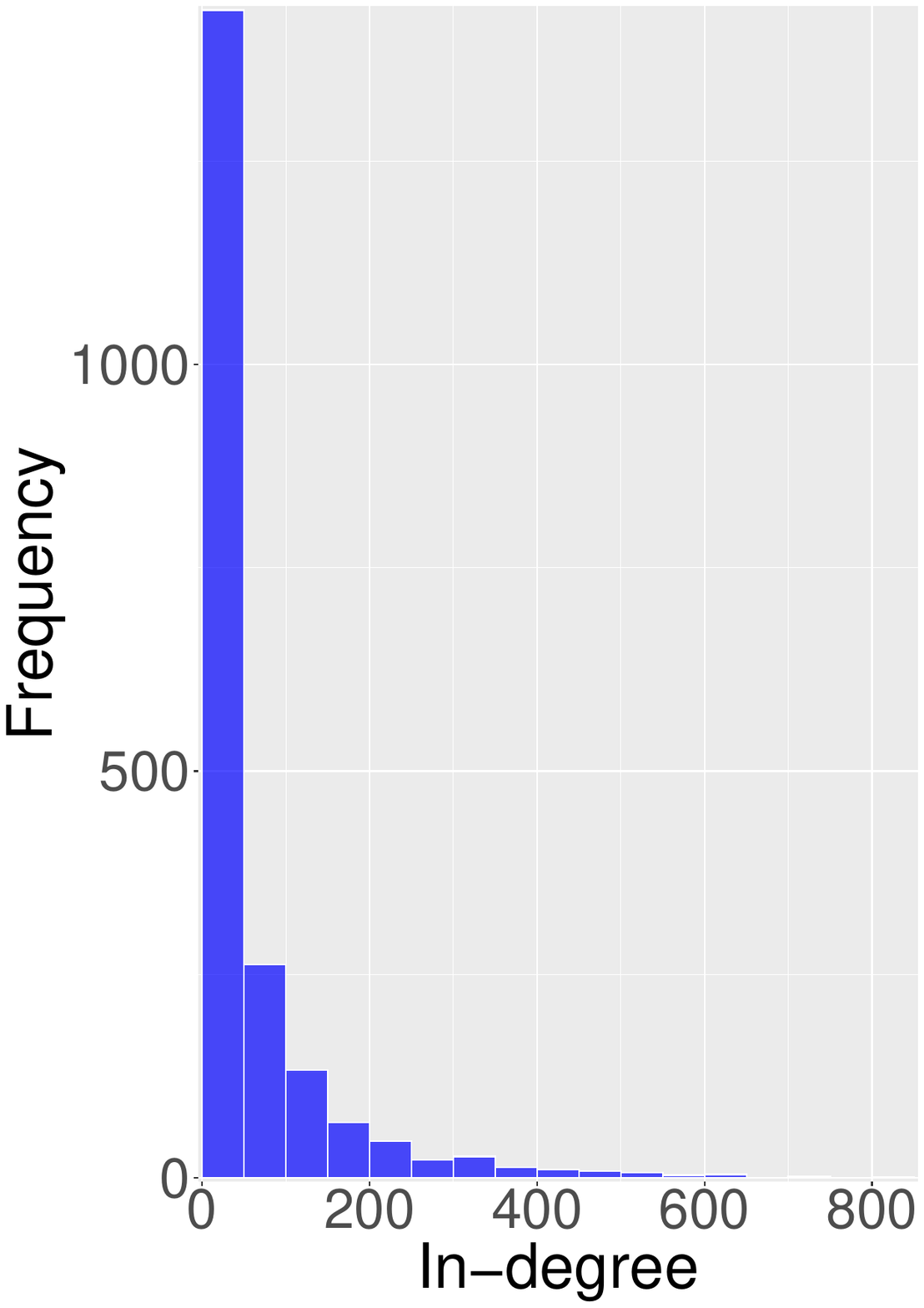}}
				\subfigure{\includegraphics[scale=0.25]{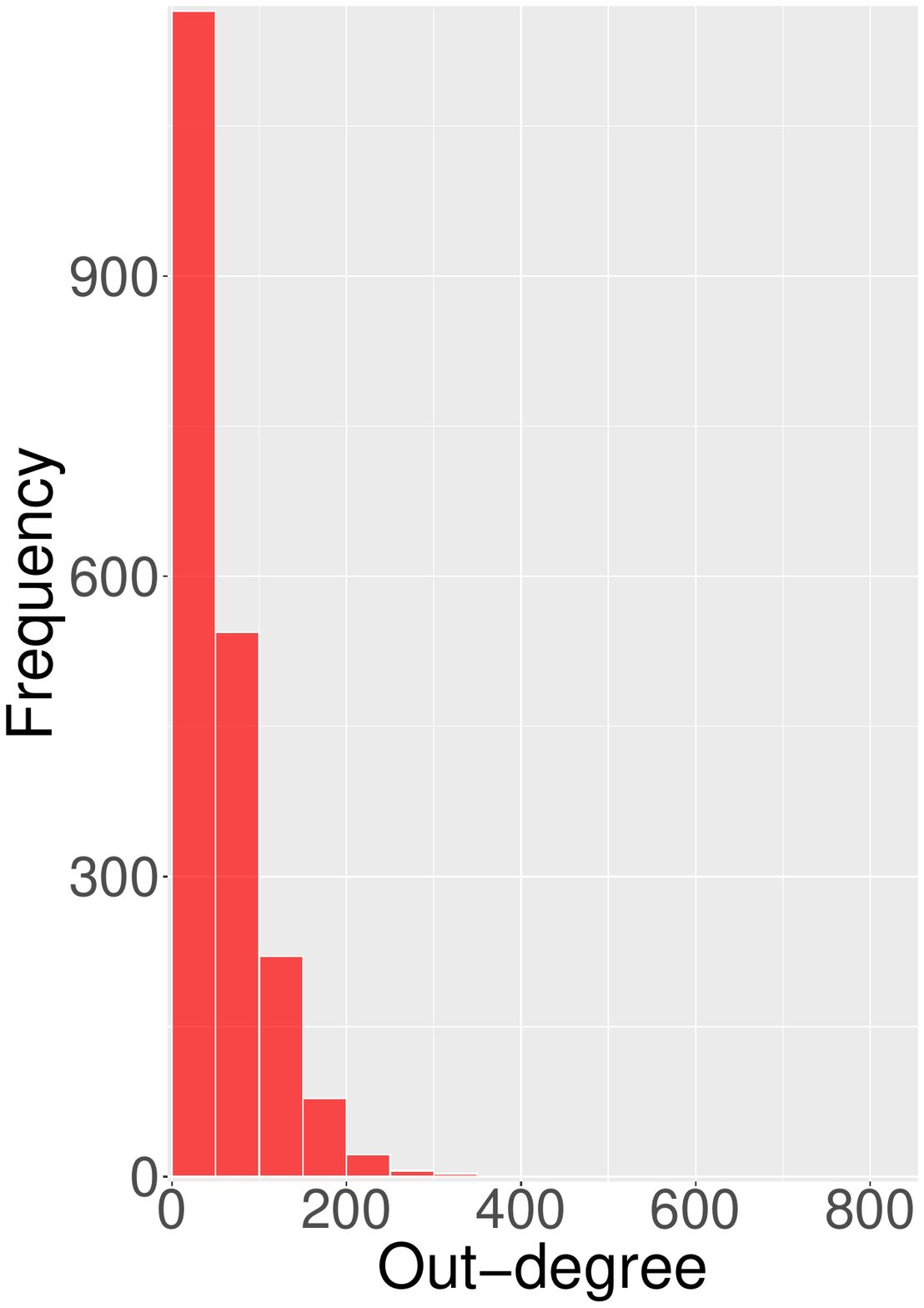}}
			\end{center}
			\caption{\small Left panel: A snapshot of James Cameron's Weibo posts; Middle panel: histogram of in-degrees; Right panel: histogram of out-degrees.} \label{degrees}
		\end{figure}

		\subsection{Model Estimation and Interpretation}
		We approximate the background intensities of the GNHP model by periodic B-spline basis with $\nk=3,\cdots,10$ equally spaced knots between $[0,\omega]$ with $\omega=24$ (hours) and the triggering functions is of the form~\eqref{truncEXP} with $b=5$ (hours). For better numerical stability, we put an upper bound $100$ on parameters $\eta_g$'s and $\gamma_g$'s.  Using the EM algorithm proposed in Section~\ref{sec:em} with $200$ initial membership and parameter estimates obtained from the algorithm in Section A.2 of the supplementary material, we first use the proposed LIC given in~\eqref{LIC} to choose the number of latent groups with $\lambda_{mT} = ({15 T})^{-1} \left(\median_{1\le i\le m}n_i\right)^{0.6}\bar d^{0.25}$. {From Figure~\ref{base_intense}, we can see that the LICs have steep increases when $G$ changes from $1$ to $2$ for all choices of $\nk$'s,} suggesting suitability to use a latent group structure to model the heterogeneity among network nodes. The optimal number of groups chosen by LIC is $G=4$ {for all choices of $\nk$'s}.
		
		{The next step is to choose the best $\nk$, for which we propose the following BIC type criterion:
\be
\label{sel:bic}
{\rm BIC}(\nk,G)=-2 mT \ell(\wh{\ut\bpsi})+\log(mT/\omega)\times G\nk T/\omega,
\ee	
where $mT\ell(\wh{\ut\bpsi})$ is the log-likelihood~\eqref{comp_lik} evaluated at the MLE $\wh{\ut\bpsi}$ given in~\eqref{mle}.

Heuristically, $T/\omega$ can be roughly interpreted as the number of weakly dependent replicates over time and $G\nk T/\omega$ can be viewed as the total number of knots placed over $[0,T]$ and $G$ groups.  Fixing $G=4$, Figure~\ref{base_intense} shows that the minimum BIC value is achieved with $\nk=6$.
The estimated background intensities using $\nk=6$ and $G=4$ are illustrated in Figure~\ref{base_intense} and the resulting parameter estimates are summarized in  Table~\ref{weibo_est}.
}

		\begin{figure}[ht]
			\centering
			\includegraphics[width = 0.325\textwidth]{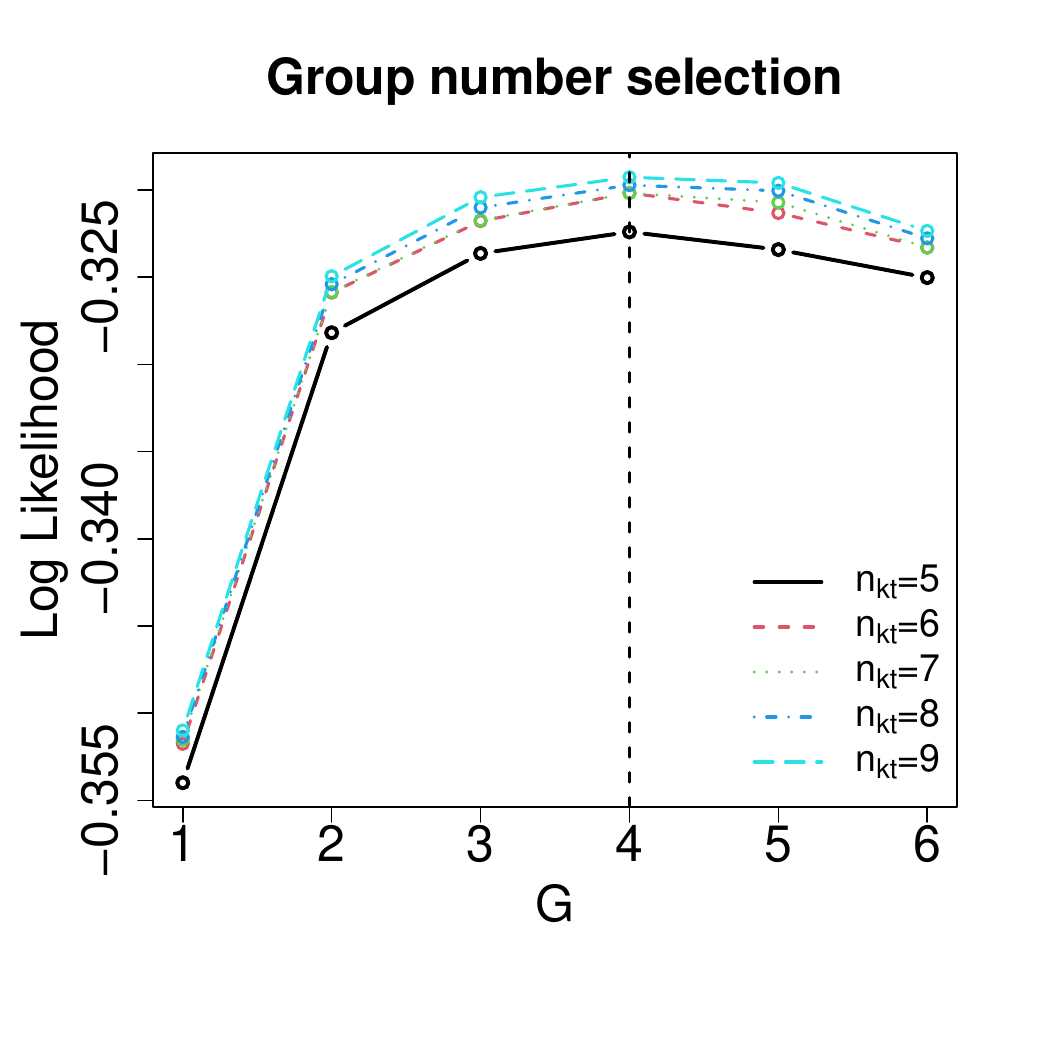}
			\includegraphics[width = 0.325\textwidth]{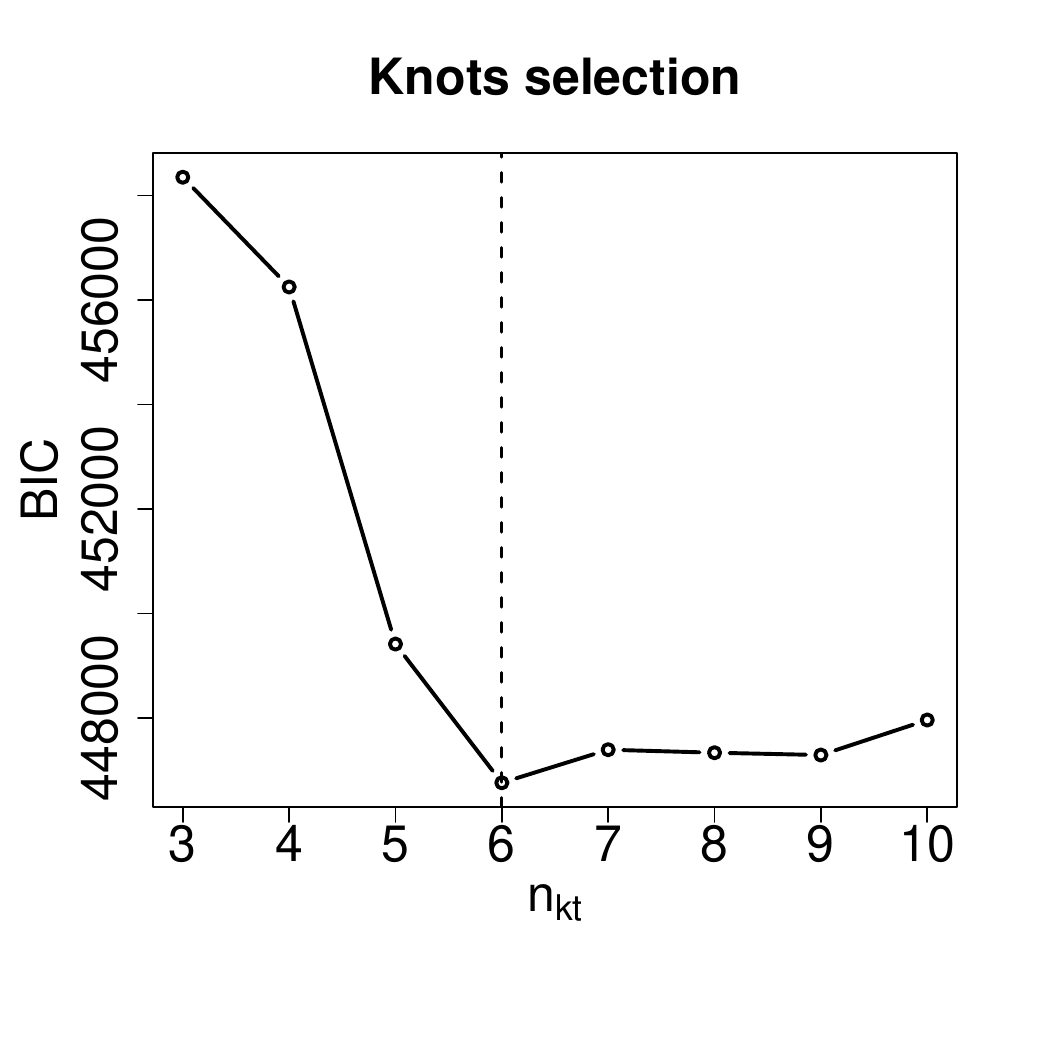}
			\includegraphics[width = 0.325\textwidth]{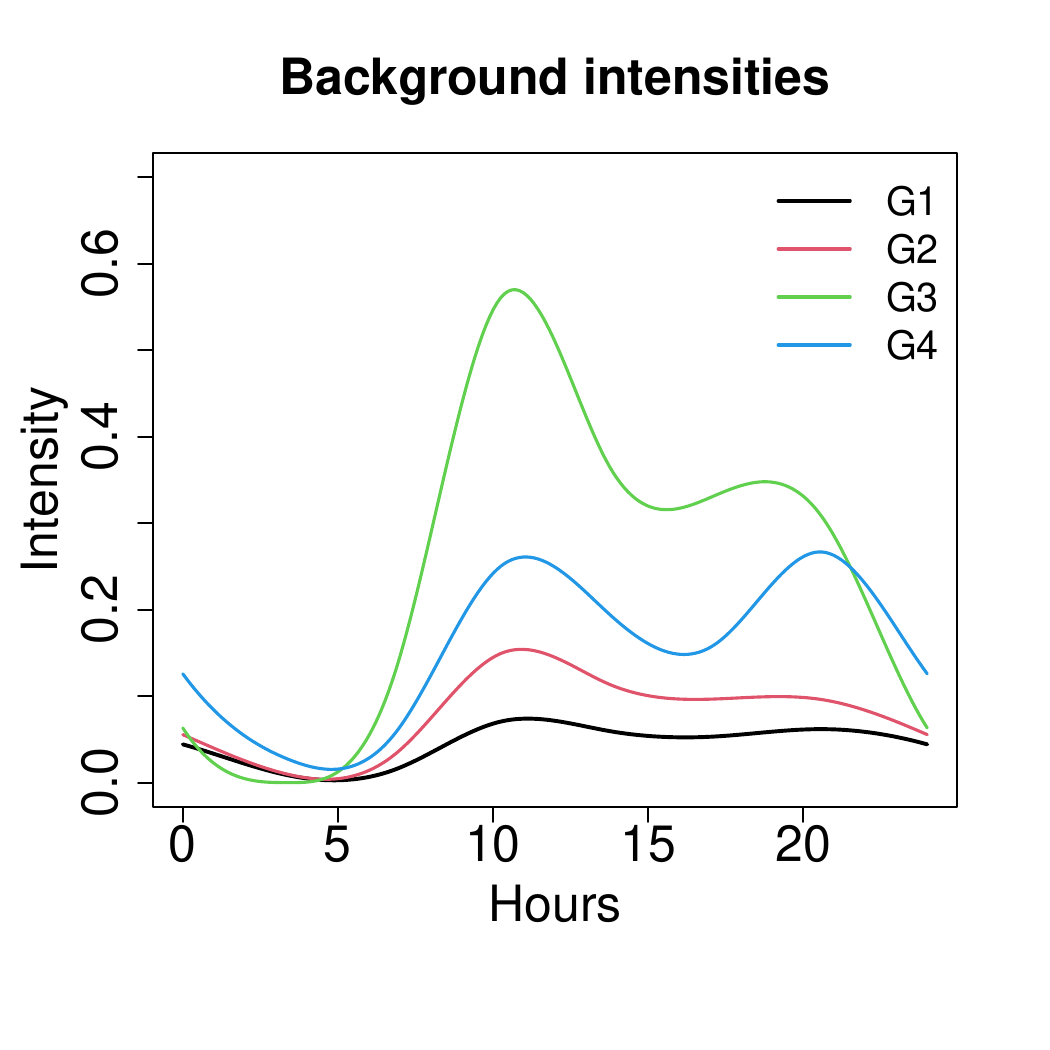}
			\vspace{-2em}
			\caption{Left panel: maximized log-likelihood of various $G$'s; Middle panel: LIC scores of various $G$'s; Right panel: estimated background intensities.}\label{base_intense}
		\end{figure}
		
				\begin{table}[ht]
			\caption{Parameter estimates for Weibo data (p-values are given in the parentheses).}\label{weibo_est}
			\centering
			\scalebox{0.65}{
				\begin{tabular}{c|c|cc|c|cccc}
					\hline
					\hline
					& Percent (\%) & $\beta_g$  & $\eta_g$ & $\gamma_g$ &  \multicolumn{4}{c}{ $\phi_{gg'}$} \\
					\hline
					Group ($g/g'$)	 &-&-   &- &-  &  1&2&3&4 \\
					\hline
					1 & 49.8& 0.346 ($<$ 0.001) & 9.69 ($<$ 0.001) & 2.73 ($<$ 0.001) & 0.0213 (0.0018) & 0.0589 ($<$ 0.001) & 0.00461 (0.0044) & 0.0189 ($<$ 0.001) \\
					2 & 28.2& 0.343 ($<$ 0.001) & 1.35 ($<$ 0.001) & 2.49 ($<$ 0.001) & 0.0474 ($<$ 0.001) & 0.144 ($<$ 0.001) & 0.0568 ($<$ 0.001) & 0.00796 (0.24) \\
					3 & 12.2& 0.585 ($<$ 0.001) & 0.603 ($<$ 0.001) & 100 ($<$ 0.001) & 0.00592 (0.55) & 0.0344 ($<$ 0.001) & 0.18 ($<$ 0.001) & 0.00512 (0.53) \\
					4 & 9.9& 0.64 ($<$ 0.001) & 7 ($<$ 0.001) & 2.31 ($<$ 0.001) & 0.0532 (0.1) & 0.147 ($<$ 0.001) & 0.0682 ($<$ 0.001) & 0.181 ($<$ 0.001) \\
					\hline
				\end{tabular}
			}
		\end{table}
	
	Figure~\ref{base_intense} shows that all estimated background intensities have two peaks around 10:00 am and 8:00 pm, suggesting that users are more active around these times. Based on the estimated GNHP model and users' information in the different groups, we summarize groups as following.
		
		%	\begin{figure}[H]
			%		\centering
			%		\includegraphics[width = 0.45\textwidth]{images/phi_heat.pdf}
			%		\caption{\small Heatmap for estimated $\phi$'s.} \label{alpha_heat}
			%	\end{figure}

		\begin{itemize}
			\item {\bf Group 1} is largest group that includes $49.8\%$ of users, who have the lowest background intensity throughout the day. Judging from $\wh\phi_{g1}$'s, users from this group have relatively low impact on users from other groups. This group also has the largest $\wh\eta$ value, suggesting that the user's past posts have the shortest time impacts on his/her future posting behavior.
			
			\item {\bf Group 2} include many users that are playing leading roles in various communities such as entertainment, business, education, and social sciences, who typically do not post very frequently throughout the day. This group has the second smallest $\hat \eta$ value, indicating that users' past posts have a relatively long time effect. The estimated $\wh\phi_{g2}$'s suggests that this group of users have the largest impacts on all groups except Group 3. Our subsequent analysis in Section~\ref{sec:influ} reveals that this group is the second most influential group.

			\item {\bf Group 3} mainly contains official accounts of some news outlets.
			Figure \ref{base_intense} shows that this group is the most active one on average, which is probably due to their mission to deliver information promptly. It has the smallest $\wh \eta$, suggesting the longest temporal dependence on the posting histories. At the same time, it has an extremely large $\wh \gamma$, which indicates that this group is very unlikely to be influenced by past posts from other groups. Although the estimated $\wh\phi_{g3}$'s are not as large as those of Group 2, the largest background intensity of this group makes it the most influential group, as we shall demonstrate in  Section~\ref{sec:influ}.

			\item {\bf Group 4} consists of users who post quite frequently as suggested by Figure~\ref{base_intense} but have rather limited impacts on other groups based on the estimated $\wh\phi_{g4}$'s. It has a large $\wh\phi_{42}=0.147$, indicating that users in this group are heavily influenced by users in Group 2.

		\end{itemize}

		\subsection{Group Interaction and Influential User Analysis}
		\label{sec:influ}
		
		In social network analysis, identifying influential users is an important task, as it may help improve the efficiency of news propagation, product release, and promotional campaign launches. To identify the most influential users, we first define the influential power of the user $i$ as the sum of the $i$th column of $(\I-\wh\B)^{-1}$, which is the {\sc Node-to-network influence} given by Theorem \ref{thmfam}.
		The bar plots of users with top 20 and 100 influential powers
		are illustrated in Figure \ref{influence}, which shows that top 20 influential users only consist of members from Groups 2 and 3, with the latter group being the most influential one. This observation is further confirmed in the barplot of the top 100 influential users, where the majority of users are from Groups 2 and 3.

		%	\begin{figure}[H]
			%		\centering
			%	\includegraphics[width = 0.33\textwidth]{images/group1.pdf}
			%		\includegraphics[width = 0.32\textwidth]{images/group2.pdf}
			%		\includegraphics[width = 0.32\textwidth]{images/group3.pdf}
			%		\includegraphics[width = 0.32\textwidth]{images/group4.pdf}
			%		\caption{The {\sc dynamic group-group influences} defined in (\ref{E_Sgg_count}) of the four groups. The asymmetric influential patterns can be observed.}\label{g2g-dynamic}
			%		\label{impact:baseline}
			%	\end{figure}

		\begin{figure}[ht!]
			\centering
			\includegraphics[width = 0.32\textwidth]{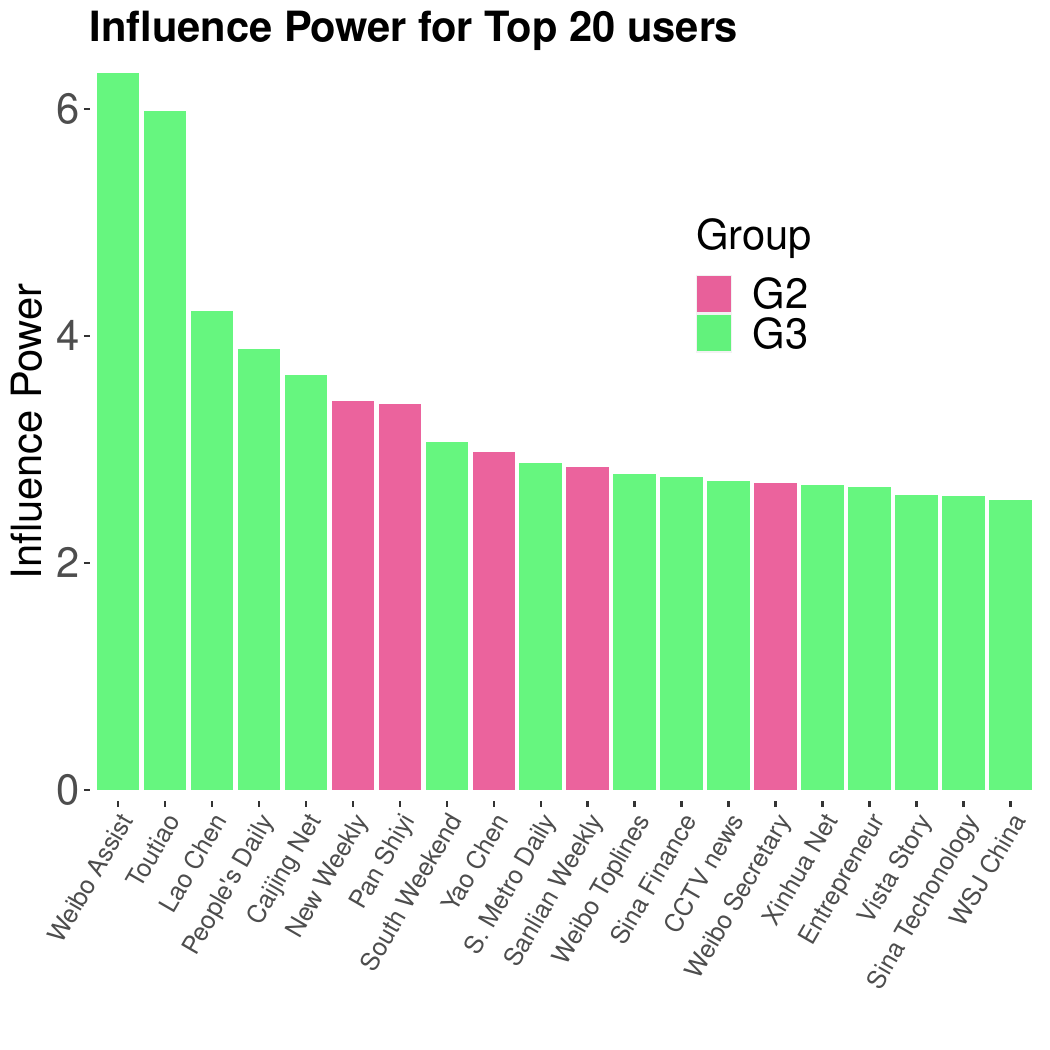}
			\includegraphics[width = 0.32\textwidth]{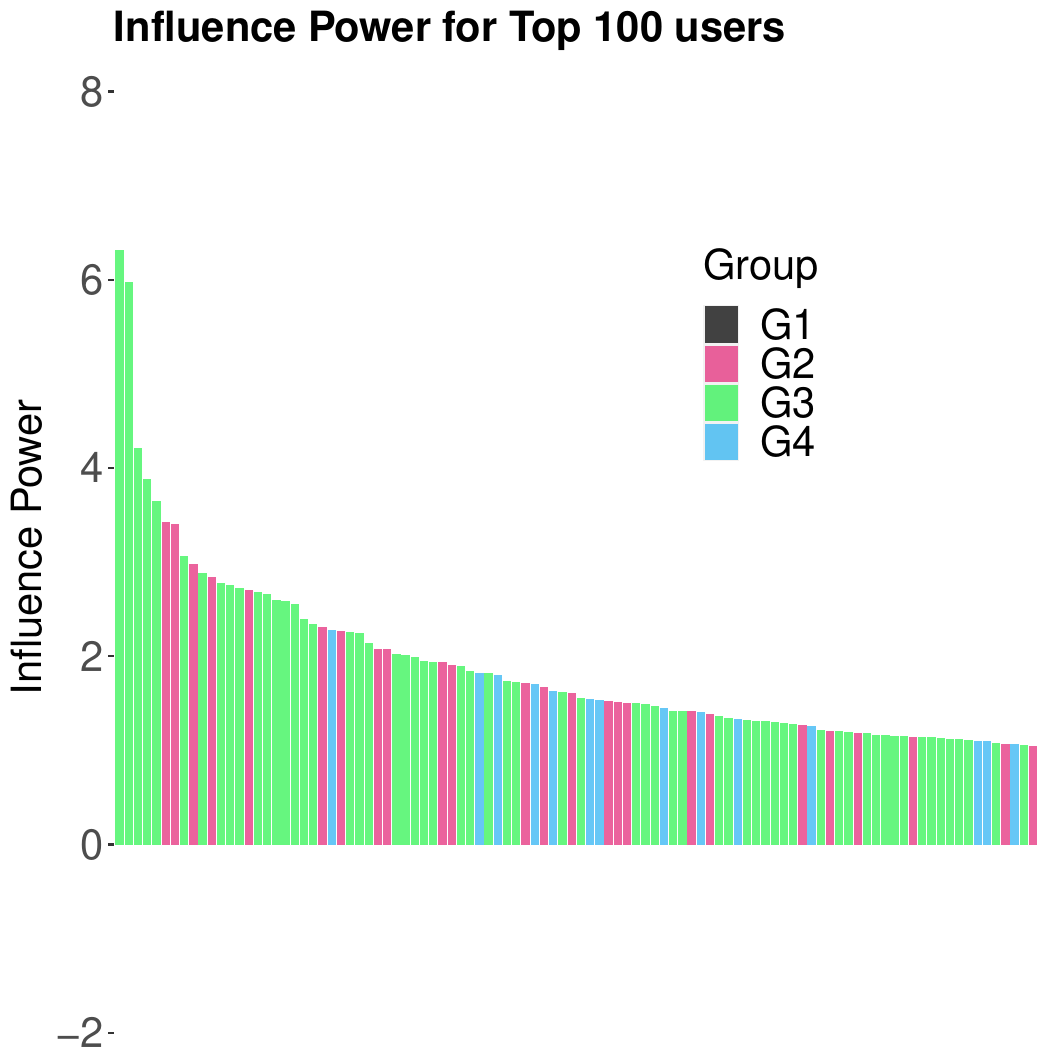}
			\includegraphics[width = 0.32\textwidth]{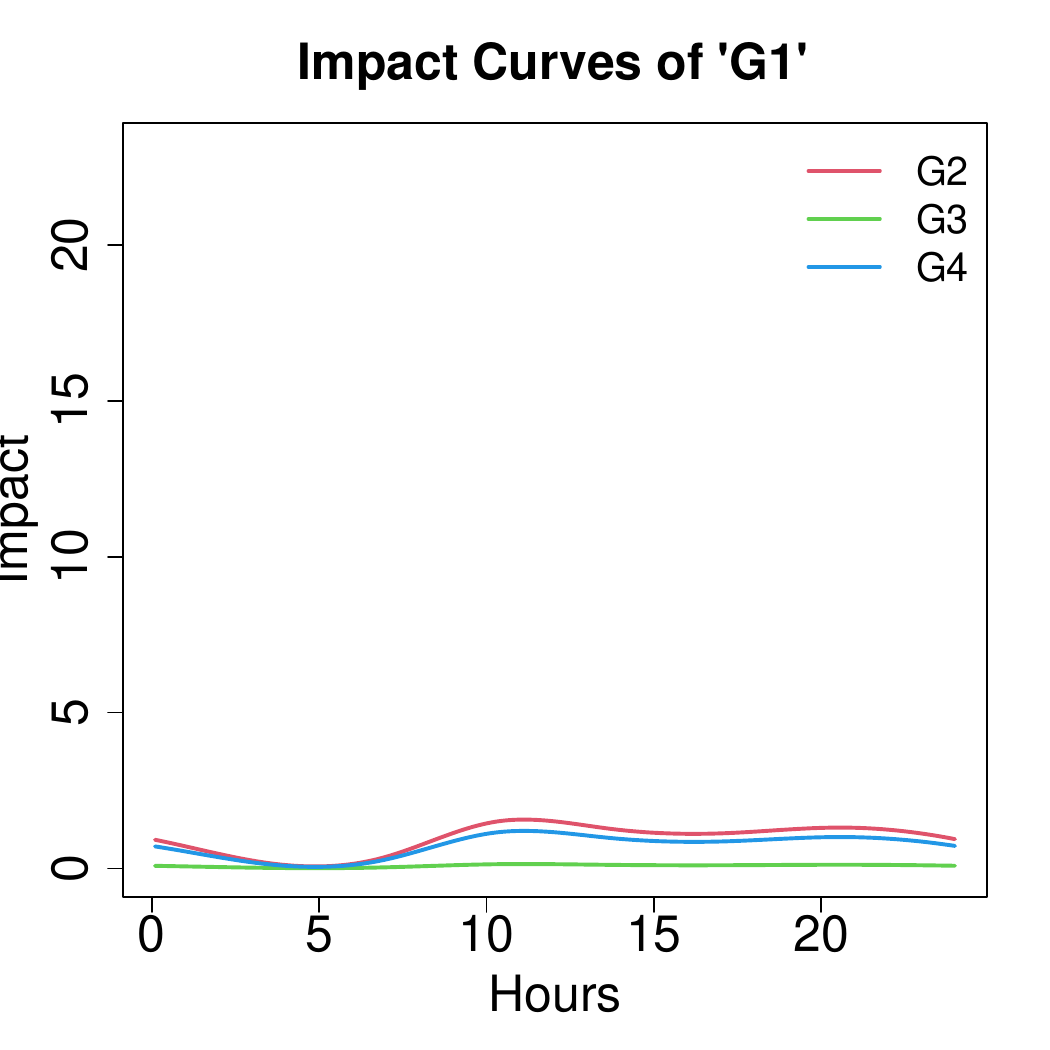}
			\includegraphics[width = 0.32\textwidth]{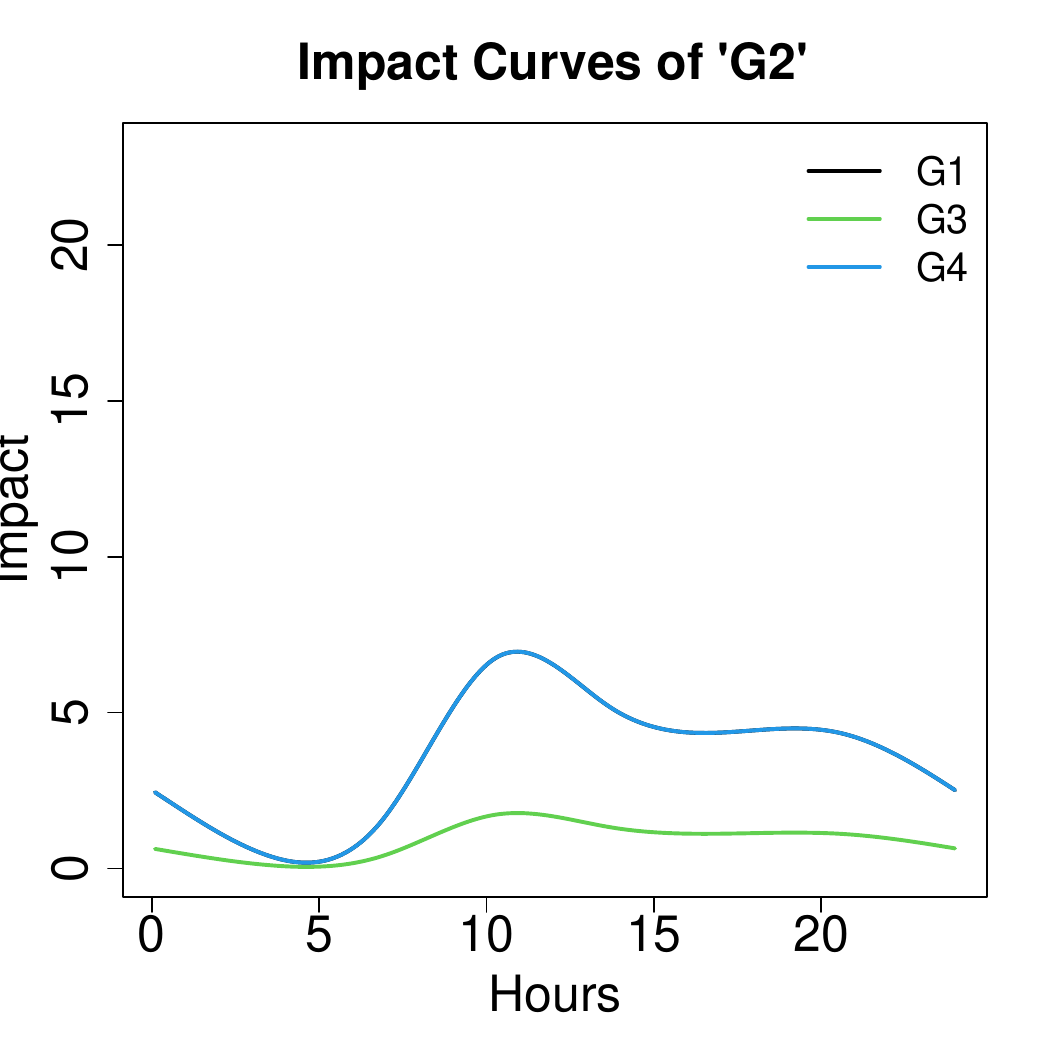}
			\includegraphics[width = 0.32\textwidth]{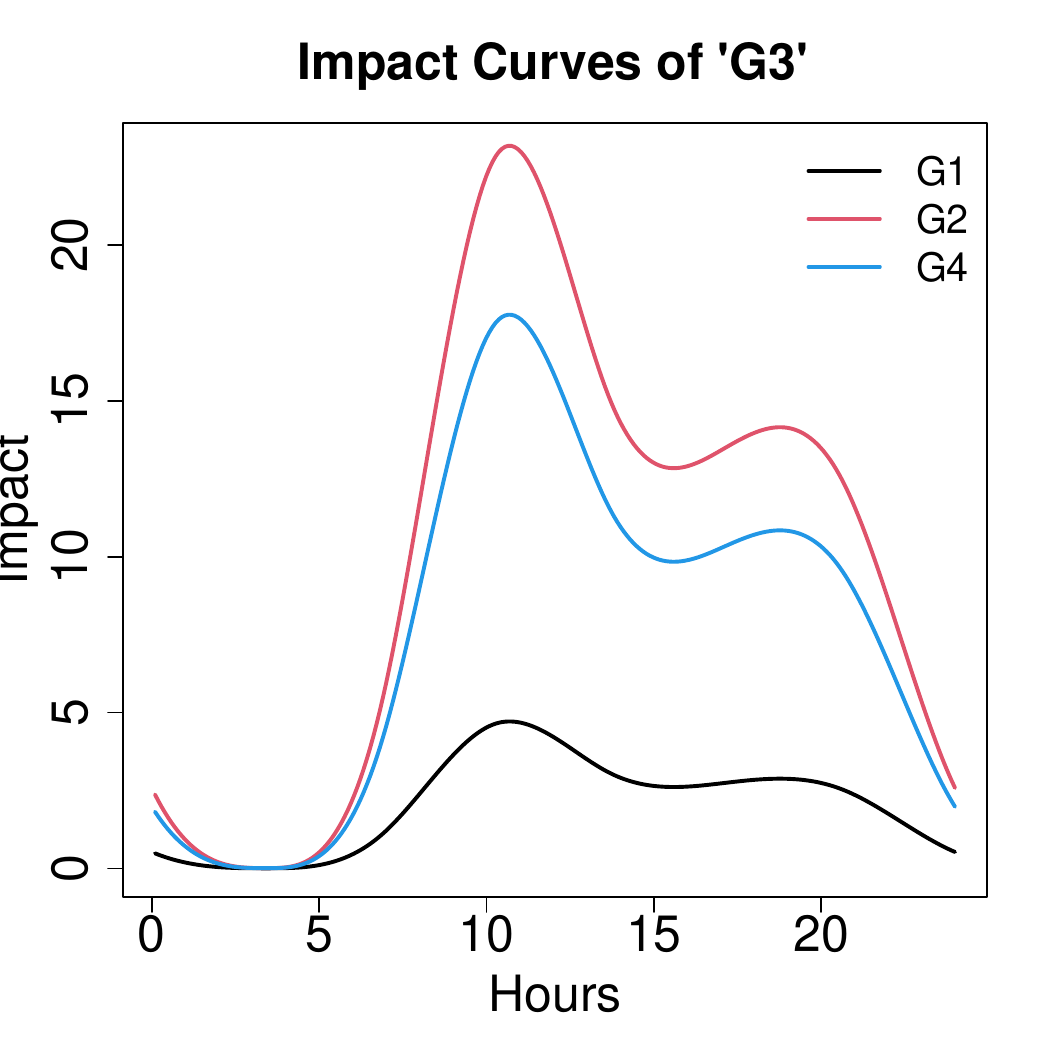}
			\includegraphics[width = 0.32\textwidth]{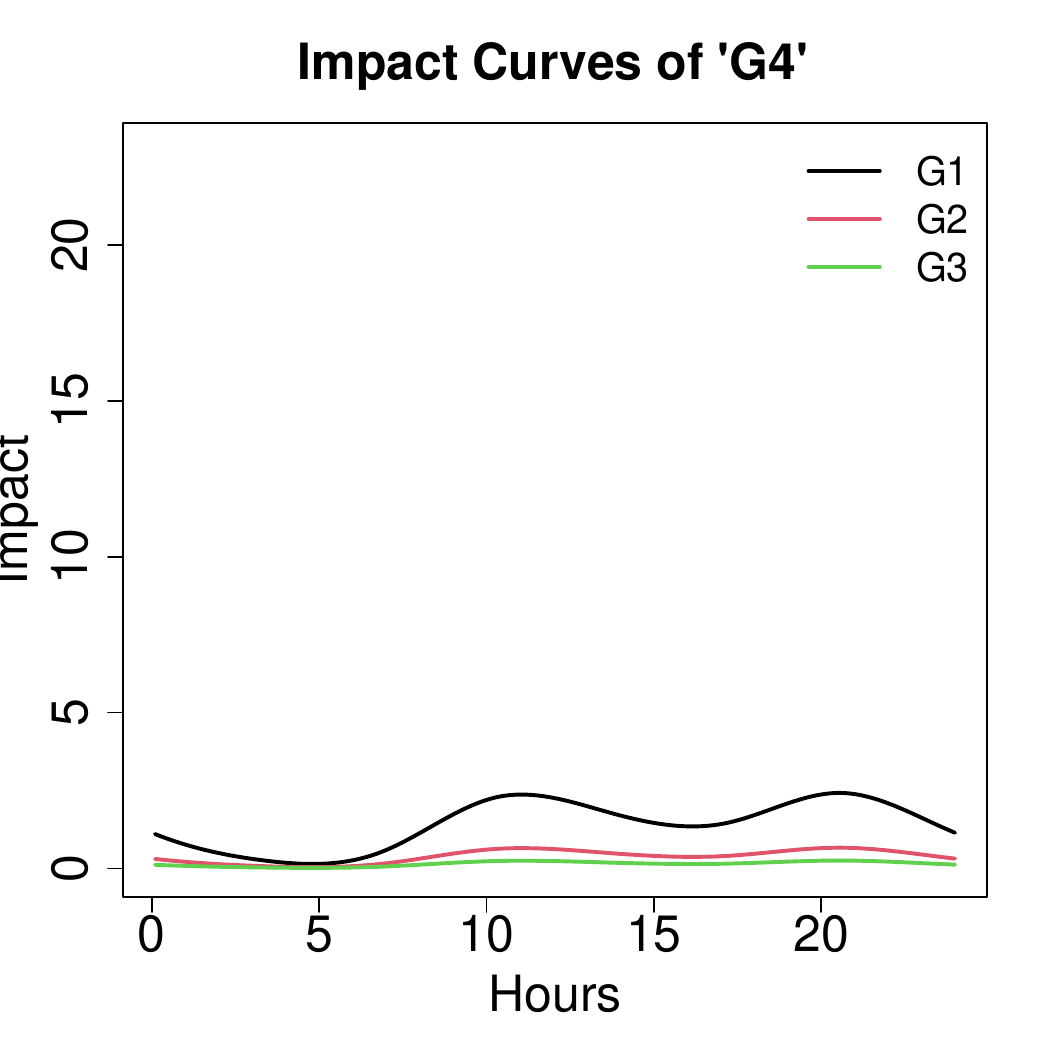}
			\caption{\small Barplot of top 20 and top 100 influential users.} \label{influence}
		\end{figure}

		To shed more light on interactions among different groups, Figure~\ref{influence} gives impact curves of each group on the other groups as defined in~\eqref{E_Sgg_count}.  As expected, both Groups 1 and 4 have rather limited impacts on other groups throughout the day. It appears that the most influential Group 3 has the largest impact on the Group 2, suggesting that users in Group 2 may have the greatest need for timely information provided by users in Group 3. Group 4 is impacted more heavily than Group 1 by Group 3, probably because users in Group 4 are relatively more active and hence can react to information from Group 1 more quickly. Lastly, the impact of Group 3 on Groups 1 and 4 are rather similar, an interesting phenomenon that may require a deeper look. Overall, the influence power plots and impact curve plots reveal some interesting interactive patterns among social network users, suggesting potential usefulness of the proposed GNHP model.
		
{Finally, in Section~B.5 of the supplementary material, we show the differences between groups obtained from GNHP and some simple alternative algorithms. More importantly, we remark that GNHP provides numerical quantification of user influence powers while others cannot.}
		
		\csection{DISCUSSION}
		
		\label{sec::discuss}
		We propose a group network Hawkes process (GNHP) that is suitable for analyzing the dynamic behavior patterns of heterogeneous users in a large network.
The GNHP model extends the classical Hawkes model by utilizing the network structure and introduces a latent group structure to account for heterogeneity among network users.
Theoretical properties are thoroughly investigated and a computationally efficient EM algorithm is proposed.
The GNHP is highly interpretable, as we have demonstrated through an application to a Sina Weibo dataset.
%		To conclude the article, we provide a discussion about potential applications and future research topics as follows.

The proposed model is suitable for abundant applications such as crime pattern analysis and financial risk management, as long as the network structure can be identified. Several research topics can be pursued for future studies. {First, in many applications, the background intensities may be more complicated with multi-scale seasonalities such as daily, weekly, and annual seasonalities, in which case we can consider an additive background intensity $\mu(t)=\sum_{j=1}^{J}\mu_{g,j}(t-\floor{t/\omega_j}{\omega_j})$ with an increasing $\omega_1<\omega_2<\cdots<\omega_J$ and each $\mu_{g,j}$ can be approximated by periodic splines. An alternative remedy is to introduce some time-dependent covariates such as ``day of the week" or categorical variables such as ``weekend or not?" and ``holiday or not?". Our theory can be easily modified for both extensions. One can further introduce user-specific covariates such as ``age", ``gender" and ``occupation" into the background intensities, but such extensions require non-trivial modifications of the current theory. Second, it will also be interesting to find a data-driven method to detect the period parameter $\omega$ if it is unknown.}
%\blue{In addition, we focus on the Hawkes process with excitation in this work. It is interesting to consider more generalized
	%Hawkes process with both excitation and inhibition \citep{chen2017multivariate,cai2020latent} as a future research topic.}
Next, while we have only considered one type of user behavior, it would be interesting to incorporate multi-type user behaviors into the model for analyzing data collected from a social network with more complicated structures.
Finally, the network structure in the current framework is assumed to be known and fixed. It is also of great interest to extend the current model to be suitable for networks whose topological structures are evolving over time.

\section*{Acknowledgment}		
	Xuening Zhu is supported by the National Natural Science Foundation of China (nos. 72222009, 71991472).
Guanhua Fang is partly supported by the National Natural Science Foundation of China (nos. 12301376) and he acknowledges the Terremoto high performance computing cluster service of Columbia University.	The authors report there are no competing interests to declare.

		\baselineskip 14.2pt
		
		\bibliographystyle{asa}
		\bibliography{Hawkes}
		
	\end{document}